\documentclass[longauth]{aaEC}

\usepackage{graphicx}
\usepackage{natbib}
\usepackage{scalerel}

\usepackage[table]{xcolor}

\usepackage{txfonts}
\usepackage[pdfencoding=auto,psdextra]{hyperref}
\hypersetup{
    colorlinks=true,
    linkcolor=blue,
    filecolor=magenta,      
    urlcolor=blue,
    citecolor=blue
}
\usepackage{pageslts}
\makeatletter
\renewcommand*\aa@pageof{, page \thepage{} of \pageref*{LastPage}}
\makeatother

\usepackage[utf8]{inputenc}

\usepackage[switch, modulo]{lineno}

\usepackage{euclid}

\usepackage{csquotes}

\usepackage{amsmath}

\newcommand{\flcgs}{\si{{\rm erg}\,{\rm s^{-1}}\,{\rm cm^{-2}}}\xspace}

\newcommand{\flcgsA}{\si{{\rm erg}\,{\rm s^{-1}}\,{\rm cm^{-2}}\,{\AA^{-1}}}\xspace}

\newcommand{\halpha}{H$\alpha$\xspace}
\newcommand{\hbeta}{H$\beta$\xspace}
\newcommand{\pabeta}{Pa$\beta$\xspace}
\newcommand{\pagamma}{Pa$\gamma$\xspace}

\newcommand{\oii}{[O~\textsc{ii}]\xspace}
\newcommand{\oiiia}{[O~\textsc{iii}]$\lambda 4959$\xspace}
\newcommand{\oiii}{[O~\textsc{iii}]\xspace}
\newcommand{\nii}{[N~\textsc{ii}]\xspace}
\newcommand{\niib}{[N~\textsc{ii}]$\lambda6584$\xspace}
\newcommand{\sii}{[S~\textsc{ii}]\xspace}
\newcommand{\siiia}{[S~\textsc{iii}]$\lambda 9069$\xspace}
\newcommand{\siiib}{[S~\textsc{iii}]$\lambda 9531$\xspace}

\newcommand{\hanii}{H$\alpha$+[N~\textsc{ii}]\xspace}

\newcommand{\ewssf}{EWS-SF$_\textrm{sim}$\xspace}
\newcommand{\edssf}{EDS-SF$_\textrm{sim}$\xspace}

\newcommand{\ztrue}{$z_\textrm{true}$\xspace}
\newcommand{\zmeas}{$z_\textrm{meas}$\xspace}

\newcommand{\zprob}{$z_\textrm{prob}$\xspace}
\newcommand{\dz}{$\Delta z/(1+z)$\xspace}
\newcommand{\dzabs}{$|\Delta z|/(1+z)$\xspace}

\newcommand{\srate}{\( \mathcal{SR} \)\xspace}
\newcommand{\cont}{\( \mathcal{C} \)\xspace}
\newcommand{\bm}{\( \mathcal{BM} \)\xspace}

\newcommand{\stsp}{{\small\texttt{SpectraPyle}}\xspace}
\newcommand{\fssp}{{\small\texttt{FastSpec}}\xspace}
\newcommand{\mambo}{{\small\texttt{MAMBO}}\xspace}

\begin{document}
%
%


\title{\Euclid preparation}
\subtitle{Predicting star-forming galaxy scaling relations with the spectral stacking code {\texttt{SpectraPyle}}}

\newcommand{\orcid}[1]{} 
\author{Euclid Collaboration: S.~Quai\orcid{0000-0002-0449-8163}\thanks{\email{salvatore.quai@unibo.it}}\inst{\ref{aff1},\ref{aff2}}
\and L.~Pozzetti\orcid{0000-0001-7085-0412}\inst{\ref{aff2}}
\and M.~Talia\orcid{0000-0003-4352-2063}\inst{\ref{aff1},\ref{aff2}}
\and C.~Mancini\orcid{0000-0002-4297-0561}\inst{\ref{aff3}}
\and P.~Cassata\orcid{0000-0002-6716-4400}\inst{\ref{aff4},\ref{aff5}}
\and L.~Gabarra\orcid{0000-0002-8486-8856}\inst{\ref{aff6}}
\and V.~Le~Brun\orcid{0000-0002-5027-1939}\inst{\ref{aff7}}
\and M.~Bolzonella\orcid{0000-0003-3278-4607}\inst{\ref{aff2}}
\and E.~Rossetti\orcid{0000-0003-0238-4047}\inst{\ref{aff8}}
\and S.~Kruk\orcid{0000-0001-8010-8879}\inst{\ref{aff9}}
\and B.~R.~Granett\orcid{0000-0003-2694-9284}\inst{\ref{aff10}}
\and C.~Scarlata\orcid{0000-0002-9136-8876}\inst{\ref{aff11}}
\and M.~Moresco\orcid{0000-0002-7616-7136}\inst{\ref{aff1},\ref{aff2}}
\and G.~Zamorani\orcid{0000-0002-2318-301X}\inst{\ref{aff2}}
\and D.~Vergani\orcid{0000-0003-0898-2216}\inst{\ref{aff2}}
\and X.~Lopez~Lopez\orcid{0009-0008-5194-5908}\inst{\ref{aff1},\ref{aff2}}
\and A.~Enia\orcid{0000-0002-0200-2857}\inst{\ref{aff8},\ref{aff2}}
\and E.~Daddi\orcid{0000-0002-3331-9590}\inst{\ref{aff12}}
\and V.~Allevato\orcid{0000-0001-7232-5152}\inst{\ref{aff13}}
\and I.~A.~Zinchenko\orcid{0000-0002-2944-2449}\inst{\ref{aff14}}
\and M.~Magliocchetti\orcid{0000-0001-9158-4838}\inst{\ref{aff15}}
\and M.~Siudek\orcid{0000-0002-2949-2155}\inst{\ref{aff16},\ref{aff17}}
\and L.~Bisigello\orcid{0000-0003-0492-4924}\inst{\ref{aff5}}
\and G.~De~Lucia\orcid{0000-0002-6220-9104}\inst{\ref{aff18}}
\and H.~J.~Dickinson\orcid{0000-0003-0475-008X}\inst{\ref{aff19}}
\and E.~Lusso\orcid{0000-0003-0083-1157}\inst{\ref{aff20},\ref{aff21}}
\and M.~Hirschmann\orcid{0000-0002-3301-3321}\inst{\ref{aff22}}
\and A.~Cimatti\inst{\ref{aff23}}
\and L.~Wang\orcid{0000-0002-6736-9158}\inst{\ref{aff24},\ref{aff25}}
\and J.~G.~Sorce\orcid{0000-0002-2307-2432}\inst{\ref{aff26},\ref{aff27}}
\and N.~Aghanim\orcid{0000-0002-6688-8992}\inst{\ref{aff27}}
\and A.~Amara\inst{\ref{aff28}}
\and S.~Andreon\orcid{0000-0002-2041-8784}\inst{\ref{aff10}}
\and N.~Auricchio\orcid{0000-0003-4444-8651}\inst{\ref{aff2}}
\and C.~Baccigalupi\orcid{0000-0002-8211-1630}\inst{\ref{aff29},\ref{aff18},\ref{aff30},\ref{aff31}}
\and M.~Baldi\orcid{0000-0003-4145-1943}\inst{\ref{aff8},\ref{aff2},\ref{aff32}}
\and S.~Bardelli\orcid{0000-0002-8900-0298}\inst{\ref{aff2}}
\and A.~Biviano\orcid{0000-0002-0857-0732}\inst{\ref{aff18},\ref{aff29}}
\and E.~Branchini\orcid{0000-0002-0808-6908}\inst{\ref{aff33},\ref{aff34},\ref{aff10}}
\and M.~Brescia\orcid{0000-0001-9506-5680}\inst{\ref{aff35},\ref{aff13}}
\and J.~Brinchmann\orcid{0000-0003-4359-8797}\inst{\ref{aff36},\ref{aff37},\ref{aff38}}
\and S.~Camera\orcid{0000-0003-3399-3574}\inst{\ref{aff39},\ref{aff40},\ref{aff41}}
\and G.~Ca\~nas-Herrera\orcid{0000-0003-2796-2149}\inst{\ref{aff42},\ref{aff43},\ref{aff44}}
\and V.~Capobianco\orcid{0000-0002-3309-7692}\inst{\ref{aff41}}
\and C.~Carbone\orcid{0000-0003-0125-3563}\inst{\ref{aff3}}
\and J.~Carretero\orcid{0000-0002-3130-0204}\inst{\ref{aff45},\ref{aff46}}
\and S.~Casas\orcid{0000-0002-4751-5138}\inst{\ref{aff47}}
\and M.~Castellano\orcid{0000-0001-9875-8263}\inst{\ref{aff48}}
\and G.~Castignani\orcid{0000-0001-6831-0687}\inst{\ref{aff2}}
\and S.~Cavuoti\orcid{0000-0002-3787-4196}\inst{\ref{aff13},\ref{aff49}}
\and K.~C.~Chambers\orcid{0000-0001-6965-7789}\inst{\ref{aff50}}
\and C.~Colodro-Conde\inst{\ref{aff51}}
\and G.~Congedo\orcid{0000-0003-2508-0046}\inst{\ref{aff52}}
\and C.~J.~Conselice\orcid{0000-0003-1949-7638}\inst{\ref{aff53}}
\and L.~Conversi\orcid{0000-0002-6710-8476}\inst{\ref{aff54},\ref{aff9}}
\and Y.~Copin\orcid{0000-0002-5317-7518}\inst{\ref{aff55}}
\and F.~Courbin\orcid{0000-0003-0758-6510}\inst{\ref{aff56},\ref{aff57}}
\and H.~M.~Courtois\orcid{0000-0003-0509-1776}\inst{\ref{aff58}}
\and A.~Da~Silva\orcid{0000-0002-6385-1609}\inst{\ref{aff59},\ref{aff60}}
\and H.~Degaudenzi\orcid{0000-0002-5887-6799}\inst{\ref{aff61}}
\and S.~de~la~Torre\inst{\ref{aff7}}
\and H.~Dole\orcid{0000-0002-9767-3839}\inst{\ref{aff27}}
\and M.~Douspis\orcid{0000-0003-4203-3954}\inst{\ref{aff27}}
\and F.~Dubath\orcid{0000-0002-6533-2810}\inst{\ref{aff61}}
\and X.~Dupac\inst{\ref{aff9}}
\and S.~Dusini\orcid{0000-0002-1128-0664}\inst{\ref{aff62}}
\and A.~Ealet\orcid{0000-0003-3070-014X}\inst{\ref{aff55}}
\and S.~Escoffier\orcid{0000-0002-2847-7498}\inst{\ref{aff63}}
\and M.~Farina\orcid{0000-0002-3089-7846}\inst{\ref{aff15}}
\and R.~Farinelli\inst{\ref{aff2}}
\and F.~Faustini\orcid{0000-0001-6274-5145}\inst{\ref{aff48},\ref{aff64}}
\and S.~Ferriol\inst{\ref{aff55}}
\and F.~Finelli\orcid{0000-0002-6694-3269}\inst{\ref{aff2},\ref{aff65}}
\and N.~Fourmanoit\orcid{0009-0005-6816-6925}\inst{\ref{aff63}}
\and M.~Frailis\orcid{0000-0002-7400-2135}\inst{\ref{aff18}}
\and E.~Franceschi\orcid{0000-0002-0585-6591}\inst{\ref{aff2}}
\and S.~Galeotta\orcid{0000-0002-3748-5115}\inst{\ref{aff18}}
\and K.~George\orcid{0000-0002-1734-8455}\inst{\ref{aff14}}
\and W.~Gillard\orcid{0000-0003-4744-9748}\inst{\ref{aff63}}
\and B.~Gillis\orcid{0000-0002-4478-1270}\inst{\ref{aff52}}
\and C.~Giocoli\orcid{0000-0002-9590-7961}\inst{\ref{aff2},\ref{aff32}}
\and J.~Gracia-Carpio\inst{\ref{aff66}}
\and A.~Grazian\orcid{0000-0002-5688-0663}\inst{\ref{aff5}}
\and F.~Grupp\inst{\ref{aff66},\ref{aff14}}
\and L.~Guzzo\orcid{0000-0001-8264-5192}\inst{\ref{aff67},\ref{aff10},\ref{aff68}}
\and S.~V.~H.~Haugan\orcid{0000-0001-9648-7260}\inst{\ref{aff69}}
\and W.~Holmes\inst{\ref{aff70}}
\and I.~M.~Hook\orcid{0000-0002-2960-978X}\inst{\ref{aff71}}
\and F.~Hormuth\inst{\ref{aff72}}
\and A.~Hornstrup\orcid{0000-0002-3363-0936}\inst{\ref{aff73},\ref{aff74}}
\and P.~Hudelot\inst{\ref{aff75}}
\and K.~Jahnke\orcid{0000-0003-3804-2137}\inst{\ref{aff76}}
\and M.~Jhabvala\inst{\ref{aff77}}
\and B.~Joachimi\orcid{0000-0001-7494-1303}\inst{\ref{aff78}}
\and E.~Keih\"anen\orcid{0000-0003-1804-7715}\inst{\ref{aff79}}
\and S.~Kermiche\orcid{0000-0002-0302-5735}\inst{\ref{aff63}}
\and A.~Kiessling\orcid{0000-0002-2590-1273}\inst{\ref{aff70}}
\and B.~Kubik\orcid{0009-0006-5823-4880}\inst{\ref{aff55}}
\and M.~K\"ummel\orcid{0000-0003-2791-2117}\inst{\ref{aff14}}
\and M.~Kunz\orcid{0000-0002-3052-7394}\inst{\ref{aff80}}
\and H.~Kurki-Suonio\orcid{0000-0002-4618-3063}\inst{\ref{aff81},\ref{aff82}}
\and A.~M.~C.~Le~Brun\orcid{0000-0002-0936-4594}\inst{\ref{aff83}}
\and S.~Ligori\orcid{0000-0003-4172-4606}\inst{\ref{aff41}}
\and P.~B.~Lilje\orcid{0000-0003-4324-7794}\inst{\ref{aff69}}
\and V.~Lindholm\orcid{0000-0003-2317-5471}\inst{\ref{aff81},\ref{aff82}}
\and I.~Lloro\orcid{0000-0001-5966-1434}\inst{\ref{aff84}}
\and G.~Mainetti\orcid{0000-0003-2384-2377}\inst{\ref{aff85}}
\and D.~Maino\inst{\ref{aff67},\ref{aff3},\ref{aff68}}
\and E.~Maiorano\orcid{0000-0003-2593-4355}\inst{\ref{aff2}}
\and O.~Mansutti\orcid{0000-0001-5758-4658}\inst{\ref{aff18}}
\and S.~Marcin\inst{\ref{aff86}}
\and O.~Marggraf\orcid{0000-0001-7242-3852}\inst{\ref{aff87}}
\and M.~Martinelli\orcid{0000-0002-6943-7732}\inst{\ref{aff48},\ref{aff88}}
\and N.~Martinet\orcid{0000-0003-2786-7790}\inst{\ref{aff7}}
\and F.~Marulli\orcid{0000-0002-8850-0303}\inst{\ref{aff1},\ref{aff2},\ref{aff32}}
\and R.~J.~Massey\orcid{0000-0002-6085-3780}\inst{\ref{aff89}}
\and E.~Medinaceli\orcid{0000-0002-4040-7783}\inst{\ref{aff2}}
\and S.~Mei\orcid{0000-0002-2849-559X}\inst{\ref{aff90},\ref{aff91}}
\and M.~Melchior\inst{\ref{aff92}}
\and Y.~Mellier\inst{\ref{aff93},\ref{aff75}}
\and M.~Meneghetti\orcid{0000-0003-1225-7084}\inst{\ref{aff2},\ref{aff32}}
\and E.~Merlin\orcid{0000-0001-6870-8900}\inst{\ref{aff48}}
\and G.~Meylan\inst{\ref{aff94}}
\and A.~Mora\orcid{0000-0002-1922-8529}\inst{\ref{aff95}}
\and L.~Moscardini\orcid{0000-0002-3473-6716}\inst{\ref{aff1},\ref{aff2},\ref{aff32}}
\and C.~Neissner\orcid{0000-0001-8524-4968}\inst{\ref{aff96},\ref{aff46}}
\and S.-M.~Niemi\orcid{0009-0005-0247-0086}\inst{\ref{aff42}}
\and C.~Padilla\orcid{0000-0001-7951-0166}\inst{\ref{aff96}}
\and S.~Paltani\orcid{0000-0002-8108-9179}\inst{\ref{aff61}}
\and F.~Pasian\orcid{0000-0002-4869-3227}\inst{\ref{aff18}}
\and K.~Pedersen\inst{\ref{aff97}}
\and W.~J.~Percival\orcid{0000-0002-0644-5727}\inst{\ref{aff98},\ref{aff99},\ref{aff100}}
\and V.~Pettorino\inst{\ref{aff42}}
\and S.~Pires\orcid{0000-0002-0249-2104}\inst{\ref{aff12}}
\and G.~Polenta\orcid{0000-0003-4067-9196}\inst{\ref{aff64}}
\and M.~Poncet\inst{\ref{aff101}}
\and L.~A.~Popa\inst{\ref{aff102}}
\and F.~Raison\orcid{0000-0002-7819-6918}\inst{\ref{aff66}}
\and R.~Rebolo\orcid{0000-0003-3767-7085}\inst{\ref{aff51},\ref{aff103},\ref{aff104}}
\and A.~Renzi\orcid{0000-0001-9856-1970}\inst{\ref{aff4},\ref{aff62}}
\and J.~Rhodes\orcid{0000-0002-4485-8549}\inst{\ref{aff70}}
\and G.~Riccio\inst{\ref{aff13}}
\and E.~Romelli\orcid{0000-0003-3069-9222}\inst{\ref{aff18}}
\and M.~Roncarelli\orcid{0000-0001-9587-7822}\inst{\ref{aff2}}
\and R.~Saglia\orcid{0000-0003-0378-7032}\inst{\ref{aff14},\ref{aff66}}
\and Z.~Sakr\orcid{0000-0002-4823-3757}\inst{\ref{aff105},\ref{aff106},\ref{aff107}}
\and A.~G.~S\'anchez\orcid{0000-0003-1198-831X}\inst{\ref{aff66}}
\and D.~Sapone\orcid{0000-0001-7089-4503}\inst{\ref{aff108}}
\and B.~Sartoris\orcid{0000-0003-1337-5269}\inst{\ref{aff14},\ref{aff18}}
\and P.~Schneider\orcid{0000-0001-8561-2679}\inst{\ref{aff87}}
\and T.~Schrabback\orcid{0000-0002-6987-7834}\inst{\ref{aff109}}
\and M.~Scodeggio\inst{\ref{aff3}}
\and A.~Secroun\orcid{0000-0003-0505-3710}\inst{\ref{aff63}}
\and E.~Sefusatti\orcid{0000-0003-0473-1567}\inst{\ref{aff18},\ref{aff29},\ref{aff30}}
\and G.~Seidel\orcid{0000-0003-2907-353X}\inst{\ref{aff76}}
\and M.~Seiffert\orcid{0000-0002-7536-9393}\inst{\ref{aff70}}
\and S.~Serrano\orcid{0000-0002-0211-2861}\inst{\ref{aff110},\ref{aff111},\ref{aff17}}
\and P.~Simon\inst{\ref{aff87}}
\and C.~Sirignano\orcid{0000-0002-0995-7146}\inst{\ref{aff4},\ref{aff62}}
\and G.~Sirri\orcid{0000-0003-2626-2853}\inst{\ref{aff32}}
\and L.~Stanco\orcid{0000-0002-9706-5104}\inst{\ref{aff62}}
\and J.-L.~Starck\orcid{0000-0003-2177-7794}\inst{\ref{aff12}}
\and J.~Steinwagner\orcid{0000-0001-7443-1047}\inst{\ref{aff66}}
\and P.~Tallada-Cresp\'{i}\orcid{0000-0002-1336-8328}\inst{\ref{aff45},\ref{aff46}}
\and D.~Tavagnacco\orcid{0000-0001-7475-9894}\inst{\ref{aff18}}
\and A.~N.~Taylor\inst{\ref{aff52}}
\and H.~I.~Teplitz\orcid{0000-0002-7064-5424}\inst{\ref{aff112}}
\and I.~Tereno\orcid{0000-0002-4537-6218}\inst{\ref{aff59},\ref{aff113}}
\and S.~Toft\orcid{0000-0003-3631-7176}\inst{\ref{aff114},\ref{aff115}}
\and R.~Toledo-Moreo\orcid{0000-0002-2997-4859}\inst{\ref{aff116}}
\and F.~Torradeflot\orcid{0000-0003-1160-1517}\inst{\ref{aff46},\ref{aff45}}
\and I.~Tutusaus\orcid{0000-0002-3199-0399}\inst{\ref{aff17},\ref{aff110},\ref{aff106}}
\and L.~Valenziano\orcid{0000-0002-1170-0104}\inst{\ref{aff2},\ref{aff65}}
\and J.~Valiviita\orcid{0000-0001-6225-3693}\inst{\ref{aff81},\ref{aff82}}
\and T.~Vassallo\orcid{0000-0001-6512-6358}\inst{\ref{aff14},\ref{aff18}}
\and G.~Verdoes~Kleijn\orcid{0000-0001-5803-2580}\inst{\ref{aff25}}
\and A.~Veropalumbo\orcid{0000-0003-2387-1194}\inst{\ref{aff10},\ref{aff34},\ref{aff33}}
\and D.~Vibert\orcid{0009-0008-0607-631X}\inst{\ref{aff7}}
\and Y.~Wang\orcid{0000-0002-4749-2984}\inst{\ref{aff112}}
\and J.~Weller\orcid{0000-0002-8282-2010}\inst{\ref{aff14},\ref{aff66}}
\and E.~Zucca\orcid{0000-0002-5845-8132}\inst{\ref{aff2}}
\and M.~Ballardini\orcid{0000-0003-4481-3559}\inst{\ref{aff117},\ref{aff118},\ref{aff2}}
\and E.~Bozzo\orcid{0000-0002-8201-1525}\inst{\ref{aff61}}
\and C.~Burigana\orcid{0000-0002-3005-5796}\inst{\ref{aff119},\ref{aff65}}
\and R.~Cabanac\orcid{0000-0001-6679-2600}\inst{\ref{aff106}}
\and A.~Cappi\inst{\ref{aff2},\ref{aff120}}
\and D.~Di~Ferdinando\inst{\ref{aff32}}
\and J.~A.~Escartin~Vigo\inst{\ref{aff66}}
\and M.~Huertas-Company\orcid{0000-0002-1416-8483}\inst{\ref{aff51},\ref{aff16},\ref{aff121},\ref{aff122}}
\and J.~Mart\'{i}n-Fleitas\orcid{0000-0002-8594-569X}\inst{\ref{aff123}}
\and S.~Matthew\orcid{0000-0001-8448-1697}\inst{\ref{aff52}}
\and N.~Mauri\orcid{0000-0001-8196-1548}\inst{\ref{aff23},\ref{aff32}}
\and R.~B.~Metcalf\orcid{0000-0003-3167-2574}\inst{\ref{aff1},\ref{aff2}}
\and A.~Pezzotta\orcid{0000-0003-0726-2268}\inst{\ref{aff124},\ref{aff66}}
\and M.~P\"ontinen\orcid{0000-0001-5442-2530}\inst{\ref{aff81}}
\and C.~Porciani\orcid{0000-0002-7797-2508}\inst{\ref{aff87}}
\and I.~Risso\orcid{0000-0003-2525-7761}\inst{\ref{aff125}}
\and V.~Scottez\orcid{0009-0008-3864-940X}\inst{\ref{aff93},\ref{aff126}}
\and M.~Sereno\orcid{0000-0003-0302-0325}\inst{\ref{aff2},\ref{aff32}}
\and M.~Tenti\orcid{0000-0002-4254-5901}\inst{\ref{aff32}}
\and M.~Viel\orcid{0000-0002-2642-5707}\inst{\ref{aff29},\ref{aff18},\ref{aff31},\ref{aff30},\ref{aff127}}
\and M.~Wiesmann\orcid{0009-0000-8199-5860}\inst{\ref{aff69}}
\and Y.~Akrami\orcid{0000-0002-2407-7956}\inst{\ref{aff128},\ref{aff129}}
\and I.~T.~Andika\orcid{0000-0001-6102-9526}\inst{\ref{aff130},\ref{aff131}}
\and S.~Anselmi\orcid{0000-0002-3579-9583}\inst{\ref{aff62},\ref{aff4},\ref{aff132}}
\and M.~Archidiacono\orcid{0000-0003-4952-9012}\inst{\ref{aff67},\ref{aff68}}
\and F.~Atrio-Barandela\orcid{0000-0002-2130-2513}\inst{\ref{aff133}}
\and P.~Bergamini\orcid{0000-0003-1383-9414}\inst{\ref{aff67},\ref{aff2}}
\and D.~Bertacca\orcid{0000-0002-2490-7139}\inst{\ref{aff4},\ref{aff5},\ref{aff62}}
\and M.~Bethermin\orcid{0000-0002-3915-2015}\inst{\ref{aff134}}
\and A.~Blanchard\orcid{0000-0001-8555-9003}\inst{\ref{aff106}}
\and L.~Blot\orcid{0000-0002-9622-7167}\inst{\ref{aff135},\ref{aff83}}
\and S.~Borgani\orcid{0000-0001-6151-6439}\inst{\ref{aff136},\ref{aff29},\ref{aff18},\ref{aff30},\ref{aff127}}
\and M.~L.~Brown\orcid{0000-0002-0370-8077}\inst{\ref{aff53}}
\and S.~Bruton\orcid{0000-0002-6503-5218}\inst{\ref{aff137}}
\and A.~Calabro\orcid{0000-0003-2536-1614}\inst{\ref{aff48}}
\and B.~Camacho~Quevedo\orcid{0000-0002-8789-4232}\inst{\ref{aff29},\ref{aff31},\ref{aff18},\ref{aff110},\ref{aff17}}
\and F.~Caro\inst{\ref{aff48}}
\and C.~S.~Carvalho\inst{\ref{aff113}}
\and T.~Castro\orcid{0000-0002-6292-3228}\inst{\ref{aff18},\ref{aff30},\ref{aff29},\ref{aff127}}
\and F.~Cogato\orcid{0000-0003-4632-6113}\inst{\ref{aff1},\ref{aff2}}
\and S.~Conseil\orcid{0000-0002-3657-4191}\inst{\ref{aff55}}
\and T.~Contini\orcid{0000-0003-0275-938X}\inst{\ref{aff106}}
\and A.~R.~Cooray\orcid{0000-0002-3892-0190}\inst{\ref{aff138}}
\and O.~Cucciati\orcid{0000-0002-9336-7551}\inst{\ref{aff2}}
\and S.~Davini\orcid{0000-0003-3269-1718}\inst{\ref{aff34}}
\and G.~Desprez\orcid{0000-0001-8325-1742}\inst{\ref{aff25}}
\and A.~D\'iaz-S\'anchez\orcid{0000-0003-0748-4768}\inst{\ref{aff139}}
\and J.~J.~Diaz\orcid{0000-0003-2101-1078}\inst{\ref{aff51}}
\and S.~Di~Domizio\orcid{0000-0003-2863-5895}\inst{\ref{aff33},\ref{aff34}}
\and J.~M.~Diego\orcid{0000-0001-9065-3926}\inst{\ref{aff140}}
\and Y.~Fang\inst{\ref{aff14}}
\and A.~G.~Ferrari\orcid{0009-0005-5266-4110}\inst{\ref{aff32}}
\and A.~Finoguenov\orcid{0000-0002-4606-5403}\inst{\ref{aff81}}
\and A.~Fontana\orcid{0000-0003-3820-2823}\inst{\ref{aff48}}
\and F.~Fontanot\orcid{0000-0003-4744-0188}\inst{\ref{aff18},\ref{aff29}}
\and A.~Franco\orcid{0000-0002-4761-366X}\inst{\ref{aff141},\ref{aff142},\ref{aff143}}
\and K.~Ganga\orcid{0000-0001-8159-8208}\inst{\ref{aff90}}
\and J.~Garc\'ia-Bellido\orcid{0000-0002-9370-8360}\inst{\ref{aff128}}
\and T.~Gasparetto\orcid{0000-0002-7913-4866}\inst{\ref{aff18}}
\and V.~Gautard\inst{\ref{aff144}}
\and E.~Gaztanaga\orcid{0000-0001-9632-0815}\inst{\ref{aff17},\ref{aff110},\ref{aff145}}
\and F.~Giacomini\orcid{0000-0002-3129-2814}\inst{\ref{aff32}}
\and F.~Gianotti\orcid{0000-0003-4666-119X}\inst{\ref{aff2}}
\and G.~Gozaliasl\orcid{0000-0002-0236-919X}\inst{\ref{aff146},\ref{aff81}}
\and M.~Guidi\orcid{0000-0001-9408-1101}\inst{\ref{aff8},\ref{aff2}}
\and C.~M.~Gutierrez\orcid{0000-0001-7854-783X}\inst{\ref{aff147}}
\and A.~Hall\orcid{0000-0002-3139-8651}\inst{\ref{aff52}}
\and S.~Hemmati\orcid{0000-0003-2226-5395}\inst{\ref{aff148}}
\and C.~Hern\'andez-Monteagudo\orcid{0000-0001-5471-9166}\inst{\ref{aff104},\ref{aff51}}
\and H.~Hildebrandt\orcid{0000-0002-9814-3338}\inst{\ref{aff149}}
\and J.~Hjorth\orcid{0000-0002-4571-2306}\inst{\ref{aff97}}
\and J.~J.~E.~Kajava\orcid{0000-0002-3010-8333}\inst{\ref{aff150},\ref{aff151}}
\and Y.~Kang\orcid{0009-0000-8588-7250}\inst{\ref{aff61}}
\and V.~Kansal\orcid{0000-0002-4008-6078}\inst{\ref{aff152},\ref{aff153}}
\and D.~Karagiannis\orcid{0000-0002-4927-0816}\inst{\ref{aff117},\ref{aff154}}
\and K.~Kiiveri\inst{\ref{aff79}}
\and C.~C.~Kirkpatrick\inst{\ref{aff79}}
\and L.~Legrand\orcid{0000-0003-0610-5252}\inst{\ref{aff155},\ref{aff156}}
\and M.~Lembo\orcid{0000-0002-5271-5070}\inst{\ref{aff75}}
\and F.~Lepori\orcid{0009-0000-5061-7138}\inst{\ref{aff157}}
\and G.~Leroy\orcid{0009-0004-2523-4425}\inst{\ref{aff158},\ref{aff89}}
\and G.~F.~Lesci\orcid{0000-0002-4607-2830}\inst{\ref{aff1},\ref{aff2}}
\and J.~Lesgourgues\orcid{0000-0001-7627-353X}\inst{\ref{aff47}}
\and L.~Leuzzi\orcid{0009-0006-4479-7017}\inst{\ref{aff2}}
\and T.~I.~Liaudat\orcid{0000-0002-9104-314X}\inst{\ref{aff159}}
\and S.~J.~Liu\orcid{0000-0001-7680-2139}\inst{\ref{aff15}}
\and A.~Loureiro\orcid{0000-0002-4371-0876}\inst{\ref{aff160},\ref{aff161}}
\and J.~Macias-Perez\orcid{0000-0002-5385-2763}\inst{\ref{aff162}}
\and G.~Maggio\orcid{0000-0003-4020-4836}\inst{\ref{aff18}}
\and F.~Mannucci\orcid{0000-0002-4803-2381}\inst{\ref{aff20}}
\and R.~Maoli\orcid{0000-0002-6065-3025}\inst{\ref{aff163},\ref{aff48}}
\and C.~J.~A.~P.~Martins\orcid{0000-0002-4886-9261}\inst{\ref{aff164},\ref{aff36}}
\and L.~Maurin\orcid{0000-0002-8406-0857}\inst{\ref{aff27}}
\and M.~Miluzio\inst{\ref{aff9},\ref{aff165}}
\and P.~Monaco\orcid{0000-0003-2083-7564}\inst{\ref{aff136},\ref{aff18},\ref{aff30},\ref{aff29}}
\and C.~Moretti\orcid{0000-0003-3314-8936}\inst{\ref{aff31},\ref{aff127},\ref{aff18},\ref{aff29},\ref{aff30}}
\and G.~Morgante\inst{\ref{aff2}}
\and S.~Nadathur\orcid{0000-0001-9070-3102}\inst{\ref{aff145}}
\and K.~Naidoo\orcid{0000-0002-9182-1802}\inst{\ref{aff145}}
\and A.~Navarro-Alsina\orcid{0000-0002-3173-2592}\inst{\ref{aff87}}
\and S.~Nesseris\orcid{0000-0002-0567-0324}\inst{\ref{aff128}}
\and F.~Passalacqua\orcid{0000-0002-8606-4093}\inst{\ref{aff4},\ref{aff62}}
\and K.~Paterson\orcid{0000-0001-8340-3486}\inst{\ref{aff76}}
\and L.~Patrizii\inst{\ref{aff32}}
\and A.~Pisani\orcid{0000-0002-6146-4437}\inst{\ref{aff63}}
\and D.~Potter\orcid{0000-0002-0757-5195}\inst{\ref{aff157}}
\and M.~Radovich\orcid{0000-0002-3585-866X}\inst{\ref{aff5}}
\and P.-F.~Rocci\inst{\ref{aff27}}
\and G.~Rodighiero\orcid{0000-0002-9415-2296}\inst{\ref{aff4},\ref{aff5}}
\and S.~Sacquegna\orcid{0000-0002-8433-6630}\inst{\ref{aff142},\ref{aff141},\ref{aff166}}
\and M.~Sahl\'en\orcid{0000-0003-0973-4804}\inst{\ref{aff167}}
\and D.~B.~Sanders\orcid{0000-0002-1233-9998}\inst{\ref{aff50}}
\and E.~Sarpa\orcid{0000-0002-1256-655X}\inst{\ref{aff31},\ref{aff127},\ref{aff30}}
\and A.~Schneider\orcid{0000-0001-7055-8104}\inst{\ref{aff157}}
\and D.~Sciotti\orcid{0009-0008-4519-2620}\inst{\ref{aff48},\ref{aff88}}
\and E.~Sellentin\inst{\ref{aff168},\ref{aff44}}
\and F.~Shankar\orcid{0000-0001-8973-5051}\inst{\ref{aff169}}
\and L.~C.~Smith\orcid{0000-0002-3259-2771}\inst{\ref{aff170}}
\and K.~Tanidis\orcid{0000-0001-9843-5130}\inst{\ref{aff6}}
\and C.~Tao\orcid{0000-0001-7961-8177}\inst{\ref{aff63}}
\and G.~Testera\inst{\ref{aff34}}
\and R.~Teyssier\orcid{0000-0001-7689-0933}\inst{\ref{aff171}}
\and S.~Tosi\orcid{0000-0002-7275-9193}\inst{\ref{aff33},\ref{aff34},\ref{aff10}}
\and A.~Troja\orcid{0000-0003-0239-4595}\inst{\ref{aff4},\ref{aff62}}
\and M.~Tucci\inst{\ref{aff61}}
\and C.~Valieri\inst{\ref{aff32}}
\and A.~Venhola\orcid{0000-0001-6071-4564}\inst{\ref{aff172}}
\and G.~Verza\orcid{0000-0002-1886-8348}\inst{\ref{aff173}}
\and P.~Vielzeuf\orcid{0000-0003-2035-9339}\inst{\ref{aff63}}
\and N.~A.~Walton\orcid{0000-0003-3983-8778}\inst{\ref{aff170}}}
										   
\institute{Dipartimento di Fisica e Astronomia "Augusto Righi" - Alma Mater Studiorum Universit\`a di Bologna, via Piero Gobetti 93/2, 40129 Bologna, Italy\label{aff1}
\and
INAF-Osservatorio di Astrofisica e Scienza dello Spazio di Bologna, Via Piero Gobetti 93/3, 40129 Bologna, Italy\label{aff2}
\and
INAF-IASF Milano, Via Alfonso Corti 12, 20133 Milano, Italy\label{aff3}
\and
Dipartimento di Fisica e Astronomia "G. Galilei", Universit\`a di Padova, Via Marzolo 8, 35131 Padova, Italy\label{aff4}
\and
INAF-Osservatorio Astronomico di Padova, Via dell'Osservatorio 5, 35122 Padova, Italy\label{aff5}
\and
Department of Physics, Oxford University, Keble Road, Oxford OX1 3RH, UK\label{aff6}
\and
Aix-Marseille Universit\'e, CNRS, CNES, LAM, Marseille, France\label{aff7}
\and
Dipartimento di Fisica e Astronomia, Universit\`a di Bologna, Via Gobetti 93/2, 40129 Bologna, Italy\label{aff8}
\and
ESAC/ESA, Camino Bajo del Castillo, s/n., Urb. Villafranca del Castillo, 28692 Villanueva de la Ca\~nada, Madrid, Spain\label{aff9}
\and
INAF-Osservatorio Astronomico di Brera, Via Brera 28, 20122 Milano, Italy\label{aff10}
\and
Minnesota Institute for Astrophysics, University of Minnesota, 116 Church St SE, Minneapolis, MN 55455, USA\label{aff11}
\and
Universit\'e Paris-Saclay, Universit\'e Paris Cit\'e, CEA, CNRS, AIM, 91191, Gif-sur-Yvette, France\label{aff12}
\and
INAF-Osservatorio Astronomico di Capodimonte, Via Moiariello 16, 80131 Napoli, Italy\label{aff13}
\and
Universit\"ats-Sternwarte M\"unchen, Fakult\"at f\"ur Physik, Ludwig-Maximilians-Universit\"at M\"unchen, Scheinerstrasse 1, 81679 M\"unchen, Germany\label{aff14}
\and
INAF-Istituto di Astrofisica e Planetologia Spaziali, via del Fosso del Cavaliere, 100, 00100 Roma, Italy\label{aff15}
\and
Instituto de Astrof\'isica de Canarias (IAC); Departamento de Astrof\'isica, Universidad de La Laguna (ULL), 38200, La Laguna, Tenerife, Spain\label{aff16}
\and
Institute of Space Sciences (ICE, CSIC), Campus UAB, Carrer de Can Magrans, s/n, 08193 Barcelona, Spain\label{aff17}
\and
INAF-Osservatorio Astronomico di Trieste, Via G. B. Tiepolo 11, 34143 Trieste, Italy\label{aff18}
\and
School of Physical Sciences, The Open University, Milton Keynes, MK7 6AA, UK\label{aff19}
\and
INAF-Osservatorio Astrofisico di Arcetri, Largo E. Fermi 5, 50125, Firenze, Italy\label{aff20}
\and
Dipartimento di Fisica e Astronomia, Universit\`{a} di Firenze, via G. Sansone 1, 50019 Sesto Fiorentino, Firenze, Italy\label{aff21}
\and
Institute of Physics, Laboratory for Galaxy Evolution, Ecole Polytechnique F\'ed\'erale de Lausanne, Observatoire de Sauverny, CH-1290 Versoix, Switzerland\label{aff22}
\and
Dipartimento di Fisica e Astronomia "Augusto Righi" - Alma Mater Studiorum Universit\`a di Bologna, Viale Berti Pichat 6/2, 40127 Bologna, Italy\label{aff23}
\and
SRON Netherlands Institute for Space Research, Landleven 12, 9747 AD, Groningen, The Netherlands\label{aff24}
\and
Kapteyn Astronomical Institute, University of Groningen, PO Box 800, 9700 AV Groningen, The Netherlands\label{aff25}
\and
Univ. Lille, CNRS, Centrale Lille, UMR 9189 CRIStAL, 59000 Lille, France\label{aff26}
\and
Universit\'e Paris-Saclay, CNRS, Institut d'astrophysique spatiale, 91405, Orsay, France\label{aff27}
\and
School of Mathematics and Physics, University of Surrey, Guildford, Surrey, GU2 7XH, UK\label{aff28}
\and
IFPU, Institute for Fundamental Physics of the Universe, via Beirut 2, 34151 Trieste, Italy\label{aff29}
\and
INFN, Sezione di Trieste, Via Valerio 2, 34127 Trieste TS, Italy\label{aff30}
\and
SISSA, International School for Advanced Studies, Via Bonomea 265, 34136 Trieste TS, Italy\label{aff31}
\and
INFN-Sezione di Bologna, Viale Berti Pichat 6/2, 40127 Bologna, Italy\label{aff32}
\and
Dipartimento di Fisica, Universit\`a di Genova, Via Dodecaneso 33, 16146, Genova, Italy\label{aff33}
\and
INFN-Sezione di Genova, Via Dodecaneso 33, 16146, Genova, Italy\label{aff34}
\and
Department of Physics "E. Pancini", University Federico II, Via Cinthia 6, 80126, Napoli, Italy\label{aff35}
\and
Instituto de Astrof\'isica e Ci\^encias do Espa\c{c}o, Universidade do Porto, CAUP, Rua das Estrelas, PT4150-762 Porto, Portugal\label{aff36}
\and
Faculdade de Ci\^encias da Universidade do Porto, Rua do Campo de Alegre, 4150-007 Porto, Portugal\label{aff37}
\and
European Southern Observatory, Karl-Schwarzschild-Str.~2, 85748 Garching, Germany\label{aff38}
\and
Dipartimento di Fisica, Universit\`a degli Studi di Torino, Via P. Giuria 1, 10125 Torino, Italy\label{aff39}
\and
INFN-Sezione di Torino, Via P. Giuria 1, 10125 Torino, Italy\label{aff40}
\and
INAF-Osservatorio Astrofisico di Torino, Via Osservatorio 20, 10025 Pino Torinese (TO), Italy\label{aff41}
\and
European Space Agency/ESTEC, Keplerlaan 1, 2201 AZ Noordwijk, The Netherlands\label{aff42}
\and
Institute Lorentz, Leiden University, Niels Bohrweg 2, 2333 CA Leiden, The Netherlands\label{aff43}
\and
Leiden Observatory, Leiden University, Einsteinweg 55, 2333 CC Leiden, The Netherlands\label{aff44}
\and
Centro de Investigaciones Energ\'eticas, Medioambientales y Tecnol\'ogicas (CIEMAT), Avenida Complutense 40, 28040 Madrid, Spain\label{aff45}
\and
Port d'Informaci\'{o} Cient\'{i}fica, Campus UAB, C. Albareda s/n, 08193 Bellaterra (Barcelona), Spain\label{aff46}
\and
Institute for Theoretical Particle Physics and Cosmology (TTK), RWTH Aachen University, 52056 Aachen, Germany\label{aff47}
\and
INAF-Osservatorio Astronomico di Roma, Via Frascati 33, 00078 Monteporzio Catone, Italy\label{aff48}
\and
INFN section of Naples, Via Cinthia 6, 80126, Napoli, Italy\label{aff49}
\and
Institute for Astronomy, University of Hawaii, 2680 Woodlawn Drive, Honolulu, HI 96822, USA\label{aff50}
\and
Instituto de Astrof\'{\i}sica de Canarias, V\'{\i}a L\'actea, 38205 La Laguna, Tenerife, Spain\label{aff51}
\and
Institute for Astronomy, University of Edinburgh, Royal Observatory, Blackford Hill, Edinburgh EH9 3HJ, UK\label{aff52}
\and
Jodrell Bank Centre for Astrophysics, Department of Physics and Astronomy, University of Manchester, Oxford Road, Manchester M13 9PL, UK\label{aff53}
\and
European Space Agency/ESRIN, Largo Galileo Galilei 1, 00044 Frascati, Roma, Italy\label{aff54}
\and
Universit\'e Claude Bernard Lyon 1, CNRS/IN2P3, IP2I Lyon, UMR 5822, Villeurbanne, F-69100, France\label{aff55}
\and
Institut de Ci\`{e}ncies del Cosmos (ICCUB), Universitat de Barcelona (IEEC-UB), Mart\'{i} i Franqu\`{e}s 1, 08028 Barcelona, Spain\label{aff56}
\and
Instituci\'o Catalana de Recerca i Estudis Avan\c{c}ats (ICREA), Passeig de Llu\'{\i}s Companys 23, 08010 Barcelona, Spain\label{aff57}
\and
UCB Lyon 1, CNRS/IN2P3, IUF, IP2I Lyon, 4 rue Enrico Fermi, 69622 Villeurbanne, France\label{aff58}
\and
Departamento de F\'isica, Faculdade de Ci\^encias, Universidade de Lisboa, Edif\'icio C8, Campo Grande, PT1749-016 Lisboa, Portugal\label{aff59}
\and
Instituto de Astrof\'isica e Ci\^encias do Espa\c{c}o, Faculdade de Ci\^encias, Universidade de Lisboa, Campo Grande, 1749-016 Lisboa, Portugal\label{aff60}
\and
Department of Astronomy, University of Geneva, ch. d'Ecogia 16, 1290 Versoix, Switzerland\label{aff61}
\and
INFN-Padova, Via Marzolo 8, 35131 Padova, Italy\label{aff62}
\and
Aix-Marseille Universit\'e, CNRS/IN2P3, CPPM, Marseille, France\label{aff63}
\and
Space Science Data Center, Italian Space Agency, via del Politecnico snc, 00133 Roma, Italy\label{aff64}
\and
INFN-Bologna, Via Irnerio 46, 40126 Bologna, Italy\label{aff65}
\and
Max Planck Institute for Extraterrestrial Physics, Giessenbachstr. 1, 85748 Garching, Germany\label{aff66}
\and
Dipartimento di Fisica "Aldo Pontremoli", Universit\`a degli Studi di Milano, Via Celoria 16, 20133 Milano, Italy\label{aff67}
\and
INFN-Sezione di Milano, Via Celoria 16, 20133 Milano, Italy\label{aff68}
\and
Institute of Theoretical Astrophysics, University of Oslo, P.O. Box 1029 Blindern, 0315 Oslo, Norway\label{aff69}
\and
Jet Propulsion Laboratory, California Institute of Technology, 4800 Oak Grove Drive, Pasadena, CA, 91109, USA\label{aff70}
\and
Department of Physics, Lancaster University, Lancaster, LA1 4YB, UK\label{aff71}
\and
Felix Hormuth Engineering, Goethestr. 17, 69181 Leimen, Germany\label{aff72}
\and
Technical University of Denmark, Elektrovej 327, 2800 Kgs. Lyngby, Denmark\label{aff73}
\and
Cosmic Dawn Center (DAWN), Denmark\label{aff74}
\and
Institut d'Astrophysique de Paris, UMR 7095, CNRS, and Sorbonne Universit\'e, 98 bis boulevard Arago, 75014 Paris, France\label{aff75}
\and
Max-Planck-Institut f\"ur Astronomie, K\"onigstuhl 17, 69117 Heidelberg, Germany\label{aff76}
\and
NASA Goddard Space Flight Center, Greenbelt, MD 20771, USA\label{aff77}
\and
Department of Physics and Astronomy, University College London, Gower Street, London WC1E 6BT, UK\label{aff78}
\and
Department of Physics and Helsinki Institute of Physics, Gustaf H\"allstr\"omin katu 2, University of Helsinki, 00014 Helsinki, Finland\label{aff79}
\and
Universit\'e de Gen\`eve, D\'epartement de Physique Th\'eorique and Centre for Astroparticle Physics, 24 quai Ernest-Ansermet, CH-1211 Gen\`eve 4, Switzerland\label{aff80}
\and
Department of Physics, P.O. Box 64, University of Helsinki, 00014 Helsinki, Finland\label{aff81}
\and
Helsinki Institute of Physics, Gustaf H{\"a}llstr{\"o}min katu 2, University of Helsinki, 00014 Helsinki, Finland\label{aff82}
\and
Laboratoire d'etude de l'Univers et des phenomenes eXtremes, Observatoire de Paris, Universit\'e PSL, Sorbonne Universit\'e, CNRS, 92190 Meudon, France\label{aff83}
\and
SKAO, Jodrell Bank, Lower Withington, Macclesfield SK11 9FT, United Kingdom\label{aff84}
\and
Centre de Calcul de l'IN2P3/CNRS, 21 avenue Pierre de Coubertin 69627 Villeurbanne Cedex, France\label{aff85}
\and
University of Applied Sciences and Arts of Northwestern Switzerland, School of Computer Science, 5210 Windisch, Switzerland\label{aff86}
\and
Universit\"at Bonn, Argelander-Institut f\"ur Astronomie, Auf dem H\"ugel 71, 53121 Bonn, Germany\label{aff87}
\and
INFN-Sezione di Roma, Piazzale Aldo Moro, 2 - c/o Dipartimento di Fisica, Edificio G. Marconi, 00185 Roma, Italy\label{aff88}
\and
Department of Physics, Institute for Computational Cosmology, Durham University, South Road, Durham, DH1 3LE, UK\label{aff89}
\and
Universit\'e Paris Cit\'e, CNRS, Astroparticule et Cosmologie, 75013 Paris, France\label{aff90}
\and
CNRS-UCB International Research Laboratory, Centre Pierre Bin\'etruy, IRL2007, CPB-IN2P3, Berkeley, USA\label{aff91}
\and
University of Applied Sciences and Arts of Northwestern Switzerland, School of Engineering, 5210 Windisch, Switzerland\label{aff92}
\and
Institut d'Astrophysique de Paris, 98bis Boulevard Arago, 75014, Paris, France\label{aff93}
\and
Institute of Physics, Laboratory of Astrophysics, Ecole Polytechnique F\'ed\'erale de Lausanne (EPFL), Observatoire de Sauverny, 1290 Versoix, Switzerland\label{aff94}
\and
Telespazio UK S.L. for European Space Agency (ESA), Camino bajo del Castillo, s/n, Urbanizacion Villafranca del Castillo, Villanueva de la Ca\~nada, 28692 Madrid, Spain\label{aff95}
\and
Institut de F\'{i}sica d'Altes Energies (IFAE), The Barcelona Institute of Science and Technology, Campus UAB, 08193 Bellaterra (Barcelona), Spain\label{aff96}
\and
DARK, Niels Bohr Institute, University of Copenhagen, Jagtvej 155, 2200 Copenhagen, Denmark\label{aff97}
\and
Waterloo Centre for Astrophysics, University of Waterloo, Waterloo, Ontario N2L 3G1, Canada\label{aff98}
\and
Department of Physics and Astronomy, University of Waterloo, Waterloo, Ontario N2L 3G1, Canada\label{aff99}
\and
Perimeter Institute for Theoretical Physics, Waterloo, Ontario N2L 2Y5, Canada\label{aff100}
\and
Centre National d'Etudes Spatiales -- Centre spatial de Toulouse, 18 avenue Edouard Belin, 31401 Toulouse Cedex 9, France\label{aff101}
\and
Institute of Space Science, Str. Atomistilor, nr. 409 M\u{a}gurele, Ilfov, 077125, Romania\label{aff102}
\and
Consejo Superior de Investigaciones Cientificas, Calle Serrano 117, 28006 Madrid, Spain\label{aff103}
\and
Universidad de La Laguna, Departamento de Astrof\'{\i}sica, 38206 La Laguna, Tenerife, Spain\label{aff104}
\and
Institut f\"ur Theoretische Physik, University of Heidelberg, Philosophenweg 16, 69120 Heidelberg, Germany\label{aff105}
\and
Institut de Recherche en Astrophysique et Plan\'etologie (IRAP), Universit\'e de Toulouse, CNRS, UPS, CNES, 14 Av. Edouard Belin, 31400 Toulouse, France\label{aff106}
\and
Universit\'e St Joseph; Faculty of Sciences, Beirut, Lebanon\label{aff107}
\and
Departamento de F\'isica, FCFM, Universidad de Chile, Blanco Encalada 2008, Santiago, Chile\label{aff108}
\and
Universit\"at Innsbruck, Institut f\"ur Astro- und Teilchenphysik, Technikerstr. 25/8, 6020 Innsbruck, Austria\label{aff109}
\and
Institut d'Estudis Espacials de Catalunya (IEEC),  Edifici RDIT, Campus UPC, 08860 Castelldefels, Barcelona, Spain\label{aff110}
\and
Satlantis, University Science Park, Sede Bld 48940, Leioa-Bilbao, Spain\label{aff111}
\and
Infrared Processing and Analysis Center, California Institute of Technology, Pasadena, CA 91125, USA\label{aff112}
\and
Instituto de Astrof\'isica e Ci\^encias do Espa\c{c}o, Faculdade de Ci\^encias, Universidade de Lisboa, Tapada da Ajuda, 1349-018 Lisboa, Portugal\label{aff113}
\and
Cosmic Dawn Center (DAWN)\label{aff114}
\and
Niels Bohr Institute, University of Copenhagen, Jagtvej 128, 2200 Copenhagen, Denmark\label{aff115}
\and
Universidad Polit\'ecnica de Cartagena, Departamento de Electr\'onica y Tecnolog\'ia de Computadoras,  Plaza del Hospital 1, 30202 Cartagena, Spain\label{aff116}
\and
Dipartimento di Fisica e Scienze della Terra, Universit\`a degli Studi di Ferrara, Via Giuseppe Saragat 1, 44122 Ferrara, Italy\label{aff117}
\and
Istituto Nazionale di Fisica Nucleare, Sezione di Ferrara, Via Giuseppe Saragat 1, 44122 Ferrara, Italy\label{aff118}
\and
INAF, Istituto di Radioastronomia, Via Piero Gobetti 101, 40129 Bologna, Italy\label{aff119}
\and
Universit\'e C\^{o}te d'Azur, Observatoire de la C\^{o}te d'Azur, CNRS, Laboratoire Lagrange, Bd de l'Observatoire, CS 34229, 06304 Nice cedex 4, France\label{aff120}
\and
Universit\'e PSL, Observatoire de Paris, Sorbonne Universit\'e, CNRS, LERMA, 75014, Paris, France\label{aff121}
\and
Universit\'e Paris-Cit\'e, 5 Rue Thomas Mann, 75013, Paris, France\label{aff122}
\and
Aurora Technology for European Space Agency (ESA), Camino bajo del Castillo, s/n, Urbanizacion Villafranca del Castillo, Villanueva de la Ca\~nada, 28692 Madrid, Spain\label{aff123}
\and
INAF - Osservatorio Astronomico di Brera, via Emilio Bianchi 46, 23807 Merate, Italy\label{aff124}
\and
INAF-Osservatorio Astronomico di Brera, Via Brera 28, 20122 Milano, Italy, and INFN-Sezione di Genova, Via Dodecaneso 33, 16146, Genova, Italy\label{aff125}
\and
ICL, Junia, Universit\'e Catholique de Lille, LITL, 59000 Lille, France\label{aff126}
\and
ICSC - Centro Nazionale di Ricerca in High Performance Computing, Big Data e Quantum Computing, Via Magnanelli 2, Bologna, Italy\label{aff127}
\and
Instituto de F\'isica Te\'orica UAM-CSIC, Campus de Cantoblanco, 28049 Madrid, Spain\label{aff128}
\and
CERCA/ISO, Department of Physics, Case Western Reserve University, 10900 Euclid Avenue, Cleveland, OH 44106, USA\label{aff129}
\and
Technical University of Munich, TUM School of Natural Sciences, Physics Department, James-Franck-Str.~1, 85748 Garching, Germany\label{aff130}
\and
Max-Planck-Institut f\"ur Astrophysik, Karl-Schwarzschild-Str.~1, 85748 Garching, Germany\label{aff131}
\and
Laboratoire Univers et Th\'eorie, Observatoire de Paris, Universit\'e PSL, Universit\'e Paris Cit\'e, CNRS, 92190 Meudon, France\label{aff132}
\and
Departamento de F{\'\i}sica Fundamental. Universidad de Salamanca. Plaza de la Merced s/n. 37008 Salamanca, Spain\label{aff133}
\and
Universit\'e de Strasbourg, CNRS, Observatoire astronomique de Strasbourg, UMR 7550, 67000 Strasbourg, France\label{aff134}
\and
Center for Data-Driven Discovery, Kavli IPMU (WPI), UTIAS, The University of Tokyo, Kashiwa, Chiba 277-8583, Japan\label{aff135}
\and
Dipartimento di Fisica - Sezione di Astronomia, Universit\`a di Trieste, Via Tiepolo 11, 34131 Trieste, Italy\label{aff136}
\and
California Institute of Technology, 1200 E California Blvd, Pasadena, CA 91125, USA\label{aff137}
\and
Department of Physics \& Astronomy, University of California Irvine, Irvine CA 92697, USA\label{aff138}
\and
Departamento F\'isica Aplicada, Universidad Polit\'ecnica de Cartagena, Campus Muralla del Mar, 30202 Cartagena, Murcia, Spain\label{aff139}
\and
Instituto de F\'isica de Cantabria, Edificio Juan Jord\'a, Avenida de los Castros, 39005 Santander, Spain\label{aff140}
\and
INFN, Sezione di Lecce, Via per Arnesano, CP-193, 73100, Lecce, Italy\label{aff141}
\and
Department of Mathematics and Physics E. De Giorgi, University of Salento, Via per Arnesano, CP-I93, 73100, Lecce, Italy\label{aff142}
\and
INAF-Sezione di Lecce, c/o Dipartimento Matematica e Fisica, Via per Arnesano, 73100, Lecce, Italy\label{aff143}
\and
CEA Saclay, DFR/IRFU, Service d'Astrophysique, Bat. 709, 91191 Gif-sur-Yvette, France\label{aff144}
\and
Institute of Cosmology and Gravitation, University of Portsmouth, Portsmouth PO1 3FX, UK\label{aff145}
\and
Department of Computer Science, Aalto University, PO Box 15400, Espoo, FI-00 076, Finland\label{aff146}
\and
Instituto de Astrof\'\i sica de Canarias, c/ Via Lactea s/n, La Laguna 38200, Spain. Departamento de Astrof\'\i sica de la Universidad de La Laguna, Avda. Francisco Sanchez, La Laguna, 38200, Spain\label{aff147}
\and
Caltech/IPAC, 1200 E. California Blvd., Pasadena, CA 91125, USA\label{aff148}
\and
Ruhr University Bochum, Faculty of Physics and Astronomy, Astronomical Institute (AIRUB), German Centre for Cosmological Lensing (GCCL), 44780 Bochum, Germany\label{aff149}
\and
Department of Physics and Astronomy, Vesilinnantie 5, University of Turku, 20014 Turku, Finland\label{aff150}
\and
Serco for European Space Agency (ESA), Camino bajo del Castillo, s/n, Urbanizacion Villafranca del Castillo, Villanueva de la Ca\~nada, 28692 Madrid, Spain\label{aff151}
\and
ARC Centre of Excellence for Dark Matter Particle Physics, Melbourne, Australia\label{aff152}
\and
Centre for Astrophysics \& Supercomputing, Swinburne University of Technology,  Hawthorn, Victoria 3122, Australia\label{aff153}
\and
Department of Physics and Astronomy, University of the Western Cape, Bellville, Cape Town, 7535, South Africa\label{aff154}
\and
DAMTP, Centre for Mathematical Sciences, Wilberforce Road, Cambridge CB3 0WA, UK\label{aff155}
\and
Kavli Institute for Cosmology Cambridge, Madingley Road, Cambridge, CB3 0HA, UK\label{aff156}
\and
Department of Astrophysics, University of Zurich, Winterthurerstrasse 190, 8057 Zurich, Switzerland\label{aff157}
\and
Department of Physics, Centre for Extragalactic Astronomy, Durham University, South Road, Durham, DH1 3LE, UK\label{aff158}
\and
IRFU, CEA, Universit\'e Paris-Saclay 91191 Gif-sur-Yvette Cedex, France\label{aff159}
\and
Oskar Klein Centre for Cosmoparticle Physics, Department of Physics, Stockholm University, Stockholm, SE-106 91, Sweden\label{aff160}
\and
Astrophysics Group, Blackett Laboratory, Imperial College London, London SW7 2AZ, UK\label{aff161}
\and
Univ. Grenoble Alpes, CNRS, Grenoble INP, LPSC-IN2P3, 53, Avenue des Martyrs, 38000, Grenoble, France\label{aff162}
\and
Dipartimento di Fisica, Sapienza Universit\`a di Roma, Piazzale Aldo Moro 2, 00185 Roma, Italy\label{aff163}
\and
Centro de Astrof\'{\i}sica da Universidade do Porto, Rua das Estrelas, 4150-762 Porto, Portugal\label{aff164}
\and
HE Space for European Space Agency (ESA), Camino bajo del Castillo, s/n, Urbanizacion Villafranca del Castillo, Villanueva de la Ca\~nada, 28692 Madrid, Spain\label{aff165}
\and
INAF - Osservatorio Astronomico d'Abruzzo, Via Maggini, 64100, Teramo, Italy\label{aff166}
\and
Theoretical astrophysics, Department of Physics and Astronomy, Uppsala University, Box 516, 751 37 Uppsala, Sweden\label{aff167}
\and
Mathematical Institute, University of Leiden, Einsteinweg 55, 2333 CA Leiden, The Netherlands\label{aff168}
\and
School of Physics \& Astronomy, University of Southampton, Highfield Campus, Southampton SO17 1BJ, UK\label{aff169}
\and
Institute of Astronomy, University of Cambridge, Madingley Road, Cambridge CB3 0HA, UK\label{aff170}
\and
Department of Astrophysical Sciences, Peyton Hall, Princeton University, Princeton, NJ 08544, USA\label{aff171}
\and
Space physics and astronomy research unit, University of Oulu, Pentti Kaiteran katu 1, FI-90014 Oulu, Finland\label{aff172}
\and
Center for Computational Astrophysics, Flatiron Institute, 162 5th Avenue, 10010, New York, NY, USA\label{aff173}}    

%
%
%

%
%
\abstract{
We introduce \stsp, a versatile spectral stacking pipeline developed for the \Euclid mission’s NISP spectroscopic surveys, aimed at extracting faint emission lines and spectral features from large galaxy samples in the Wide and Deep Surveys. Designed for computational efficiency and flexible configuration, \stsp supports the processing of extensive datasets critical to \Euclid’s non-cosmological science goals.
We validate the pipeline using simulated spectra processed to match \Euclid’s expected final data quality. Stacking enables robust recovery of key emission lines, including \halpha, \hbeta, \oiii, and \nii, below individual detection limits. 
However, the measurement of galaxy properties such as star formation rate, dust attenuation, and gas-phase metallicity are biased at stellar mass below $\log_{10}(M_/M_\odot) \sim 9$ due to the flux-limited nature of \Euclid spectroscopic samples, which cannot be overcome by stacking.
The SFR–stellar mass relation of the parent sample is recovered reliably only in the Deep survey for $\log_{10}(M_/M_\odot) \gtrsim 10$, whereas the metallicity–mass relation is recovered more accurately over a wider mass range. 
These limitations are caused by the increased fraction of redshift measurement errors at lower masses and fluxes. 
We examine the impact of residual redshift contaminants that arises from misidentified emission lines and noise spikes, on stacked spectra.
Even after stringent quality selections, low-level contamination ($<6\%$) has minimal impact on line fluxes due to the systematically weaker emission of contaminants. Percentile-based analysis of stacked spectra provides a sensitive diagnostic for detecting contamination via coherent spurious features at characteristic wavelengths.
While our simulations include most instrumental effects, real \Euclid data will require further refinement of contamination mitigation strategies.}

%
%
   \titlerunning{Star-forming galaxy scaling relations with {\small\texttt{SpectraPyle}}.}
   \authorrunning{Euclid Collaboration: S. Quai et al.}
   
   \maketitle
%
%
%
%
   
\section{\label{sc:Intro}Introduction}
In many areas of astrophysics, particularly those concerned with the physical properties of distant or faint sources, individual spectra often lack the necessary signal-to-noise (S/N) to detect and characterize key spectral features. This limitation is especially pronounced when studying processes such as star formation, chemical enrichment, or ionization conditions, which rely on the measurement of intrinsically weak emission and absorption lines.
Spectral stacking, i.e. combining spectra of different sources, is a widely used method in astrophysics to improve the S/N of low-quality spectra \citep[e.g.,][]{Francis1991}. 
We identify two primary science cases where stacking becomes indispensable: (1) The study of scaling relations and emission line diagnostics, which often target faint lines tracing star formation, AGN activity, ionized gas conditions, and shocks, and (2) stellar population studies, which require high S/N in the continuum (typically S/N $\gtrsim$ 30 per \AA) to measure features, such as absorption lines and break strengths, that are needed to infer stellar ages, metallicities, alpha-enhancement, and star formation histories \citep[e.g.,][]{Choi2014, Citro2016}.

In this work, we explore how stacking can be effectively applied to \Euclid spectroscopy, taking into account the unique properties and challenges of slitless infrared spectra at low resolution.
\Euclid is an ongoing ESA medium-class mission \citep[][]{Laureijs11, Racca2016}, designed to demonstrate the composition and evolution of the dark Universe, specifically focusing on dark energy and dark matter. The mission employs two primary cosmological probes, namely weak lensing and galaxy clustering, that are studied through high-resolution imaging and low-resolution slitless spectroscopy, provided by a  visible imager \citep[VIS,][]{EuclidSkyVIS}, and the Near-Infrared Spectrometer and Photometer \citep[NISP,][]{EuclidSkyNISP}. The NISP spectrograph (NISP-S) uses two so-called red grisms (RGS), spanning the same \RGE passband (1.206--1.892 \micron), and one blue grism (BGS) covering the \BGE passband (0.926--1.366 \micron).
\Euclid is conducting two major surveys:  the Euclid Wide Survey \citep[EWS,][]{Scaramella-EP1}, which covers about 14\,000 deg$^2$, down to a $5\sigma$ point-like source depth of $26.2$ ($24$) in \IE (\YE, \JE, \HE), and down to a $3.5\sigma$  line flux limit of $2\times10^{-16}$ \flcgs for a target source of diameter \ang{;;0.5} \citep{Q1-TP007}; and the  Euclid Deep Survey \citep[EDS,][] {EuclidSkyOverview}, that will eventually be approximately 2 magnitudes deeper and flux limit of $6\times10^{-17}$ \flcgs over an area of 53 deg$^2$, splitted in three fields.

The unparalleled volume of spectro-photometric data, encompassing precise morphological parameters for billions of galaxies and tens of millions of spectroscopic redshifts, will enable the investigation of scaling relations, mass assembly, environmental effects, and galaxy-active galactic nucleus (AGN) co-evolution in samples that include a significant number of massive star-forming and passive galaxies at intermediate and high redshift \citep{EuclidSkyOverview,EP-Selwood}.
However, the vast majority of spectra, especially from the EWS, have low S/N in the stellar continuum, 
while, a minimum S/N of $\sim$5 is generally required to robustly detect and measure emission line fluxes \citep{Gabarra-EP31}. Consequently, the direct detection of spectral features in individual spectra is generally limited to the brightest galaxies.
Further, many diagnostics rely on multiple line detections. For example, H$\beta$ is required for dust correction and SFR estimates, but is intrinsically fainter than H$\alpha$ (typically by a factor of $>3$, accounting for extinction), or [O~\textsc{iii}]$\lambda4363$, used for direct metallicity estimates via electron temperature, is over 100 times fainter than H$\alpha$ and blended with H$\gamma$ at \Euclid{}’s resolution.
Detecting such lines requires stacking dozens to thousands of spectra, depending on the target line and the intrinsic S/N of individual observations. For instance, detecting H$\beta$ in galaxies with typical EWS-quality spectra requires stacking at least a dozen spectra; for auroral lines, up to \num{10000}–\num{40000} may be needed.

The steps of creating a composite spectrum involve several critical choices, that can have a substantial impact on the final output \citep[e.g.,][]{Francis1991, VandenBerk2001}, most of which depend on the scientific question under investigation. 
We have created \stsp, a Python code for stacking low S/N spectra, that not only provides the needed flexibility to be adapted for the study of different science cases, but also streamlines the stacking process for such a large dataset, of thousands of galaxy and AGN spectra per square degree, {as the \Euclid's one}. 

A critical aspect of leveraging stacked spectra for non-cosmological studies is recognizing and correcting for the various systematic effects inherent to the \Euclid spectroscopic dataset. First and foremost, the spectroscopic redshift sample is inherently biased: it preferentially selects galaxies with bright emission lines, as these are necessary for reliable redshift determination in slitless spectroscopy. This introduces a flux-limited selection that affects all derived statistics, from emission line diagnostics to inferred physical properties. In addition, slitless spectroscopy is prone to contamination and redshift misidentifications (interlopers), particularly in crowded fields or at specific redshift ranges where line confusion is common. These issues can skew the final stacked spectra, both by introducing fake features due to noise and contamination, and by enhancing certain spectral lines through selection effects. As such, even robust stacking techniques must be applied with caution and an understanding of the biases they may amplify.

This paper explores such issues using 
 \Euclid-like 
 mock observations 
 and focusing on the use of stacking techniques to enhance the S/N.
We examine how redshift measurement errors, the presence of interlopers (i.e., galaxies assigned incorrect redshifts), and the trade-off between sample completeness (success rate) and contamination affect the recovered spectroscopic sample. 
In particular, we show how these effects propagate into stacked spectra and influence the measurement of emission line fluxes, the construction of diagnostic line ratios, and ultimately the recovery of galaxy scaling relations such as mass–metallicity trends. We also highlight the presence of biases introduced by sample selection. 
While we explore these effects within a specific simulation and using a particular analysis pipeline, the findings have general implications for the interpretation of \Euclid spectroscopic data. Our goal is to identify and quantify systematic effects that will be present, to varying degrees, in any real Euclid spectroscopic analysis.

The structure of this paper is as follows. In Sect.\,\ref{sec:nisp}, we introduce the \Euclid spectroscopic surveys, whilst in Sect.\,\ref{sec:method} we describe the \stsp code and its functionalities.
Section~\ref{sec:sample} presents the simulation and the mock \Euclid spectra  and redshift determination.
In Sect.\,\ref{sec:contamination}, we characterize redshift contamination from various sources, and we discuss the impact of redshift contaminants on the stacked spectra.  Section~\ref{sec:science} defines the success rate and contamination rate of the sample, and we examine the biases introduced by flux-limited selection, especially in the context of recovering scaling relations and physical parameters from stacked spectra. In Sect.\,\ref{scaling_relations}, we analyse some of the aforementioned key scaling relations with \Euclid stacked spectra, and  finally, in Sect.\,\ref{sec:conclusions}, we summarize our main findings.

Throughout this paper we adopt a flat $\Lambda$CDM cosmology with $\Omega_{\rm m} = 0.3$, $\Omega_\Lambda = 0.7$, and $H_0 = 70$ \si{{\rm km}\,{\rm s^{-1}}\,{\rm Mpc^{-1}}}, and assume a \citet{Chabrier2003} initial mass function (IMF). All magnitudes are in the AB system. 

\section{\label{sec:nisp}The \Euclid spectroscopic surveys}

In the EWS spectra are obtained in the RGS alone, with a wavelength sampling of 1.37\,nm pixel$^{-1}$ and a resolving power $R\simeq 500$ \citep{Q1-TP006,Q1-TP007}. 
The spectroscopic setup in the EWS is optimised for the detection of about 2000--4800 \halpha emitters deg$^{-2}$ within the redshift range $z\in[0.84,1.88]$ \citep{Pozzetti2016}, for the galaxy clustering probe, but it enables the  detection and measurement of other redshifted optical and NIR emission lines that trace the evolution of physical processes in galaxies over the last 8--10 Gyr. More details about the EWS observing strategy can be found in \citet{Scaramella-EP1} and \citet{Q1-TP006}.
\begin{figure}
\centering
\includegraphics[width=0.9\columnwidth]{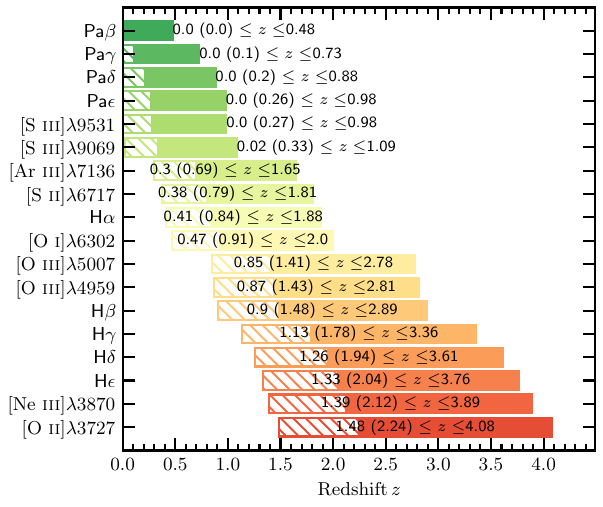}
    \caption{Redshift coverage of prominent emission lines in the EWS and EDS spectra. For each emission line, the full coloured bar represents the redshift range where the line is detectable in the RGS for the EWS, whilst the dashed bar indicates the extended redshift coverage provided by the BGS in the EDS.}
    \label{fig:lines_vs_z}
\end{figure}

For the EDS, NISP-S complement the two RGSs with the BGS, thus extending the lower H$\alpha$ redshift limit to $z = 0.41$. 
The BGS has a constant wavelength sampling of 1.24\,nm pixel$^{-1}$ and an expected resolving power $R\geq 400$. 
The exposure time ratio between the BGS and RGS will eventually be of 5:3, and the greater sensitivity achieved through repeated observations translates into  a $3.5\sigma$  flux detection limit for emission lines of $5\times10^{-17}$ \flcgs \citep{EuclidSkyOverview}.
Figure\,\ref{fig:lines_vs_z} shows the redshift extent of prominent emission lines in the EWS and EDS spectra. In Sect.\,\ref{sec:science} we will present examples of science cases related to the various sets of lines detectable in the two surveys.

The \Euclid pipeline determines redshifts using a template-fitting approach adapted from the Algorithm for Massive Automated Z Evaluation and Determination \citep[AMAZED,][]{Schmitt2019}, and optimized for \Euclid slitless spectroscopy \citep{Q1-TP007}.
For the EWS, spectra from the red grism are used, whereas for the EDS the pipeline takes into account also the blue grism to improve redshift accuracy. 
The fitting procedure produces the probability density function (zPDF), and stores up to five redshift solutions corresponding to the strongest five peaks of the zPDF. 
For each of the five redshift solutions, the \Euclid pipeline provides a probability score (hereafter, \zprob) which corresponds to the integrated probability under the zPDF peak (within $\pm3\sigma$), ranging from $0$ (lowest confidence in the redshift estimate) to $1$ (highest confidence).
In our analysis, where we perform measurements on simulated spectra using the same code, we use only the best-fit solution. 
In the EWS survey, a prior is applied in the zPDF calculation to preferentially interpret isolated emission lines in low S/N spectra as the \halpha line  \citep{Q1-TP007}. 
It particularly enhances the accuracy of redshift measurements in the range where \halpha falls within the red grism, approximately between $0.9 \lesssim z  \lesssim 1.8$.

Once the redshift has been determined, the fluxes of the detected emission lines are measured using both direct integration (DI) and Gaussian fitting (GF) methods \citep{Q1-TP007}. 
The DI method integrates line flux from the peak position after subtracting the continuum until the flux remains positive, providing flux, S/N, equivalent width (EW), and line centroid position. The GF method models emission lines with multiple Gaussians and a constant continuum. For the lines complex \halpha + [N~\textsc{ii}] doublet, that is blended at NISP-P resolution, three Gaussians are used for deconvolution. 

\section{\label{sec:method}\texttt{SpectraPyle}: a \Euclid-optimized stacking code}

\begin{figure*}
\centering
\includegraphics[width=\linewidth]
{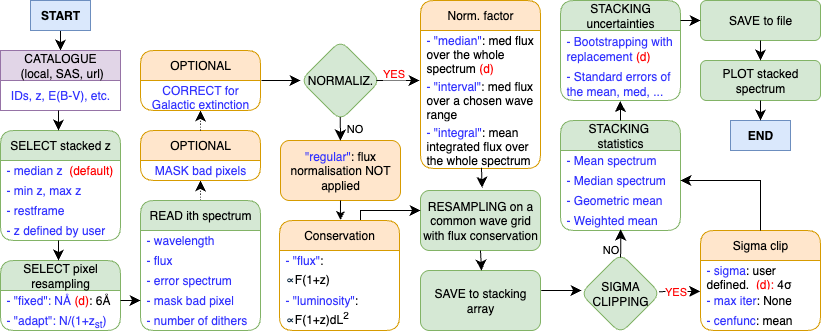}
    \caption{Flowchart of the code \stsp. The letter (d) indicates default options and parameters.}
\label{fig:flowchart}
\end{figure*}

In this paper we present \stsp, a Python-based stacking tool designed to combine several spectra from galaxy selected samples to obtain their combined spectra with increased quality. Indeed, when combining $n$ spectra with uncorrelated noise, the resulting S/N typically increases as $\sqrt{n}$, under the assumption that noise is uncorrelated and random.
This code will handle the statistical demands and flexibility needed for \Euclid-based science. 
The code is highly modular and provides tools for data preparation that preserve the underlying astrophysical information, allowing easy adaptation to user needs while remaining compatible with the Euclid data model (Fig.\,\ref{fig:flowchart}).
\stsp has been implemented as a key component of the data analysis within the ESA Datalabs \citep{Datalabscite},\footnote{The ESA Datalabs ara available at \url{https://datalabs.esa.int/}\,.} a collaborative platform that provides access to data processing tools and computational resources for researchers.\footnote{The code is currently accessible to a limited number of users and will be released to the entire Euclid Consortium prior to publication. Interested researchers in the consortium may contact the authors by email to request early access.}

\subsection{\label{sec:codestep1}Processing the individual spectra}

The code processes individual spectra consisting of three mandatory arrays: the wavelength grid (in units of \AA), flux values (in units of erg s$^{-1}$ cm$^{-2}$ \AA$^{-1}$), and the corresponding errors. Optional arrays may be included, such as quality masks for individual pixels and the number of dithered observations contributing to the co-added flux for each pixel, to exclude low-quality spectral regions from the stacking process. 
The stacking procedure can be customized through user-defined configuration parameters, such as  RGS or BGS selection, flux normalisation, sigma clipping, and bootstrapping. Users can also choose to correct for Galactic extinction on individual spectra before stacking, using the {\small\texttt{dust\_extinction}} Python package \citep{Gordon2024}.

Individual spectra are shifted to a common reference redshift, which can be set to the minimum, maximum, or median redshift of the sample, a user-defined value, or to the rest frame.
If \( z_\textrm{stack} \) is the target redshift for stacking and \( z_\textrm{gal} \) is the redshift of an individual galaxy, the wavelength of the \( i \)-th pixel (\(\lambda_i\)) in the galaxy's spectrum is shifted as
\begin{equation}
    \lambda_{\text{stack}, i} = \lambda_i \frac{1+z_\textrm{stack}}{1+z_\textrm{gal}}\,. \label{eq:lbd}
\end{equation}

\noindent \stsp allows for optional flux normalisation of spectra prior to stacking. The available methods include: \enquote{regular} (no normalisation), \enquote{median} (scaling each spectrum to its median flux across the full wavelength range), \enquote{integral} (scaling to the mean integrated flux), \enquote{interval} (scaling to the mean flux within a user-defined wavelength range), and \enquote{custom} (user-supplied normalisation factors, e.g. based on photometry, that is particularly useful for faint spectra).

If no normalisation is applied (i.e. using the \enquote{regular} option), the code offers two flux-scaling methods when shifting to the common redshift: \enquote{flux conservation} and \enquote{luminosity conservation}, depending on the intended preservation of observed or intrinsic spectral features.
\begin{itemize}
\item The flux conservation approach transforms the flux of the \( i \)-th pixel as
   \begin{equation}
       F(\lambda_{\text{stack}, i}) = F(\lambda_i) \frac{1+z_\textrm{gal}}{1+z_\textrm{stack}}\,. \label{eq:f_flux}
   \end{equation}

This transformation preserves integrated flux over a wavelength range (e.g., emission line fluxes), as the \((1+z)\) terms in Eqs.~\eqref{eq:lbd} and \eqref{eq:f_flux} cancel out. 
It is therefore useful when conserving emission line ratios is important. 
However, it does not preserve the intrinsic luminosity of spectral features, since it does not account for the dimming due to the luminosity distance. As a result, converting stacked fluxes into luminosities is an approximation, whose accuracy depends on the sample’s redshift range and distribution (Appendix \ref{app:flux_conservation}).  

\item The luminosity conservation method preserves the intrinsic luminosity of spectral features:
   \begin{equation}
   F(\lambda_{\textrm{stack}, i}) = F(\lambda_i) \frac{1+z_\textrm{gal}}{1+z_\textrm{stack}} \frac{D_\text{L}^2(z_\textrm{gal})}{D_\text{L}^2(z_\textrm{stack})}\,, \label{eq:f_lum}
   \end{equation}
   
\noindent where $D_ \text{L}$ is the luminosity distance. This method ensures that luminosity-dependent measurements, such as SFR derived from \halpha luminosity, remain correct in the composite spectrum. 
\end{itemize} 
In both cases, the spectral continuum of galaxies with $z_\textrm{gal} < z_\textrm{stack}$, contributing predominantly to the redder end of the stacked spectrum, will be dimmed, while those with $z_\textrm{gal} > z_\textrm{stack}$, contributing more to the bluer end, will be brightened. Therefore, in particular for luminosity conservation, this method should be used only with a narrow distribution of redshifts. 
Once the spectrum has been shifted, it is resampled onto a common, user-defined wavelength grid using a flux-conserving method where the  spectrum is treated as a step function, with pixel-centered wavelengths and associated flux values. 
We reconstruct the pixel edges and finely sample each pixel at a fixed resolution (0.01 \AA). The flux is then integrated over the new wavelength bins using a simple summation, and normalized by the bin widths to yield average flux densities. This approach ensures that the total flux is conserved across the resampling process and avoids interpolation artifacts.

\subsection{\label{sec:codestep2}Combining the spectra}

The steps described in the previous subsection are applied to each spectrum individually, resulting in a set of spectra aligned to a common redshift and normalised according to the chosen scheme. A Gaussian sigma-clipping procedure is then applied to the flux distribution at each pixel to identify and discard outliers.
By default, pixels with flux values exceeding four standard deviations from the mean are excluded. 
This threshold can be adjusted by the user as needed. A future upgrade will offer an alternative clipping scheme based on the interquartile range, which may be more suitable for asymmetric or non-Gaussian flux distributions.

The final step is the actual stacking, where the appropriate statistical method applied to combine the spectra depends upon the spectral quantities of interest: users can opt for the arithmetic mean to preserve the broad spectral features, the median to minimise the influence of outliers and preserve relative fluxes of the emission features (i.e., optimal for tracing physical processes traced by emission line ratios), the weighted mean for giving more influence to data points with higher reliability or significance, or the geometric mean for preserving multiplicative relationships within the spectra and to accurately determine the average spectra of objects with fluxes per \AA\ spanning a large dynamic range of several magnitudes (Appendix \ref{app:geom_mean}).
By default, \stsp estimates statistical uncertainties on the stacked spectrum using bootstrap resampling with replacement, performing 350 iterations. This number can be adjusted by the user to balance precision and computational cost (Appendix \ref{app:codestep3}). This approach provides a robust characterisation of the variability in the stacked signal. Alternatively, users may choose to compute standard errors based on the applied statistical estimator (e.g., mean or median), though this method does not account for sample variance as comprehensively as bootstrapping. 

\section{\label{sec:sample}The simulated dataset}

\subsection{\label{sec:simulation}The \mambo simulation}
In this work, we constructed a mock catalogue using Mocks with Abundance Matching in Bologna \citep[\mambo,][see also \citealp{LopezLopez24}]{Girelli2021}, that we applied to the Millennium Simulation outputs \citep{Springel2005}, tailored to the \Planck cosmology \citep{Angulo2010}, and specifically to a lightcone derived by \citet{Henriques2015} that covers a redshift range from $z = 0$ to $z =10$. 
This lightcone includes halos with masses greater than $M_{200} > 1.7 \times 10^{10} M_\odot$ over an area of $3.14$ deg$^2$, corresponding to approximately 6 independent \Euclid pointing of $0.5$ deg$^2$ each.
In the lightcone, \mambo 
assignes to each dark matter halo 
a galaxy with stellar mass $M_*$, 
following the stellar-to-halo mass relation (SHMR) by  \cite{Girelli2021}, calibrated against observed stellar mass functions (SMFs) from surveys such as SDSS \citep{York2000}, COSMOS \citep{Scoville2007}, and CANDELS fields. 
Galaxies are then classified into two main populations: quiescent and star-forming, using a probabilistic way, following the relative ratio of the blue and red populations in observed SMF \citep[][]{Peng2010, Ilbert2013, Girelli2019}. Our \mambo catalogue does not include AGN.\footnote{A \mambo catalogue including AGN is presented in \citet{LopezLopez24}.}
Stellar mass $M_*$, redshift $z$ and galaxy classification are the fundamental parameters from which all other observables are statistically derived using empirical relations, and generated using the Empirical Galaxy Generator \citep[EGG,][]{Schreiber2017}. The resulting catalogues have realistic fluxes and galaxy properties that match current observations from redshifts 0 to 6 by construction, and that are extrapolated up to redshift 10. 
Controlled random scatter is added to most observables.

In EGG, following the approach of \citet{Lang2014}, galaxies are modelled as having two components: a bulge with a Sérsic index $n=4$ and a disk with a Sérsic index $n= 1$. 
The bulge-to-total ratio (B/T) defines the mass distribution between these components, described by parameters including the projected axis ratio ($b/a$), half-light radius ($R_{50}$), and position angle ($\theta$). 
Position angles for both bulge and disk components are drawn from a uniform distribution.
SFRs are assigned based on the empirical main sequence relation between SFR and $M_*$ by \citet[][]{Schreiber2015}. 
Physical properties such as size, velocity dispersion, dust attenuation, optical rest-frame colours, and metallicity are also modelled. 
Then, an appropriate panchromatic spectral energy distribution (SED), for both the disk and bulge components, is assigned to each galaxy.
Based on its redshift, $M_*$ and quiescent/star-forming classification, each source is randomly placed on the UVJ diagram, where an SED from a pre-built stellar library of \citet{Bruzual2003} is pre-assigned to each position.
Synthetic photometry is finally generated by integrating the redshifted SED in commonly used broad-band filters from UV to submillimeter wavelengths, including the four \Euclid VIS and NISP-P bands (i.e., \IE, \YE, \JE, and \HE). 

The next step, needed to create incident spectra, is to add a set of prominent recombination and forbidden emission lines to the SED of the disk component of star-forming galaxies, namely (from redder to bluer):  
\pabeta, \pagamma, \siiib, \siiia, the [S~\textsc{ii}]$\lambda 6731$--[S~\textsc{ii}]$\lambda 6717$ doublet (hereafter \sii), [N~\textsc{ii}]$\lambda 6584$--[N~\textsc{ii}]$\lambda 6548$ (hereafter \nii), \halpha, [O~\textsc{iii}]$\lambda 5007$ (hereafter \oiii), \oiiia, \hbeta, [Ne~\textsc{iii}]$\lambda3869$, and the [O~\textsc{ii}]$\lambda 3727$ doublet (hereafter \oii). 
These spectral features are well-established as highly sensitive tracers of the physical conditions within the ionized gas of galaxies and the nature of the ionizing radiation \citep{Osterbrock1989, Kewley2019}, and their simulated luminosities are generated following empirical relations, starting from their stellar mass and SFR.\footnote{The \mambo procedure and recipes used to assign fluxes for different emission lines are described here: \url{https://github.com/xalolo/MAMBO}.}
Intrinsic emission line luminosities were adjusted to include dust attenuation using the Balmer decrement and the \cite{Calzetti2000} extinction law,  while we did not include attenuation due to Galactic extinction to the simulated SEDs. 
All the lines were initially assumed to have the same velocity dispersion, which was set to the resolution of the \citet{Bruzual2003} template (a FWHM of approximately 0.27\,nm) to align with the resolution of the stellar component.
We finally applied a Gaussian broadening kernel to the composite stellar-emission SED of the disk component and to the stellar SED of the bulge component, using a wavelength-dependent $\sigma(\lambda)$ for each pixel that corresponds to the galaxy's velocity dispersion using the mass-dependent $\sigma_\textrm{gas}$ from \citet{Bezanson2018}, converted to the pixel scale. This was done after subtracting the intrinsic resolution of the \citet{Bruzual2003} templates in quadrature.

\subsection{\label{sec:mambolines}The Euclid-like sample}
From the catalogue obtained at the end of the above procedure, we selected all \mambo galaxies up to a redshift of $z \leq 4$.
This \enquote{parent sample}, consisting of 6.56 million galaxies (98.2\% being star-forming and 1.8\% passive), is complete approximately down to \HE$=28$ and stellar masses down to $\sim 10^7$ $M_\odot$, and it is the benchmark for the analyses conducted in this paper. 
As this sample is deeper than the \Euclid sensitivity in the \HE band, we did not generate mock spectra for all galaxies; instead, we did it only for two subsamples of star-forming galaxies:
\bi{}
\item \enquote{EWS-SF$_\textrm{sim}$}: 255\,743 star-forming \mambo galaxies with \HE $\leq 24$ and at least one emission line with flux $\geq 2 \times 10^{-17}$ \flcgs within the observable wavelength range. 

\item \enquote{EDS-SF$_\textrm{sim}$}: 366\,927 star-forming \mambo galaxies with \HE $\leq 26$ and the same emission-line flux threshold as above.
\ei{}

The limiting magnitudes are those at which the \Euclid pipeline extracts 1D spectra in the two surveys, while the line flux limit is to avoid processing featureless spectra, hence to reduce storage and runtime. However, we note that our simulation flux floor is ten times fainter than the nominal \ewssf limit and three times fainter than the \edssf limit, enabling exploration of \Euclid's faintest detectable regime, particularly useful for stacking analyses discussed in this paper.

\begin{figure}
\centering
\includegraphics[width=0.9\linewidth]
{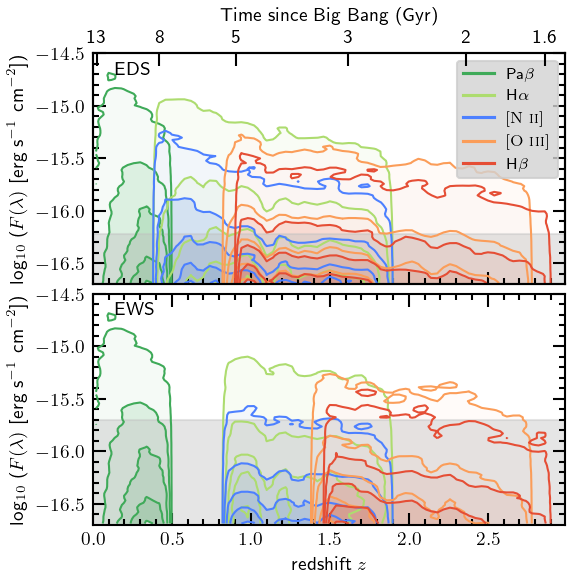}\\
\caption{True flux vs. redshift for key emission lines in the mock \edssf (top) and \ewssf (bottom) samples. Contours in dark green, light green, blue, orange, and red correspond to \pabeta, \halpha, \niib, \oiii, and \hbeta, respectively, selected to span a broad redshift range and wavelength coverage. Each line's redshift range reflects its detectability within the grism coverage. Contours from lighter to darker shades trace the 99th, 84th, 50th, 16th, and 1st percentiles of the distribution. Grey areas mark fluxes below the survey sensitivity limits. 
}
    \label{fig:counts_vs_z}
\end{figure}
Figure\,\ref{fig:counts_vs_z} shows the distribution of true fluxes as a function of redshift for several strong emission lines in the \ewssf and \edssf samples. 
Each line is shown in a distinct colour, with its redshift coverage reflecting the wavelength limits of the red grism (\ewssf) or the combined red and blue grisms (\edssf). 
The figure demonstrates how Euclid primarily detects the brightest fraction of the emission-line population, especially in the EWS, thus highlighting the impact of flux limits on the observed sample and related scaling relations. 
A more detailed analysis of the resulting number counts and selection effects is provided in Cassata et al. (in prep.) and in \citet{EP-Scharre} within the Gaea light cone framework \citep{DeLucia2014,Hirschmann2016}.


\subsection{\label{sec:fastspec}\texttt{FastSpec} spectral by-pass simulations}

We converted the incident spectra of \mambo galaxies into EWS and EDS mock 1D spectra using \fssp (de la Torre et al., in preparation), a Python-based code that simulates the instrumental and environmental effects on \Euclid NISP spectra, bypassing the image-pixel-level approach.
Briefly, the \fssp simulator takes into account effects such as the point-spread function (PSF) model, the spectral extraction window and resulting flux loss, environmental noise from zodiacal background and stray light contribution, as well as instrumental noise factors including readout noise and dark current that have been characterised during NISP ground-based tests \citep{Maciaszek_2022}. 
It also accounts for transmission functions and quantum efficiency, effective collecting surface area, and exposure time of each observation.

The strength of a bypass code simulator such as \fssp lies in its efficiency and short processing times. 
However, some limitations come into play, and we highlight here the most significant ones:

(1) In slitless spectroscopy, overlapping spectra from nearby objects contribute to noise. \Euclid uses a specific observing sequence at four different angles, while \fssp simplifies this by arranging galaxies on a grid without interference and simulating only first-order spectra. This results in simulated \Euclid-like spectra, free from contamination, ensuring fully decontaminated spectral products.
(2) The \fssp simulator processes a single incident SED per galaxy, limiting its ability to differentiate between bulge and disk components. To accommodate this, we combine bulge and disk \mambo incident SEDs into a single incident SED. The simulator then convolves this merged SED with the galaxy's surface brightness profile and instrumental PSF, including background contributions. It models the 2D surface brightness as a composite of two S\'ersic profiles, preserving total luminosity and adjusting contributions based on B/T ratio, inclination, and position angle.
Consequently, emission lines are scattered across the galaxy's 2D spectrum, neglecting spatial variations and spectral contributions from older, centrally-concentrated stellar populations, to disk outskirts where younger stars dominate.

\begin{figure*}
\centering
\includegraphics[width=0.9\linewidth]{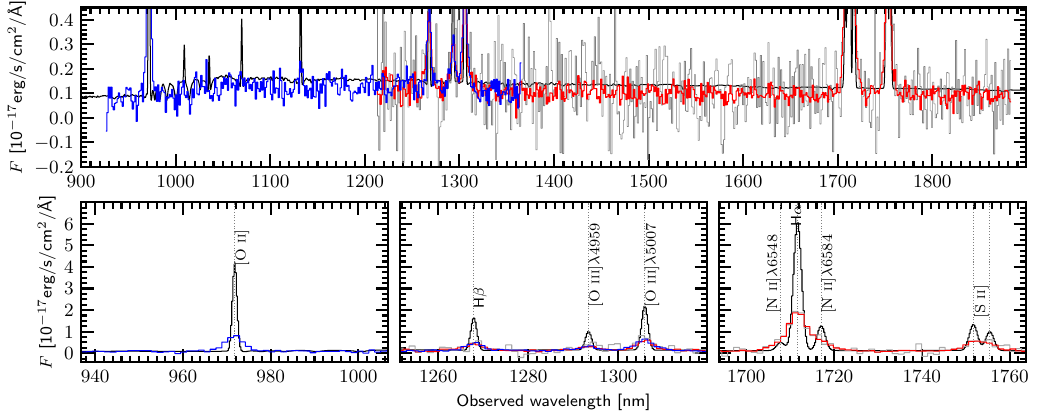}
\caption{An example of a simulated spectrum from \mambo of a galaxy at redshift $z \sim 1.6$, stellar mass of $\log_{10}(M_\ast/M_\odot) = 10.8$, an \HE $\sim$ 21.1, a SFR of 232.7 $M_\odot$/yr, an effective disk radius of \ang{;;0.66}, and a B/T of 0.33. The black spectrum represents the incident model spectrum at its native resolution ($\sim 0.27$\,nm). The EWS simulated spectrum is shown in grey, whilst the blue and red spectra correspond to the EDS blue and red grism simulated observations, respectively. The top panel focuses on the stellar continuum, whilst the bottom panels highlights key emission lines spectral regions: note how the increased survey depth of the EDS allows for the detection of fainter emission lines, such as \oiiia.
}
    \label{fig:mambo-fastsp_spec}
\end{figure*}

At this stage, we generated red grism 1D spectra for both the EWS and EDS, as well as blue spectra for the DEEP survey, by integrating the individual exposures over the cross-dispersion dimension. 
 We imposed a fixed extraction window of five pixels (equivalent to roughly $1.5$ arcseconds), regardless of the galaxy's size. 
Consequently, some flux loss is inevitable, that is proportional to the fraction of the galaxy's light distribution excluded from the extraction aperture. 
This is slightly different from the current implementation in the \Euclid pipeline, where the extraction window for extended objects is set from the semi-major axis of the source \citep{Q1-TP004}, but capped between 5 and 31 pixels \citep{Q1-TP006}.

Figure\,\ref{fig:mambo-fastsp_spec} shows an example spectrum at $z\sim 1.6$ from our \mambo simulation,  with both \ewssf and \edssf realizations. 
Both \ewssf and \edssf spectra show a flux loss relative to the incident spectrum, with an average offset of roughly 20\% per \AA$^{-1}$, due to the synthetic extraction window we used to derive the 1D spectrum.

\begin{figure}
\centering
\includegraphics[width=0.9\linewidth]{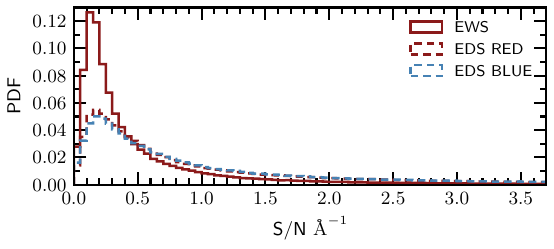}
    \caption{
    Distribution of the continuum S/N per \AA\ for the sample of simulated \mambo spectra (see Sect.\,\ref{sec:simulation}), providing a reference for typical input spectra quality. 
    The red dotted line represents the EWS \mambo sample, selecting galaxies with \HE $<24$. The red solid and blue dashed lines correspond to the EDS red and blue \mambo samples, respectively, selecting galaxies at the survey limit of \HE $<26$.    
    }
    \label{fig:sn}
\end{figure}

We note that the continua of the \edssf and \ewssf realizations in Fig.\,\ref{fig:mambo-fastsp_spec} appear noisy, with the \ewssf spectrum being noisier than the \edssf one, as expected. Defining the S/N as the flux divided by its associated error, Fig.\,\ref{fig:sn} shows the distribution of the average S/N of the stellar continuum (measured after masking emission lines) for spectra in both samples, including the blue and red grisms for \edssf. In both cases, the stellar continuum has an average S/N below 1 per \AA, emphasizing the necessity of stacking \Euclid spectra to extract reliable information for non-cosmological studies.

\subsection{\label{sec:redshift}Redshift and spectral features measurements}
Redshifts and emission line properties are measured in simulated spectra with the official \Euclid pipeline (Sect.\,\ref{sec:nisp}).
\begin{figure}
\centering
\includegraphics[width=0.9\linewidth]{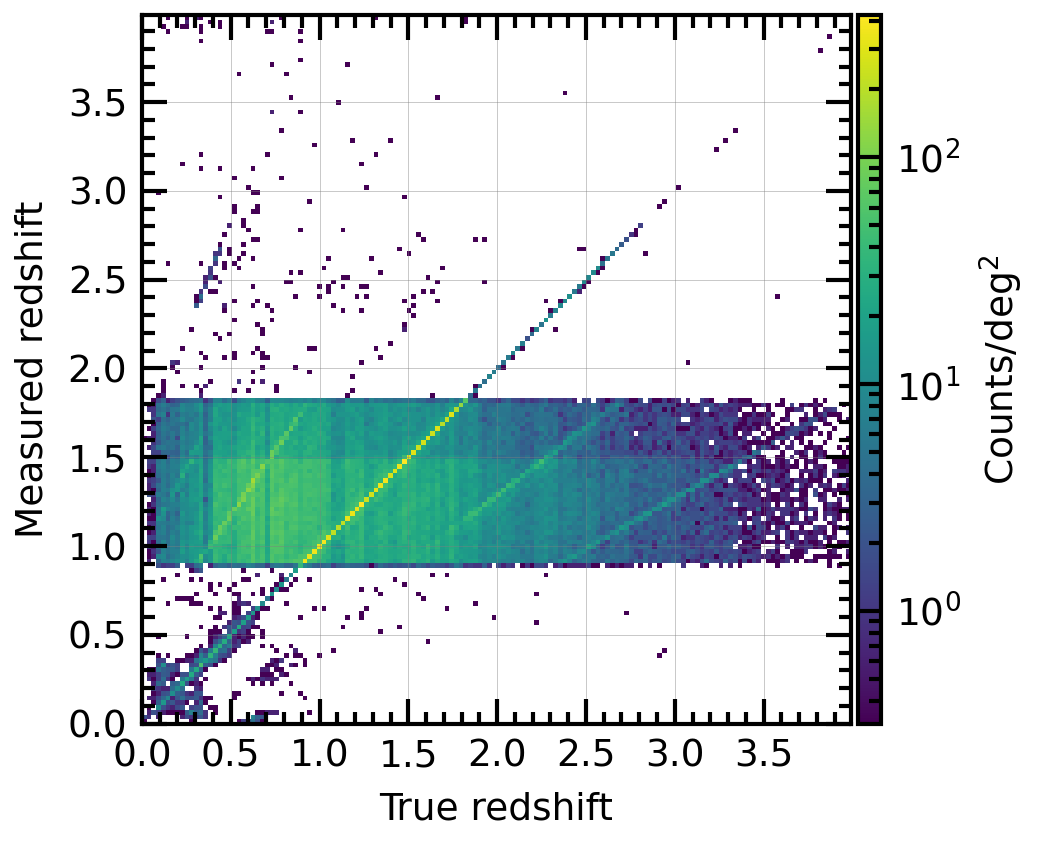}\\
\vspace{0.005mm}
\includegraphics[width=0.9\linewidth]{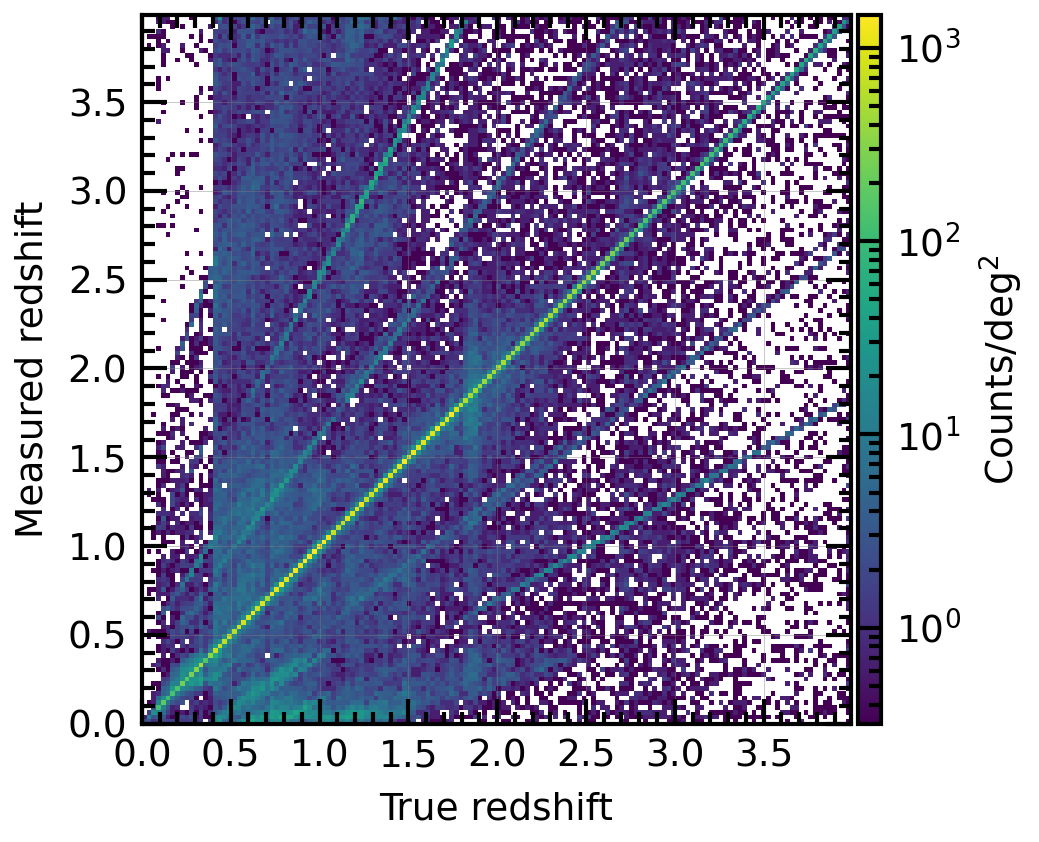}
\caption{Measured redshift vs. true redshift for the mock \ewssf (top panel) and the \edssf (bottom panel). Bins are colour-coded according to the number counts per deg$^2$.}
    \label{fig:zmeas_ztrue}
\end{figure}
Figure\,\ref{fig:zmeas_ztrue} shows the measured redshift as a function of the true galaxy redshift for both the \ewssf (left panel) and \edssf (right panel) simulations. 
The distribution of galaxies forms a characteristic spider-web pattern, within which we can qualitatively identify three categories of solutions: (1) accurate redshift measurements, which lie along the bisector of the diagram, (2) misclassified redshift solutions, where the fitting algorithm incorrectly interprets one spectral line as another, producing overdensities along straight lines with slopes distinct from the bisector, and (3) wrong redshift solutions, where noise spikes are mistaken for emission lines, resulting in a scattered distribution with no clear trend all over the parameter space.
It is important to note that the EWS results preferentially favour redshift solutions in the range $0.9 \lesssim z  \lesssim 1.8$, due to the prior constraints on redshift imposed in the fitting algorithm.
We note that, by construction, sources with intrinsically faint or absent emission lines, which are more susceptible to catastrophic redshift failures, are excluded from our mock samples (see Sect.\,\ref{sec:mambolines}). As a consequence, the contamination fraction inferred from these simulations,  should be interpreted as lower limits, and are further discussed in the following sections.


\section{\label{sec:contamination}Redshift contaminants}

Stacking is an effective method to create high-S/N spectra for galaxy evolution studies, but its effectivness heavily relies on the good alignment of the individual spectra, i.e. on the redshift accuracy (\dzabs).
We performed dedicated tests to verify that a redshift accuracy of $\leq 0.003$ represents a suitable compromise for stacking-based galaxy evolution studies. This threshold allows the inclusion of fainter galaxy populations, despite their lower redshift precision, while still contributing meaningfully to the composite spectra. As a result, we can effectively leverage the capabilities of both the EWS and EDS surveys.
A threshold of \dzabs$\leq 0.003$ ensures that, for a galaxy at $z \sim 1.68$, the \halpha peak is displaced by no more than $\pm 0.53$\,nm (i.e., about four pixels) from its true centroid.
This displacement keeps \halpha within the wavelength range of the blended \hanii lines, 
minimizing the risk of misinterpreting nearby noise spikes as \halpha, while also ensuring that we can probe emission line measurements 
at fainter flux levels using stacking techniques.

All galaxies with \dzabs $> 0.003$ are considered redshift contaminants (hereafter simply referred to as contaminants). 
Before stacking \Euclid spectra, we must evaluate the impact of their inclusion on the quality of stacked spectra when \dz will not be available as in real spectra. In the following, we analyse the two classes of contaminants introduced in Sect.\,\ref{sec:redshift}: spectral line misidentifications and spurious line detections where noise spikes are mistaken for emission lines.

\subsection{\label{sec:misclas_lines}Redshift contaminants from misclassified emission lines}

\begin{figure*}
\centering
\includegraphics[width=0.52\linewidth]{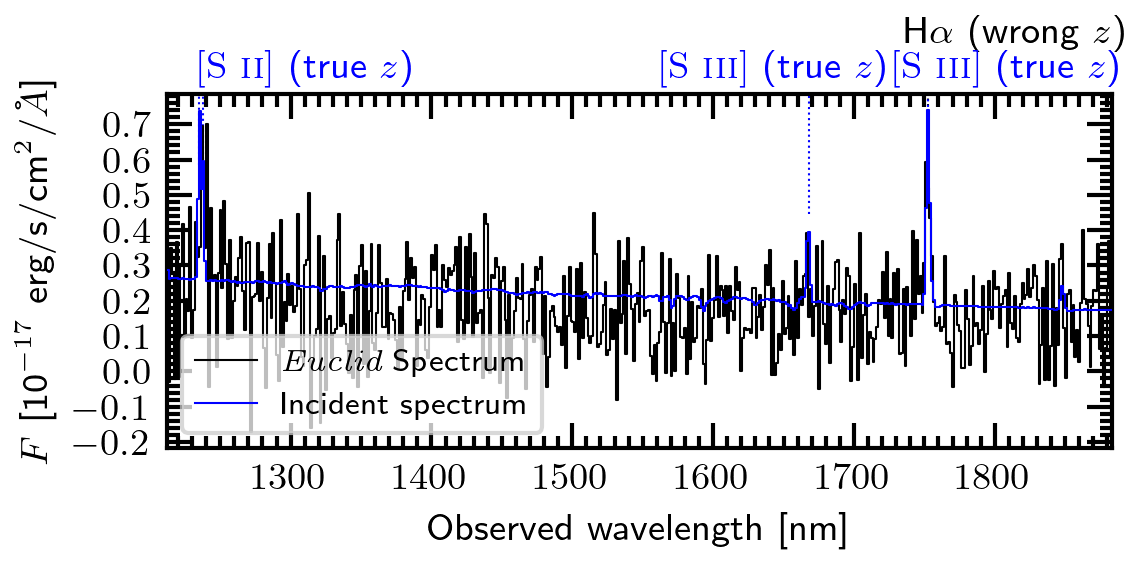}
\includegraphics[width=0.47\linewidth]{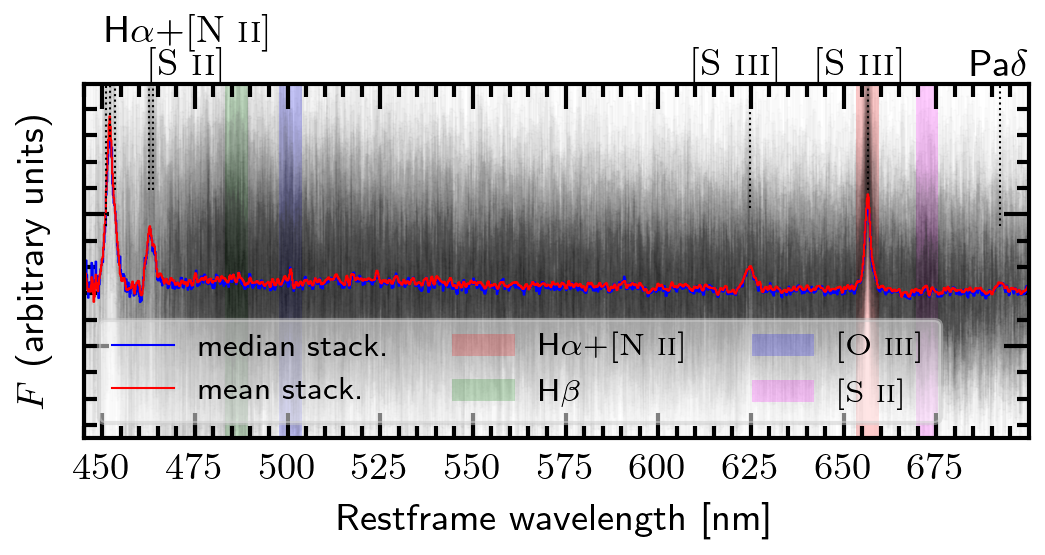}
\caption{Left panel: an example of redshift contaminants due to \siiib at 1752.7\,nm misidentified as \halpha, yielding \zmeas $=1.67$ instead of \ztrue $=0.836$. Right panel: the median (in blue) and mean (in red) stack of 200 spectra (in grey) of galaxies where the \siiib line was misinterpreted as \halpha. Vertical dotted lines indicate the \enquote{true} lines, while the coloured bands highlight the expected rest-frame positions of key optical emission lines at the wrong redshift of the stacked spectrum (\hbeta, \oiii, \halpha, and \nii).
}
\label{fig:spec_interlopers}
\end{figure*}

Some contaminant galaxies have incorrect redshift estimates due to the misidentification of emission lines by the \Euclid pipeline. An emission line at rest-frame wavelength $\lambda_\text{true}^\text{rest}$ is observed at 
$\lambda_\text{true}^\text{obs}$,  
but if it is misidentified as another line with rest-frame wavelength $\lambda_\text{wrong}^\text{rest}$, the measured redshift becomes $\lambda_\text{true}^\text{obs} = \lambda_\text{wrong}^\text{rest}\,(1+z_\textrm{meas})$. 

The resulting redshift offset is

\begin{equation}\label{eq:interlop}
     \frac{\Delta z}{1+z} = \frac{\lambda_\text{true}^\text{rest} - \lambda_\text{wrong}^\text{rest}}{\lambda_\text{wrong}^\text{rest}}\,.
\end{equation}

This implies that misclassified emission lines lead to redshift contaminants with fixed \dz values, determined solely by the line misidentification. For example, \siiib misidentified as \halpha leads to \dz $=0.452$, and \oiii as \halpha gives \dz $=-0.237$.
These contaminants align as straight lines in the \zmeas vs. \ztrue plane (see Fig.\,\ref{fig:zmeas_ztrue}), 
with 
slope 
$\lambda_\text{true}^\text{rest}/\lambda_\text{wrong}^\text{rest}$, and intercept 
$(\lambda_\text{true}^\text{rest} - \lambda_\text{wrong}^\text{rest})/\lambda_\text{wrong}^\text{rest}$\,.
To illustrate this, Fig.\,\ref{fig:spec_interlopers} (left panel) presents an example of such misidentification in the spectrum of an individual galaxy. 
Although the spectrum shows multiple (low-S/N) emission lines that could, in principle, be correctly identified, the redshift pipeline incorrectly assigns the redshift likely due to the use of the \halpha-based prior (see Sect.\,\ref{sec:nisp}), which is optimized for the low-SNR \halpha-dominated sample targeted by the EWS. While such a strategy reflects realistic expectations for automated redshift determination in survey data, it may lead to misidentifications in cases like these. Improvements could be achieved by retuning the prior for specific samples, but this was not explored in the present study, which relies on the default \Euclid pipeline.

These misclassified sources, if not removed, contaminate the stacked spectra. Since stacking relies on the measured redshift, the spectral features of contaminants are aligned incorrectly, degrading the stacked spectrum by introducing shifted and spurious features. Fig.\,\ref{fig:spec_interlopers} (right panel) shows an example, demonstrating how misclassified \siiib lines contaminate the \halpha region of a stacked spectrum generated with \stsp by combining 200 individual \enquote{contaminants}, applying the \textit{regular} normalization with \textit{luminosity} conservation, and omitting sigma clipping (Sect.\,\ref{sec:method}). 
Additional misplaced spectral features from contaminant spectra can affect regions of the stacked spectrum that are unrelated to key emission lines, thereby degrading the stellar continuum. 
For instance, 
features near 450\,nm and 625\,nm in the stacked spectrum of Fig.\,\ref{fig:spec_interlopers}, originate from interloper galaxy spectra whose emission lines  (i.e., \hanii and \sii  and \siiia), are redshifted into these observed-frame positions.

\subsection{\label{sec:noise_spike}Redshift contaminants from random noise spikes}
\begin{figure*}
\centering
\includegraphics[width=0.52\linewidth]{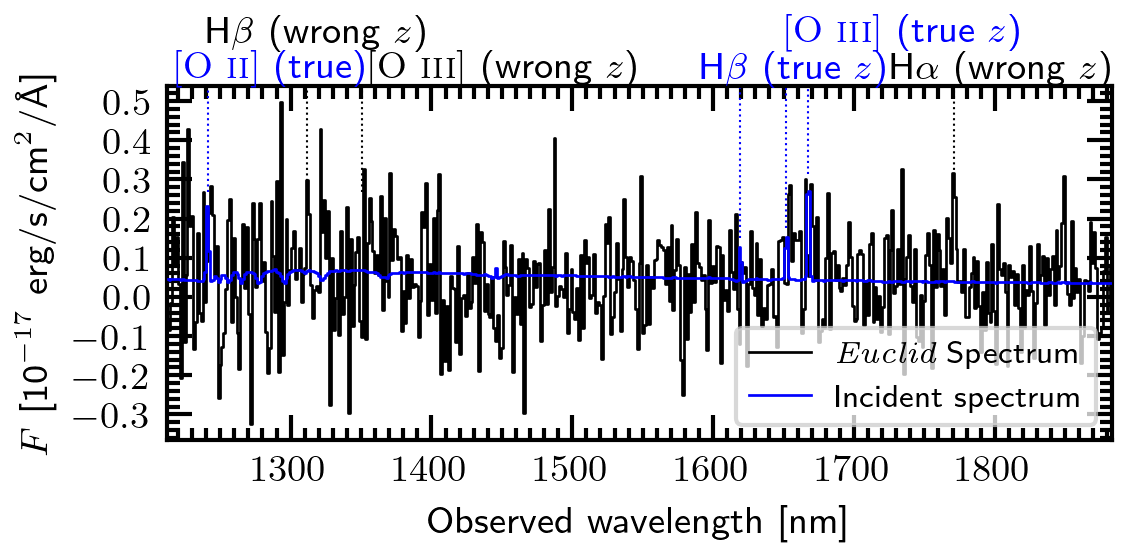}
\includegraphics[width=0.47\linewidth]{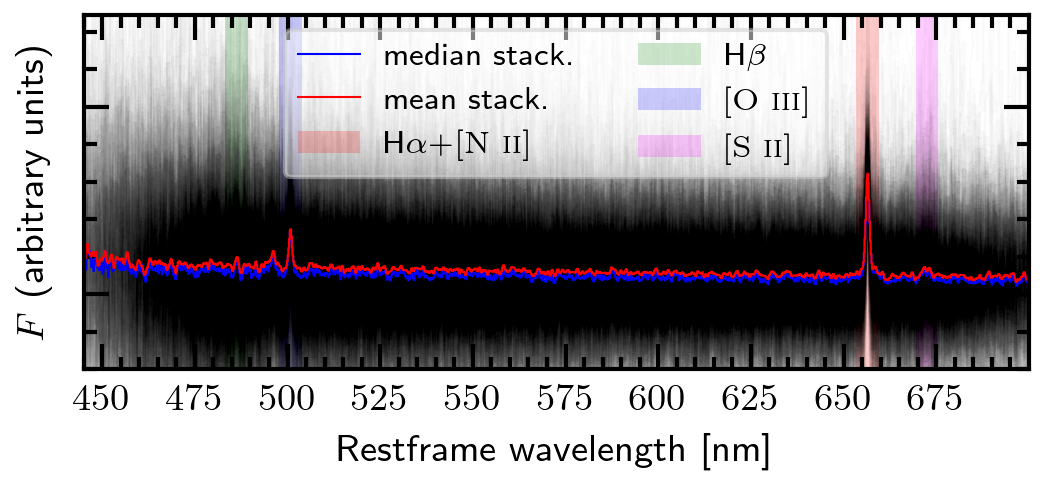}
\caption{Left panel: an example where a random noise spike is misinterpreted as the \halpha emission line, leading to an erroneous redshift measurement. Right panel: the median (in blue) and mean (in red) stacked spectrum of 500 spectra (in grey) of galaxies with misclassified redshift, illustrating the accumulation of these spurious noise spikes at notable lines positions.}
    \label{fig:stack_spikes}
\end{figure*}

A significant category of redshift contaminants arises from random noise spikes in the spectra of faint galaxies, which are erroneously interpreted as emission lines. 
These contaminants result in completely spurious redshift measurements that are unrelated to the true underlying spectra, leading to a broad distribution in \dz and a random scatter in the \zmeas vs. \ztrue plane.

Fig.\,\ref{fig:stack_spikes} shows an example of this type of line misidentification in an individual galaxy (left panel) and in the stacked spectrum of 500 such contaminants (right panel). 
The purpose of this exercise of stacking only contaminants is to demonstrate the contribution of incorrect redshifts to the overall stacked signal, if they are not properly filtered out. 
As expected, a spurious \halpha emission line emerges in the stacked spectrum. 
However, a striking feature of the stacked spectrum is the appearance of not only the spurious \halpha line but also other emission lines such as \hbeta, \oiiia, \oiii, and [S~\textsc{ii}]$\lambda\lambda$6716--6731. This is a predictable consequence of the pipeline’s redshift determination method, which fits templates with multiple emission lines. In noisy spectra, random noise spikes can coincidentally align with the expected positions of these lines, leading the algorithm to assign incorrect redshifts. When these misidentified spectra are stacked, the template features reinforce each other, causing all expected emission lines to appear coherently at the spurious redshift location.
\begin{figure}
\centering
\includegraphics[width=0.95\linewidth]{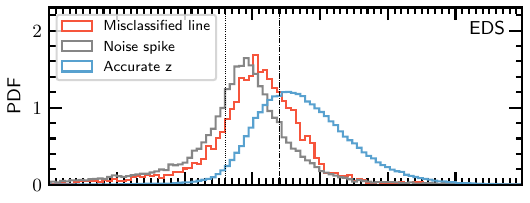}\\
\vspace{-2mm}
\includegraphics[width=0.95\linewidth]{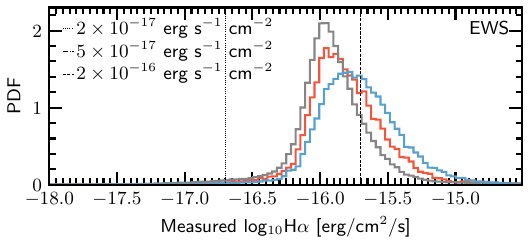}
\caption{Distribution of measured \halpha fluxes (corrected for flux loss following Cassata et al., in prep.) versus true \halpha fluxes for the \edssf (top) and \ewssf (bottom) simulated samples. Blue histograms show galaxies with accurate redshifts with \dzabs $\leq 0.003$, red histograms indicate contaminants from misidentified emission lines, and grey histograms represent contaminants from noise spikes.}
    \label{fig:HaMeas_hist}
\end{figure}

Figure\,\ref{fig:HaMeas_hist} shows the \halpha flux distributions (corrected for flux loss, following Cassata et al., in preparation)for the \edssf (top panel) and the \ewssf (bottom panel).
The figure shows the effects of redshift misclassification and noise spikes on \halpha measurements, using \halpha as a representative example for all spectral lines.
Most contaminants exhibit measured fluxes below the flux limits of their respective surveys and could, in principle, be excluded by imposing stricter flux thresholds. However, even when restricting the analysis to galaxies with \halpha fluxes above these limits, at the cost of excluding fainter regimes of the scaling relations, a significant fraction of contaminants remains. As shown in Fig.\,\ref{fig:HaMeas_hist}, these residual contaminants predominantly contribute low flux values to the stack, thereby  their impact on the composite measurements should be limited. This effect will be examined in greater detail in the following sections.

\subsection{\label{stack_mixture} Insights on contamination level from high-percentile stacked spectra}
\begin{figure}
\centering
\includegraphics[width=\linewidth]{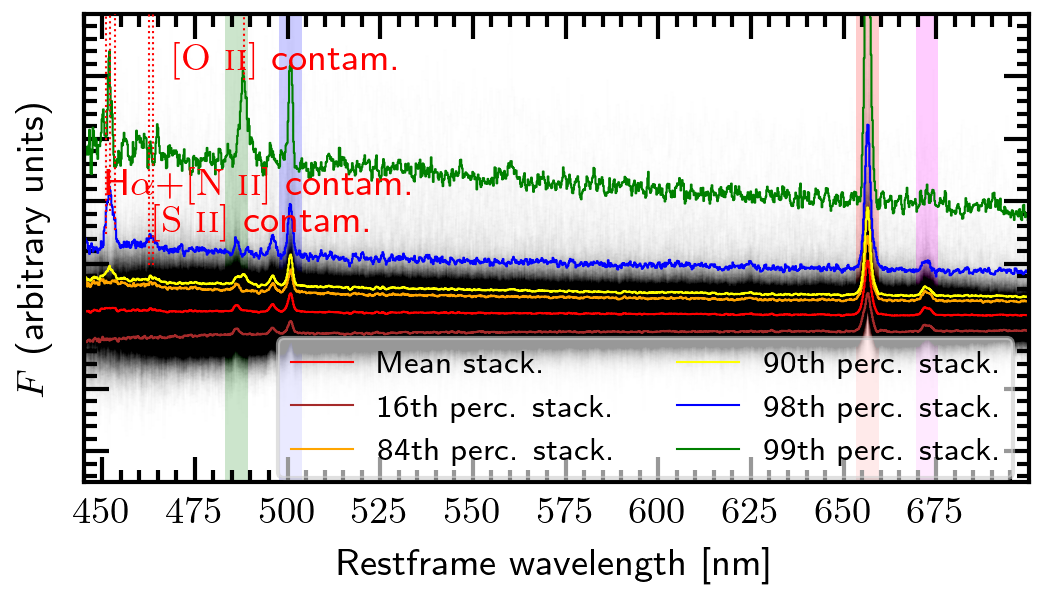}
\caption{Stacked mean spectrum of simulated \mambo galaxies from the EWS with stellar masses in the range \(10^{10} \, M_\odot \leq M_* \leq 10^{10.5} \, M_\odot\) and redshifts \(1.52 \leq z \leq 1.86\). 
The stacked is composed of a mixture of 2642 spectra, with contributions from galaxies with accurate redshifts (51\%), as well as redshift contaminants arising from noise spikes (36\%), \siiib misclassified as \halpha (8\%), and \oiii misclassified as \halpha (2\%), plus minor contributions from other misclassified emission lines ($\sim5\%$ in total).
The plot includes the mean spectrum (red) and composite spectra at the 16th, 84th, 90th, 98th, and 99th percentiles (brown, orange, yellow, blue, and green, respectively). The contribution of individual galaxy spectra is shown in black.}
\label{fig:stack_mixture}
\end{figure}
To summarise the impact of redshift contaminant galaxies in \Euclid stacked spectra, in Fig.\,\ref{fig:stack_mixture} we present a composite 
built from 2642 \ewssf individual spectra, 51\% of which with an accurate redshift measurement, while the remaining 49\% are a mixure of different types of redshift contaminants. 
The mean spectrum demonstrates the resilience of the composite stacking approach, as it is minimally affected by redshift contaminants. However, the percentile composite spectra reveal intriguing features. 
Notably, above the 90th percentile (and faintly visible also in the 84th percentile composite), spectral features attributed to redshift contaminants emerge. These features include the \halpha + [N~\textsc{ii}] complex and [S~\textsc{ii}] lines at wavelengths around 452.5\,nm and 463.0\,nm, respectively, corresponding to contaminants where \siiib has been misclassified as \halpha. Additionally, the \oii doublet appears near 490\,nm, originating from contaminants where \oiii has been misclassified as \halpha.

This has interesting implications for future studies. The presence of redshift contaminants in the high-percentile composite spectra suggests that stacking analyses may offer a potential diagnostic tool to qualitatively assess sample contamination. While still exploratory, this idea opens the possibility of using statistical or machine learning techniques to analyse flux distributions across percentile composites as a way to characterise contamination. In a future work, we plan to investigate whether specific outlier galaxies contributing to these percentiles can be identified, which may eventually support the development of methods to improve sample selection.

\section{\label{sec:science}Scientific cases}

The strength of applying stacking techniques specifically to \Euclid is, of course, the enormous total number of spectra that will be available, which will translate into the possibility of creating very high-S/N stacked spectra to explore the physical parameter space of galaxies and AGN in a very fine grid.
To fully exploit this potential, it is essential to first quantify the quality of the input samples in terms of success rate and contamination rate, which determine the balance between completeness and purity. Once these selection metrics are established, we can proceed to analyse the resulting stacked spectra, assessing how the adopted criteria influence the recovery of key spectral features and derived physical quantities.

\subsection{\label{sec:pandc}Success rate and contamination rate}

In this section, we introduce metrics that provide a direct measure of the trade-off between sample completeness and contamination, and are critical for evaluating the reliability of stacked spectra in both the \ewssf and \edssf configurations.
We define the following conditions:
\begin{itemize}
     \item z(true): galaxies that actually fall within a certain redshift interval  ($z_\textrm{true} \in [z_\textrm{min},z_\textrm{max}]$);
    \item z(meas): galaxies that are measured to be in the defined redshift interval  ($z_\textrm{meas} \in [z_\textrm{min},z_\textrm{max}]$);
    \item z(accur): galaxies satisfying the redshift accuracy requirement (\dzabs $\leq 0.003$);
    \item z(prob): galaxies with a \zprob (redshift probability) above a certain threshold (between 0 and 1); 
    \item F(true): galaxies whose line of interest has true flux greater than a given threshold ($F_\textrm{true} > $ $F_\textrm{threshold}$);   
    \item F(meas): galaxies whose line of interest has a measured flux greater than the threshold ($F_\textrm{meas} > $ $F_\textrm{threshold}$);
    \item S($\theta$): to take into account additional constraints or conditions. For example, $\theta$ could be a condition to select galaxies within a certain stellar mass (as in the following section), or a condition on the minimum S/N of a certain measured spectral feature, etc. 
\end{itemize}
With these definitions, we can express the success rate (\srate) as the ratio of galaxies in a certain population that are successfully detected to the total number of such galaxies that exist above the defined thresholds

\begin{equation}\label{eq:sr}
    \mathcal{SR} \!=\!
    \frac{\#\left[ \text{z(meas)} \,{\scriptstyle\&}\, \text{z(accur)} \,{\scriptstyle\&}\, 
    \text{z(prob)} \,{\scriptstyle\&}\, \text{F(true)} \,{\scriptstyle\&}\, 
    \text{F(meas)}\,{\scriptstyle\&}\, \text{S}(\theta)\right]}
    {\#\left[ \text{z(true)} \,{\scriptstyle\&}\, \text{F(true)} \,{\scriptstyle\&}\, 
    \text{S}(\theta)\right]}.
\end{equation}

In particular, we are interested in an EWS sample selected in the redshift interval extended between 0 and 4, including \halpha at $0.9 \lesssim z_\textrm{min} \lesssim 1.8$, as well as other emission lines, such as \pabeta at lower redshifts or \oii at higher redshifts. Finally, we aim to reach depth, with thresholds as low as $2 \times 10^{-17}$ \flcgs in both our surveys.

Similarly, we define the contamination rate (\cont) as a parameter that quantifies the fraction of misclassified sources relative to the observationally selected sample
\begin{equation}\label{eq:contam}
    \mathcal{C} = \frac{\#\left[\text{z(meas)} \,{\scriptstyle\&}\, \neg\text{z(accur)} \,{\scriptstyle\&}\, \text{z(prob)} \,{\scriptstyle\&}\,  \text{F(meas)}\,{\scriptstyle\&}\, \text{S}(\theta)\right]}{\#\left[ \text{z(meas)}\,{\scriptstyle\&}\, \text{z(prob)} \,{\scriptstyle\&}\, \text{F(meas)} \,{\scriptstyle\&}\, \text{S}(\theta)\right]}.
\end{equation}
Here, the symbol \( \neg \) denotes negation, indicating the complement of a condition (e.g., \( \neg\text{z(accur)} \) represents galaxies that do not satisfy the redshift accuracy requirement).

Finally, we define an optimal observational selection that represents the best selection sample that one can achieve with \Euclid for a given redshift range and flux threshold, and with no redshift contaminants. 
We call this the benchmark sample (\bm), defined as:
\begin{equation} \label{eq:bm}
    \mathcal{BM} = \text{z(meas)} \,{\scriptstyle\&}\, \text{z(accur)}\,{\scriptstyle\&}\,  \text{F(true)} \,{\scriptstyle\&}\,  \text{F(meas)} \,{\scriptstyle\&}\, \text{S}(\theta)\,
\end{equation}

\subsection{\label{sec:stacked_intro}Sample selection for stacking}
In this section, we examine a preliminary
finding which will serve as the basis for analysing key scaling relations relevant to galaxy evolution in the following section. 
\begin{figure*}
\centering
\includegraphics[width=0.495\linewidth]{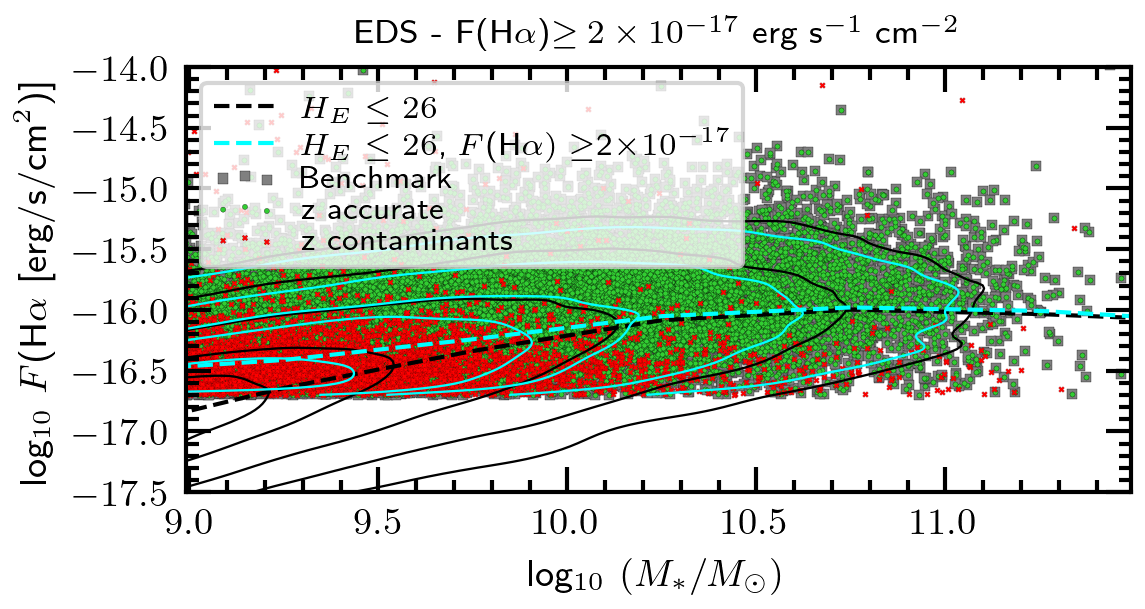}
\includegraphics[width=0.495\linewidth]{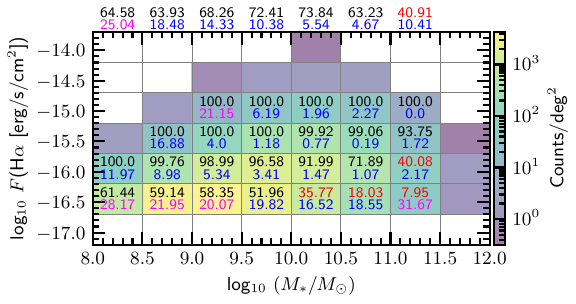}\\
\vspace{0.005cm}
\includegraphics[width=0.495\linewidth]{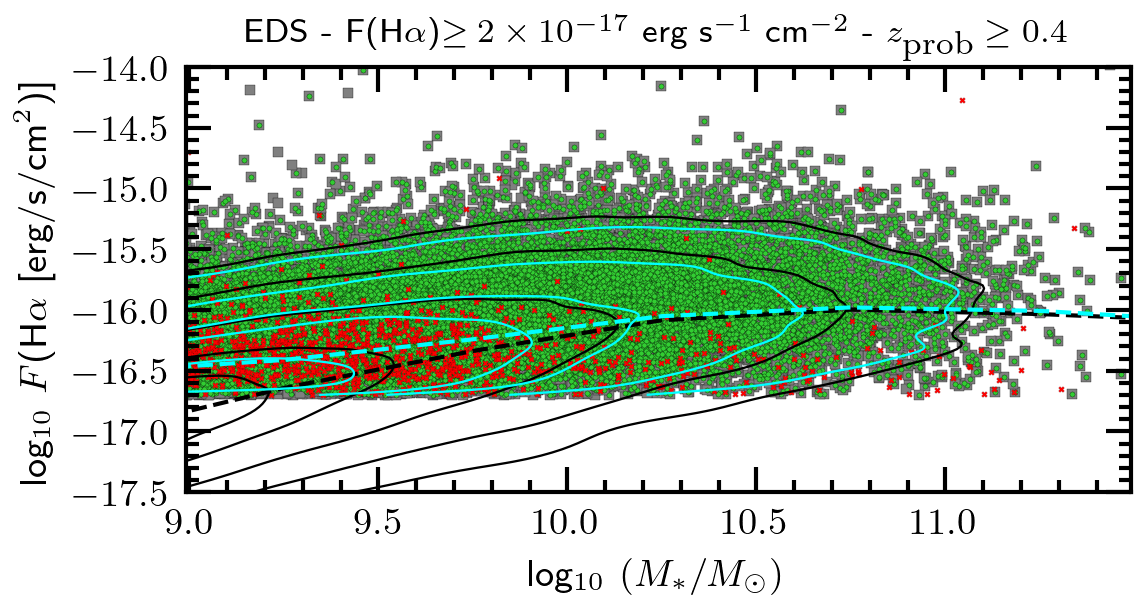}
\includegraphics[width=0.495\linewidth]{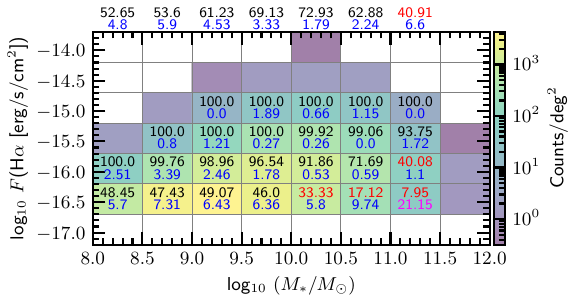}\\
\caption{
The simulated \halpha emission line as a function of stellar mass for the \edssf selection is shown for the lowest redshift range: $0.9 \leq \text{\zmeas} \leq 1.2$. The top panels present the relation at the flux limit of $F$(\halpha)$ \geq 2\times10^{-17}$ \flcgs and with no cut in \zprob, whilst the bottom panels present the improvement obtained with our fiducial selection that also includes a cut in redshift probability \zprob $\geq 0.4$.
The left panels show the distribution of different subsamples: \bm sample (grey points), observationally selected galaxies with accurate redshift (green points), and redshift contaminants (red points). Black and cyan contours/dashed lines show the ground-truth distributions for the full EDS parent sample (\HE $\leq$ 26) and a subsample selected with $F$(\halpha)$ \geq 2\times10^{-17}$ \flcgs, respectively. These contours serve as a reference to illustrate the intrinsic (ground-truth) distributions, allowing a qualitative comparison with the observation-like measurements, and to highlight the intrinsic selection bias introduced by the flux limit, which would persist even without redshift measurement errors. The right panels show galaxy number densities per deg$^2$ in bins of stellar mass and \halpha flux, with \srate (top number) and \cont (bottom number) indicated. Numbers are colour-coded: black/blue for reliable bins (\srate > 45\%, \cont < 20\%) and red/magenta for others. The numbers reported above each mass bin represent the \srate (top number) and \cont (bottom number) for the entire mass bin.
}
    \label{fig:ha_vs_m_deep}
\end{figure*}
Based on the selection criteria detailed in Appendix~\ref{sec:selection_test},  which aim to maximize \srate and minimize \cont, the left panels of Fig.\,\ref{fig:ha_vs_m_deep} show the \halpha emission line distribution as a function of stellar mass for the \edssf selection in the redshift interval $0.9 \leq \text{\zmeas} \leq 1.2$. 
The left panels show the distribution of the \bm sample, 
galaxies with accurate redshift measurements, 
and redshift contaminants. 
For contaminants, the $y$-axis reflects misidentified spectral features interpreted as \halpha by the pipeline. 
The top panel shows the relationship at the flux limit of $F$(\halpha)$ \geq 2\times10^{-17}$ \flcgs without applying a cut in redshift probability. In contrast, the bottom panel demonstrates the effects introduced by our fiducial selection, which includes a cut at \zprob $\geq 0.4$. 
The contours in both panels show the ground-truth distributions for the full \edssf parent sample (\HE $\leq 26$) and a subsample with $F$(\halpha) $\geq 2\times10^{-17}$ \flcgs, respectively, along with 
their median trends.
Their comparison highlights a critical limitation: even in the absence of instrumental effects, applying a flux threshold introduces a strong bias toward higher \halpha fluxes, particularly at stellar masses below $10^{10} M_\odot$. As a result, global scaling relations involving \halpha cannot be reliably recovered at low masses using flux-limited samples such as the \Euclid spectroscopic selection. This bias is intrinsic to the selection function and not a consequence of the stacking procedure itself.
Nevertheless, at stellar masses above $10^{10} M_\odot$, the two ground-truth curves agree within 0.05--0.1 dex, suggesting that in this regime it may still be possible, in principle, to recover unbiased global scaling relations involving \halpha.

Building on this, the bottom panel of Fig.\,\ref{fig:ha_vs_m_deep} demonstrates that applying a moderate quality cut (\zprob > 0.4) reduces the contribution of contaminants while still retaining a high level of completeness. This trade-off is critical to confirm that meaningful statistical trends can be extracted from the flux-limited sample when appropriate quality filters are applied.

To quantitatively assess the effect of this enhancement, we constructed the right panels of Fig.\,\ref{fig:ha_vs_m_deep}, where galaxies are binned in both stellar mass and \halpha flux, and colour-coded by surface density. The corresponding \srate and \cont values are reported for each bin, highlighting which regions of parameter space yield reliable measurements: bins with \srate $> 45$\% and \cont $<20$\% are shown in black and blue, while 
less reliable bins are shown in red and magenta.
This simulation highlights the potential of the full EDS survey to map the H$\alpha$–stellar mass relation with high fidelity, with most of the Mass and SFR bin with high completeness and low contamination, in particular if we include a cut on \zprob. Covering $\sim$50 deg$^2$ of sky, the EDS will enable stacking in finer bins than 0.5 dex, thereby boosting the S/N in composite spectra. However, the present mock catalogue, limited to 3.14 deg$^2$, includes fewer galaxies per bin, constraining the achievable S/N. To optimize both the statistical robustness and sample homogeneity, we therefore perform stacking in bins of stellar mass alone, rather than jointly in stellar mass and \halpha flux. This choice ensures that each bin retains a representative population while maximising the number of spectra per stack, thus enabling reliable derivation of scaling relations despite the selection effects. Note that \srate and \cont for each stellar mass bin are reported at the top of the figure, being always \srate>50--60\% but the most massive bin and \cont<7\% if a \zprob cut is used.

\begin{figure*}
\centering
\includegraphics[width=0.495\linewidth]{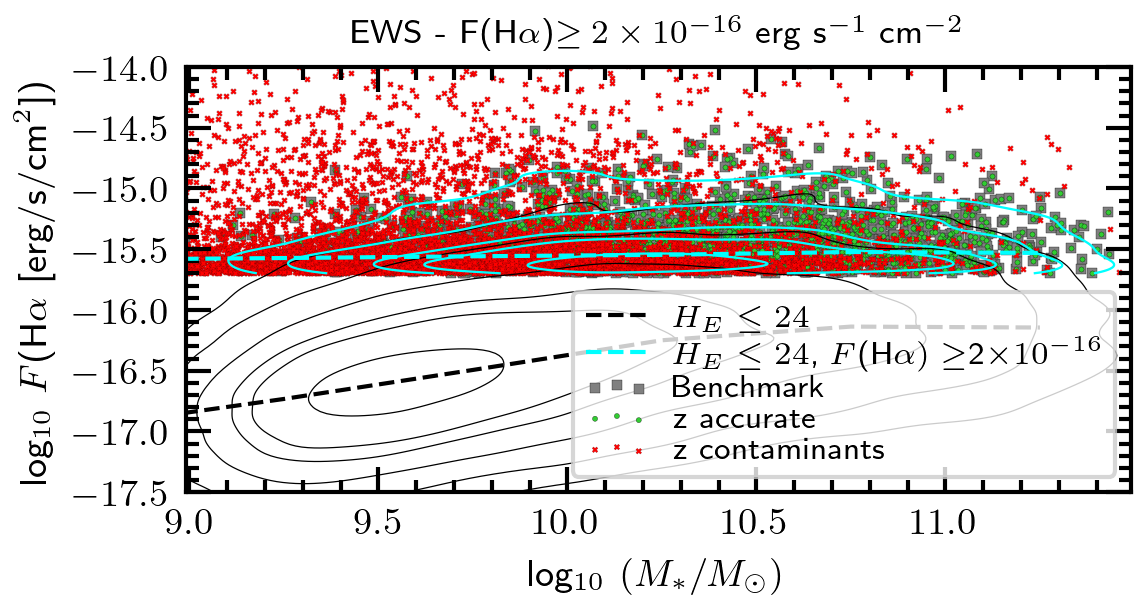}
\vspace{0.005cm}
\includegraphics[width=0.495\linewidth]{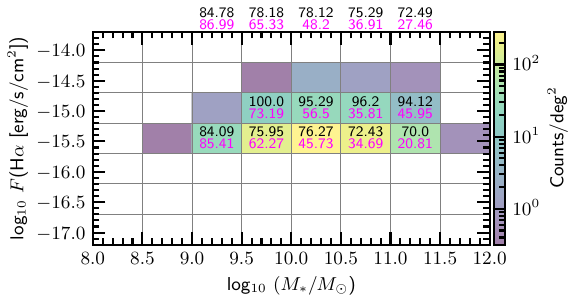}\\
\includegraphics[width=0.495\linewidth]{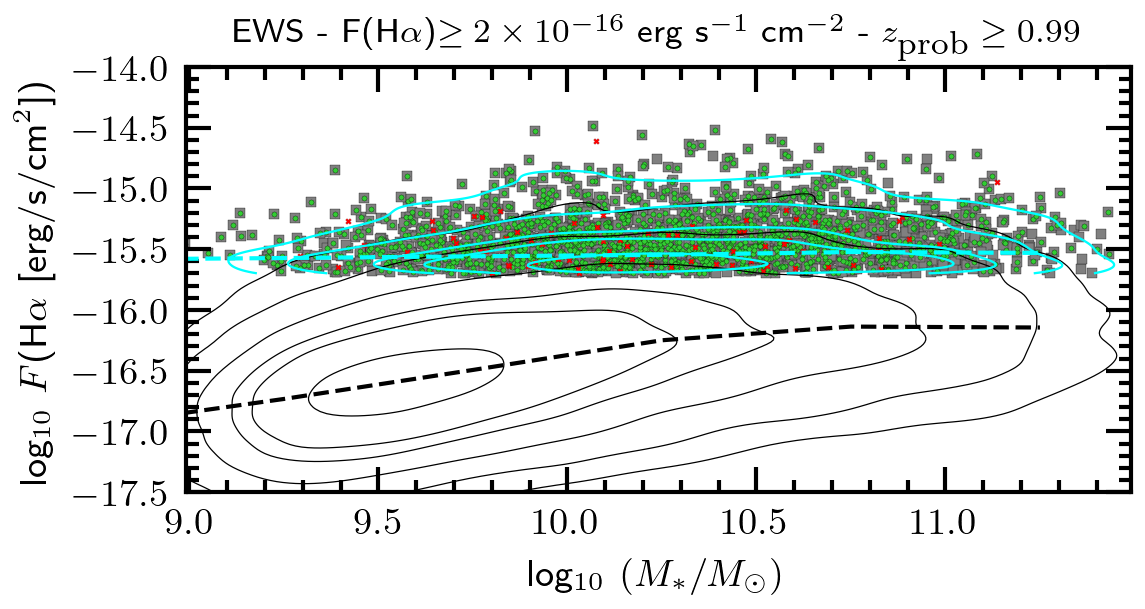}
\vspace{0.005cm}
\includegraphics[width=0.495\linewidth]{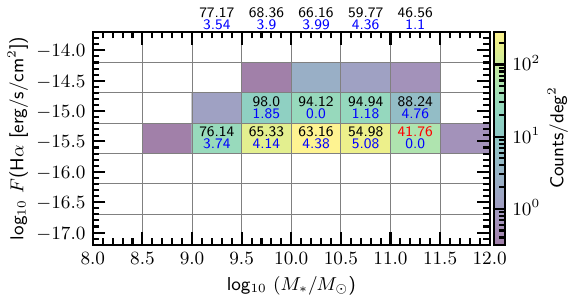}\\
\caption{The simulated \halpha emission line as a function of stellar mass for the \ewssf selection. The top panels present the relation at the flux limit of $F$(\halpha)$ \geq 2\times10^{-16}$ \flcgs and with no cut in \zprob, whilst the bottom panels present the improvement obtained with our fiducial selection that also includes also a cut in redshift probability \zprob $\geq 0.99$.
The layout is the same as in Fig.\,\ref{fig:ha_vs_m_deep}, the only difference is that the black and cyan contours/dashed lines show the ground-truth distributions for the full EWS parent sample (\HE $\leq$ 24) and a subsample selected with $F$(\halpha)$ \geq 2\times10^{-16}$ \flcgs, respectively. Note that in this case, the bias between the cut in magnitude only and the cut in magnitude and flux limit is larger than the \edssf one. Therefore, even with stacking, we will be mapping scaling relations of galaxies with \halpha flux above $2\times10^{-16}$ \flcgs.} \label{fig:ha_vs_m_wide}
\end{figure*}
Figure\,\ref{fig:ha_vs_m_wide} shows the \halpha emission line as a function of stellar mass for the \ewssf sample. The figure highlights the impact of applying a flux limit of $F$(\halpha) $> 2\times10^{-16}$ \flcgs, showing a bias between the magnitude-selected sample and the sample with the additional flux limit. 
This bias is larger than that estimated for the \edssf sample (Fig.\,\ref{fig:ha_vs_m_deep}), particularly at lower stellar masses. 
This discrepancy arises due to the brighter flux limit of the \ewssf sample, which inherently excludes a larger fraction of lower-mass, fainter galaxies. 
The top panels of Fig.\,\ref{fig:ha_vs_m_wide} show the relationship at the flux limit of $F$(\halpha) $> 2\times10^{-16}$ \flcgs without imposing a redshift probability cut. 
In comparison, the bottom panels highlight the improvements achieved using our fiducial selection (see Appendix~\ref{sec:selection_test}), which applies a cut at \zprob $\geq 0.99$, demonstrating once more that our refined approach substantially reduces contaminant contributions while preserving an acceptable level of completeness within the regime achievable with the EWS's sensitivity.
Thanks to the vast sky coverage, the EWS will provide large statistical power to investigate scaling relations for galaxies with \halpha flux above $2\times10^{-16}$ \flcgs. Similarly to the \edssf, stacking spectra in bins of stellar mass alone, rather than in bins defined by both stellar mass and \halpha flux, will optimise the number of galaxies included in each composite spectrum. This approach ensures that, even within the limited volume of our simulation, stacking can yield high-quality composite spectra at the high-flux, high-mass end of \halpha scaling relations with high statistical precision.
Note that \srate and \cont for each stellar mass bin are reported at the top of the figure, being always \srate>50--60\% but the most massive bin and \cont<5\% if a \zprob cut is used.

\subsection{\label{sec:stacked_spectra_intro}Properties of the stacked mock spectra}
\begin{figure}
\centering
\includegraphics[width=\linewidth]{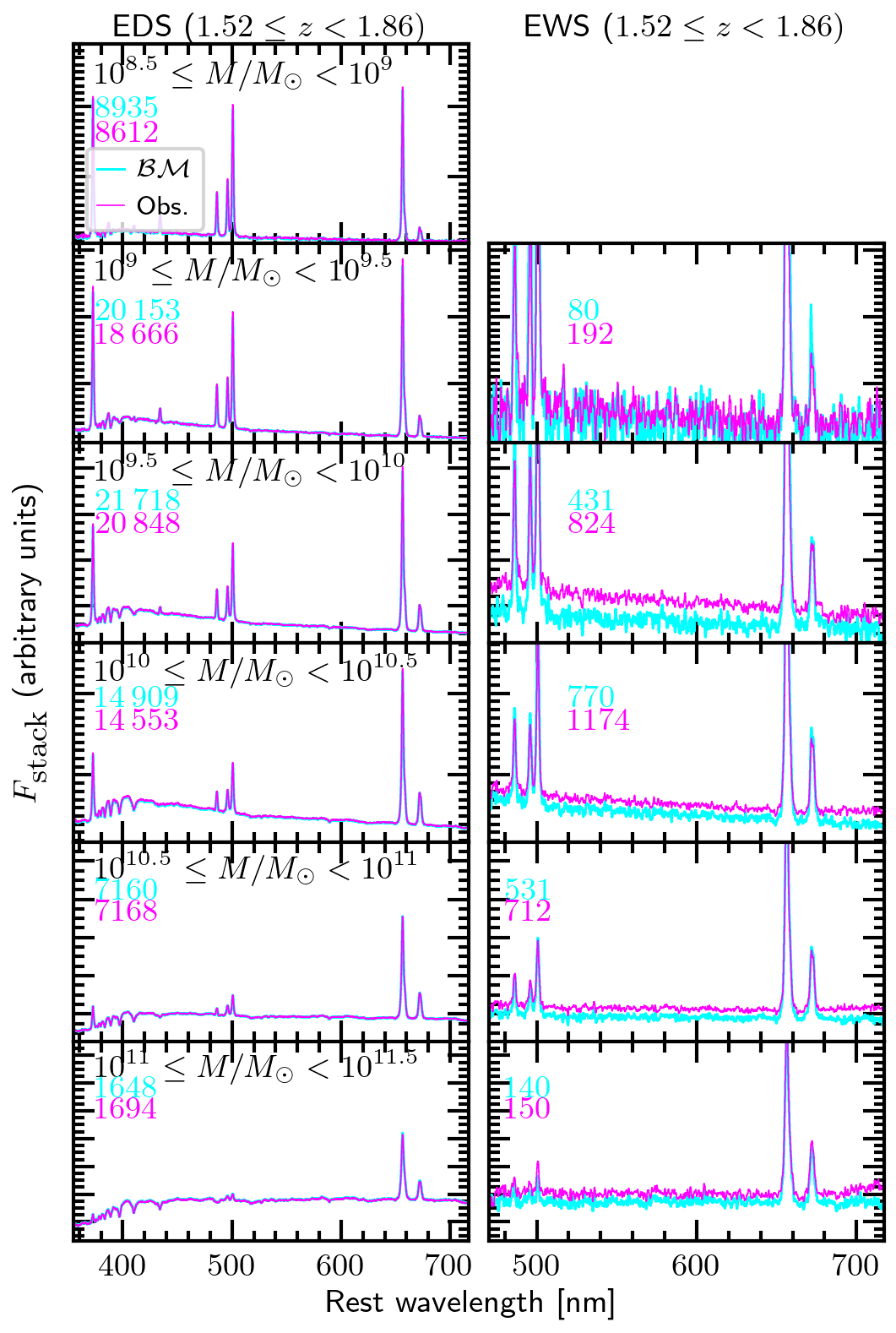}\\
\caption{Stacked mean spectra for the mock galaxies used for the scientific predictions in this paper, for the redshift bin $[1.52, 1.86]$.
The panels show stacked spectra divided by sample type and stellar mass bins: left panels show the \edssf, while right panels show the \ewssf. Rows correspond to stellar mass bins of 0.5 dex, ranging from $\log_{10}(M_*/M_\odot) \geq 8.5$ to $\leq 11.5$. Bins with fewer than 50 spectra are not shown or considered in the analysis. In each panel, cyan and magenta spectra represent the stacked spectra for the \bm and observationally selected samples, respectively, with the numbers indicating the spectra counts.
Note that the \oiii/\oiiia\ ratios of all stacks here are approximately 2.3, lower than the theoretical value of 3, due to an error in the simulated \oiiia\ emission line. Since the \oiii\ values are correct and \oiiia\ was not used in the analysis, the results remain unaffected.}
\label{fig:stack_specs}
\end{figure}
Building on the selection tests presented in Appendix~\ref{sec:selection_test} and the findings of the previous subsection, we generated stacked spectra using:
\begin{itemize}
    \item \ewssf: $F > 2 \times 10^{-16}$ \flcgs and \zprob $\geq 0.99$;
    \item\edssf: $F > 2 \times 10^{-17}$ \flcgs and \zprob $\geq 0.4$.
\end{itemize}

In the case of \edssf, stacked spectra span three redshift bins ([0.9, 1.2], [1.2, 1.52], and [1.52, 1.86]) and are further divided into stellar mass bins of 0.5 dex, ranging from \(\log_{10}(M_*/M_\odot) \geq 8.5\) to \(\leq 11.5\). For the \ewssf, stacked spectra were created focusing on the redshift bin [1.52, 1.86], where the key emission lines from \hbeta to \nii are present in all spectra. 
The stacked spectra were produced with \stsp, using the \textit{regular} normalization option with \textit{luminosity conservation} (Sect.\,\ref{sec:method}), applying $4\sigma$ sigma clipping to remove outliers, and 350 bootstrap resamplings to estimate uncertainties. All spectra were shifted to the rest frame and resampled at 6~\AA\ per pixel.

Figure\,\ref{fig:stack_specs} shows the rest-frame stacked spectra of galaxies in the redshift bin [1.52, 1.86], for qualitative inspection. 
The \edssf stacks have high S/N, whereas the \ewssf stacks, particularly in the lowest mass bins, appear noisier. This difference arises from both the intrinsically higher S/N of individual \edssf spectra and the limited simulation area of approximately 3 deg$^2$. In the actual EWS survey, the larger sky coverage will yield higher S/N in the stacked spectra.
The observationally selected \edssf spectra 
closely match their benchmark counterparts, 
indicating that the adopted selection criteria effectively recover the same underlying galaxy population. Within each stellar mass bin, the stacked spectra become progressively brighter at lower redshifts, consistent with the reduced distances to these sources from us. Moreover, the continua become redder with increasing stellar mass, which reflects the diminished contribution of blue light from younger stellar populations. The emission line ratios also evolve: the \halpha/\hbeta ratio increases with stellar mass, suggesting stronger dust attenuation, while the \oiii/\halpha ratio decreases, driven by both increased dust extinction and a lower fraction of ionizing photons from young stars in more massive galaxies \citep[e.g.,][]{Citro2017}. The high S/N of the \edssf stacks also reveals absorption features such as the NaD doublet around $0.59\,\micron$, which becomes especially prominent at $\log_{10}(M_*/M_\odot) \geq 9.5$. Once the full survey area is covered, the enhanced S/N of the stacked spectra will enable the detection of fainter emission and absorption features, expanding the range of scientific analyses that can be pursued with \Euclid.

\section{\label{scaling_relations}Scaling relations with \Euclid stacked spectra}

In this section, we explore key scaling relations using the most prominent emission lines in the optical domain. The fluxes of these lines were measured in the stacked spectra described in the previous section, using  \texttt{slinefit}\footnote{The \texttt{slinefit} code is publicly available at \url{https://github.com/cschreib/slinefit}} \citep{Schreiber2018, Talia2023}. This fitting process includes continuum subtraction and deblending of \halpha from \nii, ensuring accurate flux measurements.
For the \edssf sample, we focus on the highest redshift bin of the \edssf selection ($1.52 \leq z \leq 1.86$), which also overlaps with the redshift coverage of the \ewssf sample. Equivalent plots for \edssf in the lower redshift bins ($0.9 \leq z \leq 1.2$ and $1.2 \leq z \leq 1.52$) are provided in Appendix~\ref{fig:line_vs_M_stack_DEEP_other_z}.

\subsection{Emission lines - Stellar mass}
\begin{figure}
\centering
\includegraphics[width=0.95\linewidth]{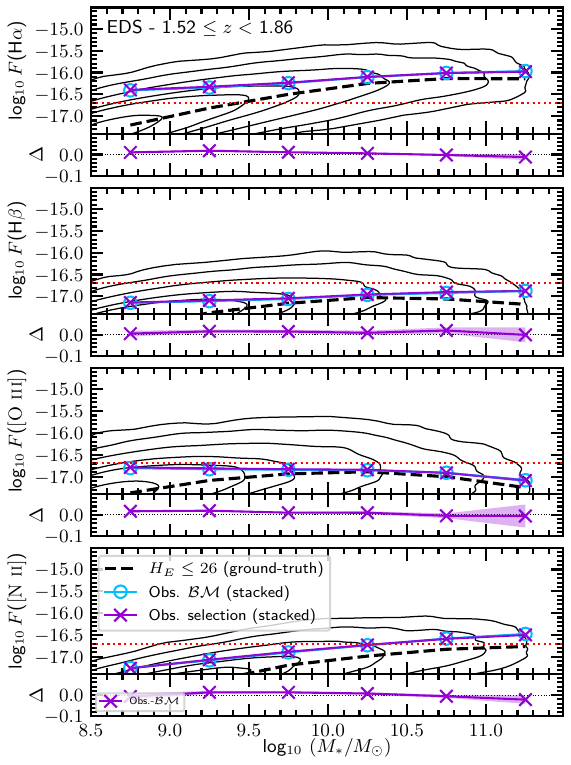}\\
\caption{Emission line fluxes (in units of \flcgs) measured on stacked spectra as a function of stellar mass for the \edssf selection corresponding to the redshift interval $1.52 \leq z \leq 1.86$. From top to bottom, panels show the fluxes of \halpha, \hbeta, \oiii, and \nii emission lines. Fluxes have been corrected for flux loss using the method by Cassata et al. (in preparation). Magenta curves represent measurements from observationally selected galaxy stacks, whilst cyan circles correspond to benchmark (\bm) stacks. 
The red dotted horizontal line in each panel indicates the flux limit of $2\times10^{-17}$ \flcgs imposed on the \edssf. Shaded regions denote the $1\sigma$ uncertainty, whilst black contours and the dashed black line in each panel trace the ground-truth distribution and median relation of the parent \edssf sample (\HE\ $\leq$ 26), respectively. The bottom insets in each panel show the difference (in dex) between the observational and benchmark flux measurements. 
}
\label{fig:line_vs_M_stack_DEEP}
\end{figure}
The fluxes of prominent emission lines measured from stacked mock spectra as a function of stellar mass provide a diagnostic of the performance of the \edssf selection across redshift and stellar mass bins. Figure\,\ref{fig:line_vs_M_stack_DEEP} presents the fluxes of \halpha, \hbeta, \oiii, and \nii, corrected for flux losses following the methodology outlined in Cassata et al. (in preparation), within the redshift interval $1.52 \leq z \leq 1.86$.
Magenta curves show flux measurements from mean stacks of observationally selected galaxies, while cyan circles represent the mean fluxes from the benchmark (\bm) selection. The difference between these is displayed in the bottom inset of each panel (in purple, in dex), along with 
the $1\sigma$ uncertainties. Across all stellar mass bins, the observational selection reproduces the benchmark fluxes with remarkable fidelity, with typical deviations ranging between 0.001 and 0.06 dex.
Notably, the agreement with benchmark values improves at higher stellar masses, particularly above $10^{10}$ M$_\odot$, where differences are reduced to between 0.001 and 0.03 dex.
The limited impact of residual contaminants on the flux arises because, as shown in Fig.\,\ref{fig:HaMeas_hist}, their fluxes generally fall below the surveys’ flux limits, contributing mainly low values to the stack and minimally affecting the composite measurements.

Black contours and the dashed black line in each panel trace the ground-truth distribution and median relation of the parent \edssf sample (\HE\ $\leq$ 26), respectively. The comparison between fluxes measured from the stacked spectra and the photometric ground-truth again highlights the intrinsic bias introduced by the flux-limited selection of \edssf, which is independent of the stacking process. This bias is especially evident at stellar masses below $10^{10}$ M$_\odot$ (see Sect.\,\ref{sec:science}).

\begin{figure}
\includegraphics[width=0.95\linewidth]{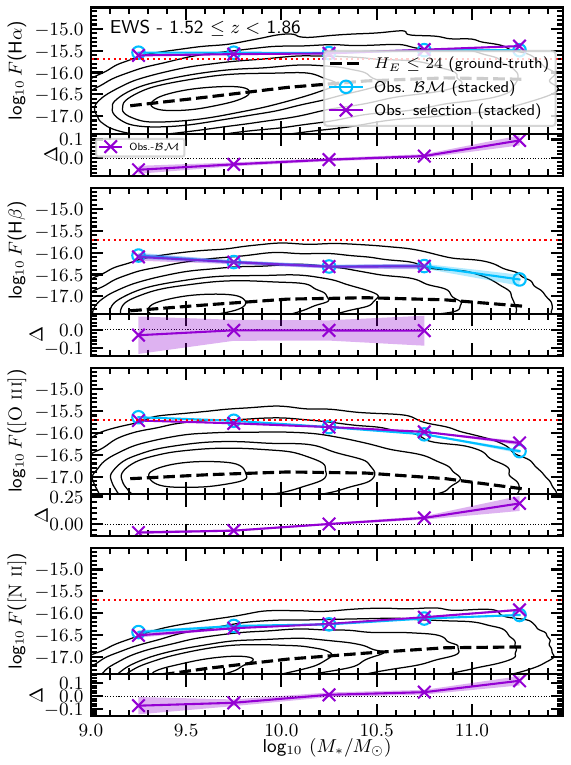}\\
\caption{Emission line fluxes measured on stacked mock spectra as a function of stellar mass for the \ewssf selection. From top to bottom, panels display the fluxes of \halpha, \hbeta, \oiii, and \nii emission lines. The red dotted horizontal line in each panel indicates the flux limit of $2\times10^{-16}$ \flcgs imposed on the \halpha line. The layout of the panels is the same as in Fig.\,\ref{fig:line_vs_M_stack_DEEP}.} 
    \label{fig:line_vs_M_stack_WIDE}
\end{figure}
For the \ewssf selection, the fluxes of prominent emission lines measured on stacked spectra as a function of the stellar mass are shown in Fig.\,\ref{fig:line_vs_M_stack_WIDE}. 
We find that all lines we analysed, such as \halpha, \hbeta, \oiii, and \nii, show a significant positive bias compared to those of the ground-truth parent sample at \HE $\leq$ 24. These elevated values reflect the more stringent flux limit of $2\times10^{-16}$ \flcgs we imposed on the \ewssf, which preferentially selects galaxies with brighter emission lines (see Sect.\,\ref{sec:science}). 
The consistently narrow $\sigma$ uncertainties across all stellar mass bins and emission lines demonstrate the strength of the \ewssf sample in probing bright systems with high statistical precision. This capability is particularly valuable for studying the most active phases of galaxy evolution, such as starbursts, where accurate measurements of emission line properties are essential for constraining physical conditions of the most active star formation phases.

\subsection{The stellar mass attenuation diagram and the Star-formation rate-Stellar mass relation}
The \enquote{main sequence} (MS) of SFGs describes a well-established, tight correlation between total stellar mass (M$_\ast$) of a galaxy and its SFR 
\citep[e.g.,][]{ Noeske2007, Daddi2007, Wuyts2011, Rodighiero2011, Rodighiero2014, Popesso2023}, 
showing a clear evolutionary trend across cosmic time.

The \halpha emission line, emitted by ionized hydrogen in star-forming regions, is a valuable tracer for measuring the SFR in star-forming galaxies, particularly because it is well-calibrated for use with local galaxies \citep{Kennicutt1998, Hopkins2003, Kennicutt2012}, 
and can be 
observed in surveys like the EWS at redshifts $0.9 < z < 1.8$ and the EDS at redshifts $0.4 < z < 1.8$. 
By leveraging these observations, we can potentially extend the main sequence characterization to earlier epochs, providing insights into the star-formation history and the evolving galaxy population over time and extending it to the most massive galaxies.
\begin{figure}
\centering
\includegraphics[width=0.95\linewidth]{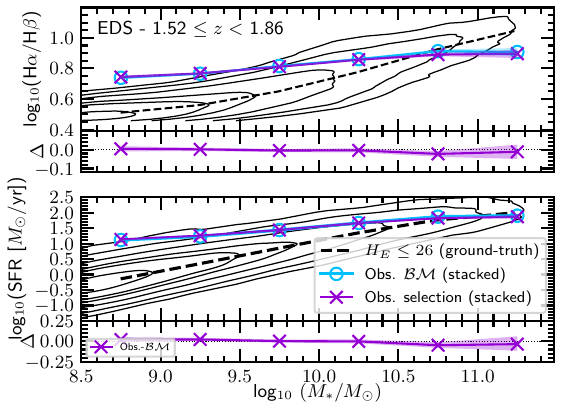}\\
\caption{The attenuation–$M_*$ relation (top panels) and the SFR–$M_*$ relation (bottom panels) derived from the \edssf stacked spectra. The attenuation is measured using the Balmer decrement (\halpha/\hbeta ratio), while the SFR is calculated from the dust-corrected \halpha luminosity, applying the \cite{Calzetti2000} extinction law and assuming case B recombination conditions (electron density of 100 cm$^{-2}$ and $10\,000$ K. 
The panel layout follows the structure of Fig.\,\ref{fig:line_vs_M_stack_DEEP}. Shaded regions represent the $1\sigma$ uncertainties in the measured quantities.
}
\label{fig:SFR_vs_M_stack_DEEP}
\end{figure}
Fig.\,\ref{fig:SFR_vs_M_stack_DEEP} shows the attenuation–$M_*$ relation (top panels) and the SFR–$M_*$ relation (bottom panels) derived from the \edssf stacked mock spectra.
The attenuation, represented by the Balmer decrement (\halpha/\hbeta ratio not corrected for underlying Balmer absorption), shows a dependence on stellar mass, with higher attenuation observed at higher masses. In parallel, the SFR–stellar mass relation highlights the ability of the EDS to probe star-forming galaxies at stellar masses above $\sim 10^{10}\, M_\odot$, where the measured SFR closely aligns with the expected relation for all galaxies with \HE $< 26$. However, for galaxies in the lower-mass regime ($<10^{10}\, M_\odot$), deviations of up to 1 dex below the expected SFR–$M_*$ relation are evident. This discrepancy suggests that while the deeper flux limits of the EDS improve the sensitivity to lower-mass systems compared to the EWS, systematic offsets remain, 
due to the flux limit of the survey as evident from Fig.\,\ref{fig:line_vs_M_stack_DEEP}.

\begin{figure}
\includegraphics[width=0.95\linewidth]{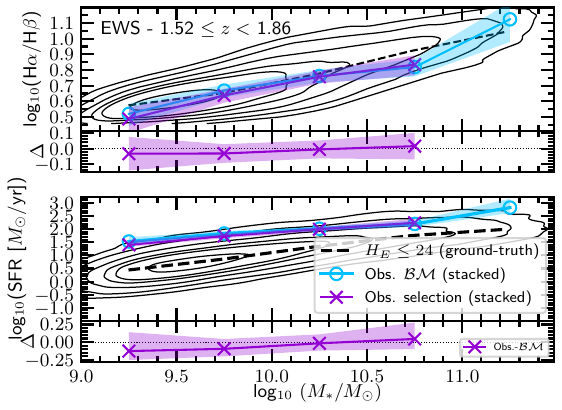}\\
\caption{The attenuation–$M_*$ relation (top panel) and the SFR–$M_*$ relation (bottom panel) derived from the \ewssf stacked mock spectra. The attenuation is measured using the Balmer decrement (\halpha/\hbeta ratio), whilst the SFR is calculated from the dust-corrected \halpha luminosity, applying the \cite{Calzetti2000} extinction law and assuming case B recombination conditions (electron density of 100 cm$^{-2}$ and temperature of $10\,000$ K.. The panel layout follows the structure of Fig.\,\ref{fig:line_vs_M_stack_DEEP}. Shaded regions represent the $1\sigma$ uncertainties in the measured quantities.}
    \label{fig:SFR_vs_M_stack_WIDE}
\end{figure}

Figure\,\ref{fig:SFR_vs_M_stack_WIDE} shows the attenuation–stellar mass relation (top panels) and the SFR–$M_*$ relation (bottom panels) derived from the \ewssf stacked spectra.
For the \ewssf sample, the attenuation–stellar mass relation derived from the Balmer decrement (\halpha/\hbeta ratio) overlaps with the reference relation of the \HE $\leq 24$ sample. 
This agreement indicates that the corrections for dust extinction applied to \halpha fluxes from the stacked spectra align closely with those of the reference sample, ensuring comparable dust attenuation estimates. 
Consequently, the SFRs derived for the \ewssf reflect similar biases to those observed in the \halpha flux versus stellar mass relation, as presented in the top panel of Fig.\,\ref{fig:line_vs_M_stack_WIDE}. 
This reinforces that while the \ewssf expands the reach of the survey to larger sky areas, the derived SFR scaling relations remain systematically affected by the high \halpha flux limit relative to the benchmark magnitude-limited sample at all stellar masses, and especially at the low-mass end.

\subsection{The BTP diagram}
Diagnostic diagrams that combine two emission-line ratios are capable of distinguishing the dominant ionizing sources in galaxies, whether from young, massive stars formed in recent star formation (SF) or from an AGN. 
The well-known Baldwin–Phillips–Terlevich (BPT) diagrams \citep{Baldwin1981, Veilleux1987}, which compare the ratios [O~\textsc{iii}]$\lambda 5007$/H$\beta$ with [N~\textsc{ii}]$\lambda 6584$/H$\alpha$, [S~\textsc{ii}]$\lambda 6724$/H$\alpha$, and [O~\textsc{i}]$\lambda 6300$/H$\alpha$, have proven effective for differentiating ionising sources in nearby galaxies \citep{Kewley2001, Kauffmann2003}, 
while it remains uncertain whether the BPT diagrams will maintain their effectiveness at higher redshifts. As galaxies evolve, their physical conditions and chemical compositions change, potentially complicating the interpretation of these traditional diagnostic diagrams in the distant Universe.  \Euclid will allow us to extend with unprecedent statistic previous studies using only small sample 
observed within the Fiber Multi-Object Spectrograph (FMOS)-COSMOS
survey \citep{Kashino19} and the MOSFIRE Deep Evolution Field (MOSDEF) survey with Keck by \cite{Kriek15} and \cite{Reddy15} in the redshift range $1.4 < z < 1.7$. Further studies are indeed required to assess whether these diagnostic tools can reliably classify ionising sources in high-redshift galaxies. 

\begin{figure}
\centering
\includegraphics[width=0.95\linewidth]{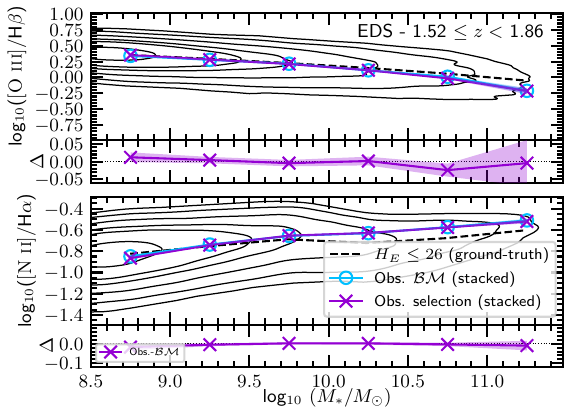}\\
\vspace{-0.9mm}
\hspace{2mm}
\includegraphics[width=0.93\linewidth]{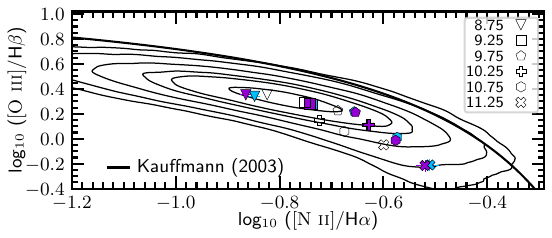}
\caption{The BPT diagram and corresponding stellar mass relations for the \oiii/\hbeta and \niib/\halpha ratios based on the \edssf stacked mock spectra. The top and middle panels show the \oiii/\hbeta–stellar mass \citep[i.e., the MEx diagnostic diagram,][]{Juneau2011}, and \niib/\halpha–stellar mass relations, respectively, with measurements from the stacked spectra in different stellar mass bins. The bottom panel displays the BPT diagram with symbols representing the positions of the stacked spectra for various mass bins. The black contours in all panels represent the \edssf parent sample cut at \HE $\leq$ 26, and the black dashed curves in the top and middle panels represent its median relations.
}
\label{fig:BPT_stack_DEEP}
\end{figure}
In Fig.\,\ref{fig:BPT_stack_DEEP}, we analyse the \edssf BPT-\niib, which relates the \oiii/\hbeta and \niib/\halpha ratios to the ionization mechanisms galaxies. In this analysis, the top and middle panels show the relations between these emission line ratios and stellar mass, based on stacked spectra from the \edssf sample, in the redshift bin $1.52 \leq z \leq 1.86$.

The \oiii/\hbeta ratio has been shown to be a reliable proxy for the ionization parameter $U$, which traces the current ionization state in star-forming regions \citep[e.g.,][]{Citro2017, Quai2018, Quai2019}. It is also useful for distinguishing between ionization driven by recent star formation and that dominated by AGN activity, especially when contrasted against stellar mass, as demonstrated by the stellar mass-excitation (MEx) diagnostic \citep[e.g.,][]{Juneau2011}, which we employ in this paper.
We observe that the \oiii/\hbeta ratio in the stacked spectra aligns closely with the median values from the EDS parent sample cut at \HE $\leq$ 26, as indicated by the black dashed curve, across nearly all mass bins. In contrast, the \niib/\halpha ratio shows a small systematic positive offset from the relation for galaxies with \HE $< 26$.
This offset might be either a genuine difference between the \niib/\halpha in the selected and reference populations, or due to a non-perfect deblending of the \halpha and the \nii lines, which may still be affecting the measurement of the \nii flux.

Now examining the \oiii/\hbeta and \niib/\halpha ratios together in the bottom panel of Fig.\,\ref{fig:BPT_stack_DEEP}, the data points in the BPT diagram remain below the \cite{Kauffmann2003} curve, placing them in the region associated with star formation. This indicates that the dominant ionization source in the stacked spectra is star formation, as expected because our \mambo selection does not include AGN contribution. While there may be biases in individual measurements, particularly in the \niib/\halpha ratio as discussed above, this does not compromise the overall conclusion. 
The data consistently support that stacking is robust enough to preserve the position of galaxies even in BPT-[N~\textsc{ii}] diagram potentially affected by deblending issues.

\begin{figure}
\includegraphics[width=0.95\linewidth]{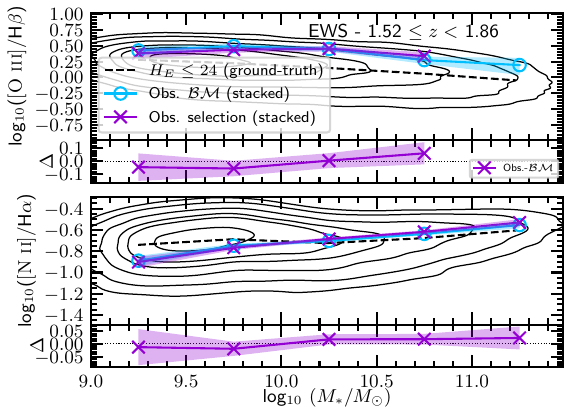}\\
\vspace{0.005mm}
\includegraphics[width=0.93\linewidth]{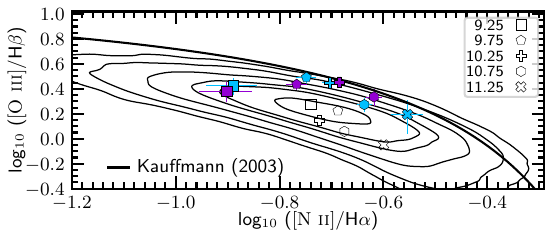}\\
\caption{The BPT diagram and corresponding stellar mass relations for the \oiii/\hbeta and \niib/\halpha ratios based on the \ewssf stacked mock spectra. The top and middle panels show the \oiii/\hbeta–$M_*$ \citep[i.e., the MEx diagnostic diagram,][]{Juneau2011}, and \niib/\halpha–$M_*$ relations, respectively, with measurements from the stacked spectra in different stellar mass bins. The bottom panel shows the BPT diagram with symbols representing the positions of the stacked spectra for various mass bins. The black contours in all panels represent the \ewssf parent sample cut at \HE $\leq$ 24, and the black dashed curves in the top and middle panels represent its median relations.}
    \label{fig:BPT_stack_WIDE}
\end{figure}

\subsection{The gas-phase metallicity}
Gas-phase metallicity, often expressed as the oxygen-to-hydrogen ratio (O/H), is a tracer of galaxy evolution, reflecting the integrated effects of star formation, gas inflows, and outflows over cosmic time. It plays a role in shaping the emission spectrum of ionized gas in galaxies, influencing the strength and ratios of nebular emission lines. Understanding gas-phase metallicity provides insights into chemical enrichment processes and the baryon cycle within galaxies \citep[e.g.][]{Kewley2019, Maiolino2019}.

Oxygen abundance is commonly estimated using strong line diagnostics calibrated against either direct temperature ($T_\textrm{e}$) measurements or photoionization models. One widely used diagnostic is the O$_3$N$_2$ parameter, defined as (\oiii/\hbeta)/(\niib/\halpha), which serves as an effective method for estimating metallicity, especially in galaxies where direct methods are impractical due to the faintness of the auroral lines required to determine $T_\textrm{e}$.  These calibrations, based on observations of local galaxies \citep[e.g.,][]{Curti2017}, should not be blindly applied to higher redshift, as several studies suggest that O/H evolves with redshift \citep[e.g.,][]{Curti2024, Sanders2024}. To address this, \citet{EP-Scharre} provided guidelines from semi-empirical methods for adjusting metallicity calibrations for high-redshift galaxies.  
However, the \mambo simulation adopts the metallicity calibration from \citet{Curti2017} for its modelled galaxies. Consequently, we derive 12+$\logten$(O/H) values for our stacked spectra and analyse how they vary with stellar mass (for both \edssf and \ewssf) and redshift (for \edssf). 

\begin{figure}
\centering
\includegraphics[width=0.95\linewidth]{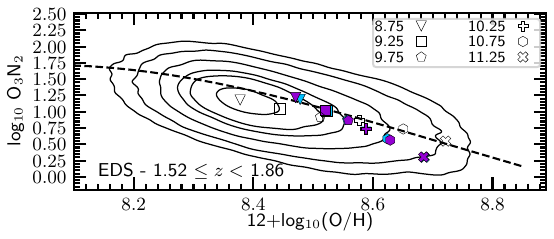}\\
\vspace{0.005mm}
\includegraphics[width=0.95\linewidth]{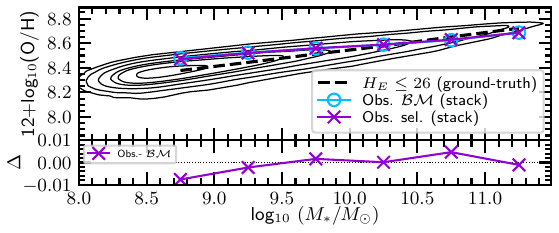}\\
\caption{The top panels show the O$_3$N$_2$ as a tracer of the gas-phase metallicity 12+$\logten$(O/H) parameter, in the redshift bin [1.52, 1.86]. 
The O$_3$N$_2$ values for stacked mock spectra in this study are calibrated using the \citet{Curti2017} relation. The layout of the panels is consistent with the bottom panel of Fig.\,\ref{fig:BPT_stack_DEEP}.
The bottom panels show the mass-metallicity relation (MZR), showing the dependence of gas-phase metallicity (inferred from O$_3$N$_2$) on stellar mass. The layout matches Fig.\,\ref{fig:line_vs_M_stack_DEEP}.}
\label{fig:Z_stack_DEEP}
\end{figure}

The results regarding the \edssf are shown in Fig.\,\ref{fig:Z_stack_DEEP}.
First, we observe an almost perfect agreement between the observationally selected samples (purple) and the corresponding benchmark samples (cyan), indicating that we can fully recover the information up to the instrument's limits.
Compared to the \HE $< 26$, our analysis reveals systematic trends in gas-phase metallicity with $M_*$, consistent with the mass-metallicity relation (MZR). 
At lower stellar masses, metallicities derived from O$_3$N$_2$ tend to be overestimated  in all the three redshift bins analysed, potentially due to biases in the calibration or contamination from emission-line ratios influenced by different ionization conditions. Conversely, at higher stellar masses (\(M_\star > 10^{10}M_\odot\)), metallicities are either consistent with expectations or slightly underestimated. 

\begin{figure}
\centering
\includegraphics[width=0.95\linewidth]{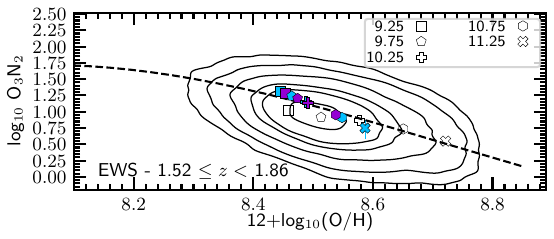}\\
\vspace{0.005mm}
\includegraphics[width=0.95\linewidth]{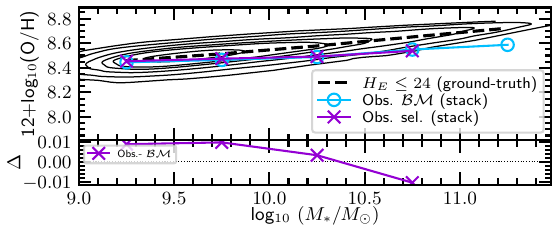}\\
\caption{The top panel show the O$_3$N$_2$ as a tracer of the gas-phase metallicity 12+$\logten$(O/H) parameter for the \ewssf. 
The O$_3$N$_2$ values for stacked mock spectra in this study are calibrated using the \citet{Curti2017} relation. 
The layout of the panels is consistent with the bottom panel of Fig.\,\ref{fig:BPT_stack_DEEP}. 
The bottom panel show the mass-metallicity relation (MZR), representing the dependence of gas-phase metallicity (inferred from O$_3$N$_2$) on stellar mass. The layout matches Fig.\,\ref{fig:line_vs_M_stack_DEEP}.}
    \label{fig:Z_stack_WIDE}
\end{figure}

\section{\label{sec:conclusions}Summary and conclusions}

In this preparatory paper for \Euclid's non-cosmological science, we have developed and tested a dedicated stacking pipeline tailored for for the \Euclid mission's NISP spectroscopic surveys, with the goal of recovering faint spectral features from large galaxy samples in the Euclid Wide and Deep Surveys (EWS, and EDS, respectively). 

To perform stacking, we released the dedicated code \stsp (see Sect.\,\ref{sec:method}) on the ESA Datalabs platform, accessible to the Euclid Consortium.  \stsp's modular design allows for flexible configurations tailored to diverse scientific objectives, while its computational efficiency supports large dataset processing, essential for \Euclid's extensive sky coverage (see Sect.\,\ref{sec:method}).
The method involves selecting suitable input spectra, aligning them to a common reference frame, applying normalization procedures chosen by the users among several possibilities, and combining the data statistically to produce high-quality composite spectra.

Our workflow has been applied to two sets of simulated spectra generated with the \mambo simulation, and processed to match \Euclid specifications using the \fssp code. The resulting datasets, \ewssf\ and \edssf, correspond to the EWS and EDS, respectively, and are designed to represent the final surveys quality, assuming that known systematics, such as zeroth-order residuals, persistence, and contamination from neighbors, are fully masked and corrected in future pipeline releases, and that flux losses are accounted for, achieving the fidelity expected from an optimal extraction of \textit{Euclid} spectra.

The analysis begins with redshift and flux measurements for each simulated spectrum using the automated fitting procedure of \Euclid pipeline described in Sect.\,\ref{sec:redshift}. We then performed tests aimed at characterising the best selection, to maximise performances in terms of retaining high success rate (\srate) whilst limiting contamination rate (\cont) for both samples.
Our results indicate that a residual percentage of redshift contaminants survive even after rigorous quality cuts (see Appendix~\ref{sec:selection_test}). This underscores the need for careful interpretation and additional refinement of selection methods when analysing stacked \Euclid spectra.
For the \edssf, we obtain a fiducial selection with \srate between $\sim \%$ and $\sim 70\%$ and \cont $< 6\%$, while for the fiducial \ewssf sample we find \srate between  $\sim 50\%$ and $\sim 80\%$ and \cont $<4.5\%$. At these levels, stacking delivers high-quality spectra of the benchmark (\bm) sample down to $\log_{10}(M_*/M_\odot) \simeq 8.5$ in the \edssf\ and $\log_{10}(M_*/M_\odot) \simeq 9.0$ in the \ewssf, allowing robust reconstruction of galaxy spectra and emission line fluxes and therefore the derivation of physical parameters such as star formation rate (SFR), dust attenuation, and gas-phase metallicity. The key to this performance is that noise contaminants have significantly lower measured \halpha fluxes than galaxies with accurate redshifts (Fig.\,\ref{fig:HaMeas_hist}), so their impact on the stacks is minimal.

We have investigated the nature of stacked spectra for different categories of redshift contaminants. Redshift contaminants arise from various sources, including spikes of noise misinterpreted as \halpha, as well as misidentifications of spectral lines such as \siiib and \oiii being erroneously attributed to \halpha. Stacking spectra consisting purely of noise spikes naturally reproduces the features of the template used for galaxy redshift measurements (Sect.\,\ref{sec:noise_spike}), confirming that such contaminants do contribute to the recovered line fluxes. However, their impact is not catastrophic, as they exhibit significantly lower measured \halpha fluxes than galaxies with accurate redshifts, as noted above.Contaminants produced by misclassified emission lines introduce spurious flux at the wavelengths of the line they were mistaken for, with the contamination amplitude scaling with the intrinsic strength of the misidentified line (a dependence that may differ between our simulations and real data), and, since the whole spectrum is shifted to the wrong redshift, additional emission features from those interloper spectra can produce further spurious signals in the stacked continuum and at other line wavelengths (see Sect.\,\ref{sec:misclas_lines}).  Misidentified-redshift contaminants, however, can be recognised in higher-order percentiles of the stack, where faint but coherent features appear at wavelengths corresponding to the misinterpreted lines (Sect.\,\ref{stack_mixture}). This could provide a useful diagnostic for their identification in real data.

By exploiting the clean benchmark stacks, we show that it is possible to recover fluxes for secondary lines such as \hbeta, \oiii, and \nii at levels well below the individual detection limit (Figs.~\ref{fig:line_vs_M_stack_DEEP}--\ref{fig:line_vs_M_stack_WIDE}). This capability can, in principle, be extended to even fainter lines, significantly expanding the range of physical diagnostics accessible through stacking.

We also quantify the important limitations due to selection effects. In particular, the flux-limited nature of the spectroscopic samples that prevents to measure the redshift at fainter fluxes, introduces biases that affect the recovery of global scaling relations, such as SFR–$M_*$ and attenuation–$M_*$. As demonstrated in Sect.\,\ref{sec:science}, these biases are not caused by the stacking technique itself, but are inherent to the input spectroscopic sample.
For the \edssf, the deeper flux limit enables a recovery of scaling relations that more closely match those of the parent sample, thanks to the higher completeness in redshift measurements, particularly at the high-mass end ($M_* \gtrsim 10^{10} \, M_\odot$), where the effects of incompleteness are minimized. 
In contrast, for the \ewssf, the selection restricts analysis to the bright upper envelope of such relations at all masses, preventing a full characterization of the underlying population. These findings underscore the need for caution when interpreting trends from flux-limited samples and emphasize the critical role of careful quality cuts and completeness assessments in deriving reliable conclusions from stacked spectra.

It is important to emphasise that these results are derived from controlled simulations designed to emulate the final \textit{Euclid} survey data, after full pipeline optimisation and removal of known artefacts. The mocks do not include truly passive galaxies or other line-poor systems, whose presence in real observations would likely increase contamination rates and amplify certain biases. Current Q1 \textit{Euclid} spectra still contain residual systematics from zeroth order, persistence, and contamination; however, these are expected to be substantially mitigated in forthcoming pipeline releases through improved extraction algorithms and more aggressive masking. Our conclusions should therefore be viewed as representative of the survey’s ultimate performance rather than of the current early data products.

\begin{acknowledgements}
\AckEC  
\AckDatalabs
S.Q., L.P., M.T., B.G., A.E., E.D., V.A., G.D.L., H.D. acknowledge support from the ELSA project. \enquote{ELSA: Euclid Legacy Science Advanced analysis tools} (Grant Agreement no. 101135203) is funded by the European Union. Views and opinions expressed are however those of the author(s) only and do not necessarily reflect those of the European Union or Innovate UK. Neither the European Union nor the granting authority can be held responsible for them. UK participation is funded through the UK HORIZON guarantee scheme under Innovate UK grant 10093177. 
V.A. and E.L. acknowledge the support from the INAF Large Grant \enquote{AGN \& Euclid: a close entanglement} Ob. Fu. 01.05.23.01.14.
C.S. acknowledges the support of NASA ROSES Grant 12-EUCLID11-0004
M.S. acknowledges support by the State Research Agency of the Spanish Ministry of Science and Innovation under the grants \enquote{Galaxy Evolution with Artificial Intelligence} (PGC2018-100852-A-I00) and \enquote{BASALT} (PID2021-126838NB-I00) and the Polish National Agency for Academic Exchange (Bekker grant BPN/BEK/2021/1/00298/DEC/1). This work was partially supported by the European Union's Horizon 2020 Research and Innovation program under the Maria Sklodowska-Curie grant agreement (No. 754510).
L.R. acknowledges support from the Next Generation EU funds within the National Recovery and Resilience Plan (PNRR), Mission 4 - Education and Research, Component 2 - From Research to Business (M4C2), Investment Line 3.1 - Strengthening and creation of Research Infrastructures, Project IR0000034 – “STILES - Strengthening the Italian Leadership in ELT and SKA”. 
\end{acknowledgements}

\bibliography{my, Euclid}

%
\begin{appendix}
  
\section{Additional notes on the stacking procedure}

\subsection{\label{app:flux_conservation} Test on the \enquote{flux conservation} method}

In Sect. \ref{sec:codestep1} we mentioned the limitations of the \enquote{flux conservation} method. For a more quantitative assessment, we conducted tests using toy RGS spectra consisting of a flat, noiseless continuum at $2\times10^{-18}$ \flcgsA (comparable to the average continuum in Fig.\,\ref{fig:mambo-fastsp_spec}) and a noiseless \halpha Gaussian line with a flux of $2\times10^{-16}$ \flcgs. Redshifts were randomly drawn from uniform distributions within various intervals in the range $0.9 \leq z \leq 1.8$, where \halpha falls in the \Euclid RGS. We then stacked these spectra at the sample average redshift and analysed the resulting composite spectra. We found that the \halpha luminosities in the stacked spectra deviated from the expected average by up to $10\%$, demonstrating the limitations of this approach for luminosity-sensitive analyses. 

\subsection{\label{app:geom_mean} Caveats on the geometric mean}
One of the spectral combination options offered by \stsp is the geometric mean, defined by
\begin{equation}
\langle F_\lambda\rangle = \left( \prod_{i=1}^{n} F_{\lambda, i} \right)^{1/n}\,,
\label{eq:geomean}
\end{equation}

\noindent where $F_{\lambda, i}$ is the flux density of the $i$-th spectrum in the bin centred on wavelength $\lambda$, and $n$ is the number of spectra contributing to the bin. 
\cite{VandenBerk2001} has shown that the geometric mean preserves the global continuum shape, which is useful, for example, when analysing spectra with power laws (often used to approximate quasar continua), whereas neither the median nor the mean typically yields a power law with the mean index.
Instead, the geometric mean correctly yields a composite power law whose spectral index is equal to the arithmetic mean of the frequency power-law indices.
Furthermore, according to the definition in Eq.\,\eqref{eq:geomean}, the geometric mean is less likely to be dominated by the brightest spectra than the arithmetic mean.
Therefore, in many astrophysical applications involving samples of galaxies with fluxes covering a large dynamical range, the geometric mean should be favoured.
However, \Euclid spectra, especially the EWS spectra, have a low S/N, and sometimes their flux values subtracted from the background are less than zero.
Therefore, the geometric mean cannot be used in these cases as it is only defined for positive values.

\subsection{\label{app:codestep3} \stsp performance evaluation}

For this \Euclid preparation paper, the \stsp code was run on simulated spectra to produce stacks for EWS and EDS spectra. 
In operational mode, the code will handle \Euclid's actual spectra, preferably read directly from the ESA Science Analysis System (SAS\footnote{The SAS is available at \url{https://eas.esac.esa.int/}.}), i.e., a software suite for processing and analysing mission data.  
However, the code is designed to work independently of the specific input data.
We also created and tested a Jupyter Notebook tutorial enabling users to prepare configuration files and source lists. These include selecting galaxy spectra for stacking by querying a provided dataset sample. Users can also run the code within the Jupyter Notebook, and they can see plots of the resulting stacked spectra directly in the Jupyter Notebook.

\begin{figure}
\centering
\includegraphics[width=0.85\linewidth]{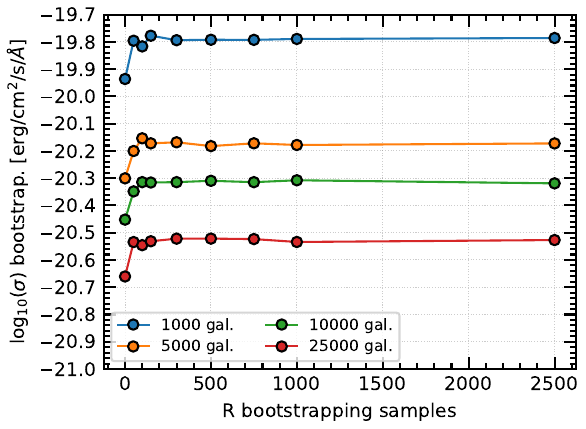}
 \caption{Uncertainties (at $1\sigma$ level) of the median stacked spectra as a function of number of bootstrap samples. Different curves and colours vary depending on the number of individual spectra in the stacking (\num{1000}, \num{5000}, \num{10000}, \num{25000} galaxies).}
    \label{fig:sig_vs_boots}
\end{figure}
Using the version v.3.1 of the code, we conducted performance tests to assess computing resource impact, varying code parameters, such as bootstrapping samples for uncertainty estimates, stacked spectra count, survey configurations, spectral normalisation type, and resampling. 
The code has three computationally intensive steps, controlled by used-defined parameters: 1) recursively reading galaxy spectra from input files for stacking; 2) pixel-scale operations (e.g., flux normalisation, resampling) on each spectrum; 3) bootstrapping resampling for estimating statistical uncertainties of stacked products. 
Through our tests we identified the parameter configurations for efficient runtime, without compromising statistical and scientific performance. 
We observed linear runtime scaling with the number of bootstrap samples and stacked spectra. However, the uncertainties converge around 200--300 bootstrap samples for any stacked galaxy count (see Figure\,\ref{fig:sig_vs_boots}).  For this reason, we cap the number of bootstrap samples at a maximum of 1000 iterations, issuing a warning to users that require a number of bootstrap samples exceeding this limit, to prevent runtime slowdowns.
The convergence test for the number of bootstrap samples will have to be replicated with real \Euclid’s spectra, though we do not expect a significant difference from the value obtained with our mock \mambo catalogue. 
Dedicated tests on the ESA Datalabs indicate that, using an environment configured for \stsp users, with 15 CPUs and 32GB of RAM, the stacking of 10\,000 \Euclid RGS spectra can be completed in under 10 minutes.

\section{\label{sec:selection_test}Sample selection for stacking}
In this section, we identify the \zprob threshold at a given flux limit that enables the selection of observationally defined samples for stacking, striking a balance between minimizing contamination (\cont) and maximizing success rate (\srate).
\begin{figure}
\centering
\includegraphics[width=0.95\linewidth]{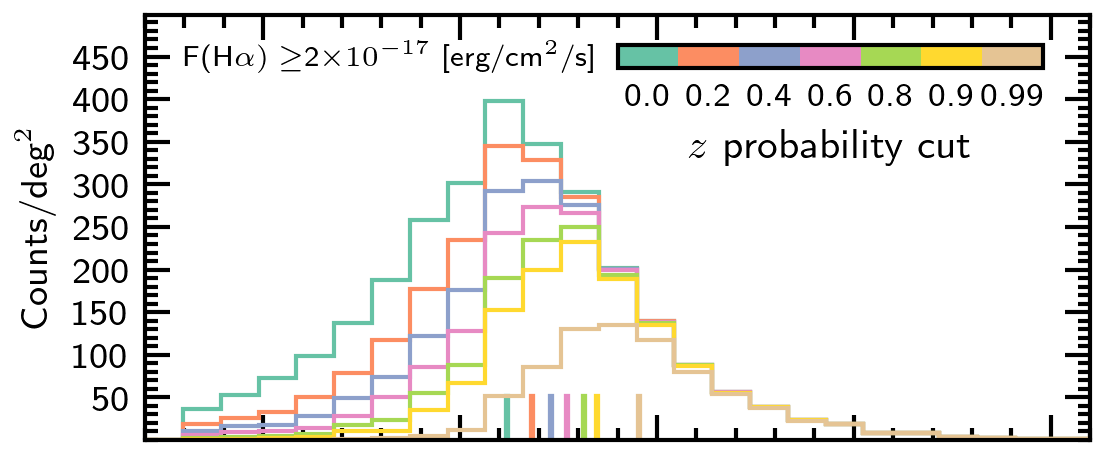}\\
\vspace{0.005mm}
\includegraphics[width=0.95\linewidth]{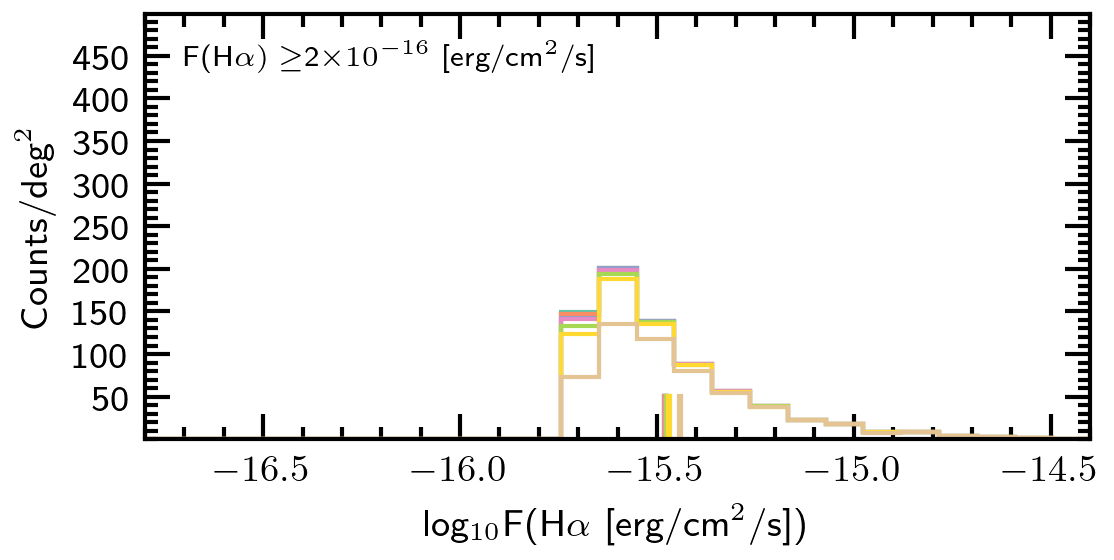}\\
\caption{Distributions of true \halpha flux for the mock \ewssf sample. The top panel shows galaxies with true \halpha flux $F$(\halpha) $\geq 2 \times 10^{-17}$ \flcgs, whereas the bottom panel shows galaxies with $F$(\halpha) $\geq 2 \times 10^{-16}$ \flcgs. }
    \label{fig:ha_distr_zprob}
\end{figure}

In Fig.\,\ref{fig:ha_distr_zprob} we show the variation of the distribution of \ewssf \halpha fluxes with increasing \zprob thresholds, to demonstrate which is the net effect of progressively increasing the \zprob threshold on the flux distribution. 
The top panel refers to the flux threshold $2 \times 10^{-17}$ \flcgs, whereas the bottom one is for $2 \times 10^{-16}$ \flcgs. 
In the top panel, we observe that higher \zprob cuts lead to net distributions at higher \halpha fluxes, progressively reducing the completeness at fainter fluxes. 
The purple distribution represents the distribution corresponding to the benchmark sample (no cut in \zprob). 
Then, as we progressively increase the \zprob threshold, we are filtering out fainter galaxies from the selection, resulting in a brighter median \halpha flux, and we expect to observe this effect also in the corresponding stacked spectra. 
This result demonstrates that our goal of probing fainter galaxy populations within the EWS through stacking is fundamentally constrained by the need to rigorously clean the sample of redshift contaminants. 
This process, whilst necessary, has the drawback of reducing the inclusion of the target population at fainter fluxes. 
As a result, attempting to push to deeper fluxes with stacking could become ineffective, as the selected sample is progressively reduced at fainter fluxes.
In contrast, the bottom panel of Fig.\,\ref{fig:ha_distr_zprob} shows that for the fiducial flux threshold of $2 \times 10^{-16}$ \flcgs, increasing the \zprob cut has a minimal impact on the \halpha distribution. Therefore, in this case we expect the quality of the stacked spectra across different \zprob cuts to depend primarily on the influence of residual redshift contaminants that are not removed by the selection cuts.

We analysed how \zprob thresholds and flux limits affect the success rate and contamination level in stacked spectra for the \ewssf. Galaxies were selected within $[z_{\rm min},z_{\rm max}]=[1.52, 1.86]$, where strong emission lines from \hbeta to \halpha lie within the red grism’s wavelength range. This selection ensures uniform coverage of key spectral lines. For stacking, we tested combinations of the following parameters
\begin{itemize}
    \item \zprob thresholds: {0, 0.2, 0.4, 0.6, 0.8, 0.9, 0.99}.
    \item Stellar mass (in bins of $\Delta$log$_{10} M_* = 0.5$): {[9, 9.5], [9.5, 10], [10, 10.5], [10.5, 11], [11, 11.5]}. Note that we did not consider higher stellar masses due to the limited number of galaxies with $M_* > 10^{11.5} \ M_\odot$ in the 3 deg$^2$ volume covered by our simulation.
    \item Flux thresholds for the \halpha line:  $2 \times 10^{-17}$ and $2 \times 10^{-16}$ \flcgs.
\end{itemize}
For each combination, a stacked spectrum was generated applying Eq.\,\eqref{eq:contam}. Stacking was performed using \stsp with the following process: individual spectra were shifted to $z = 1.68$ (the median redshift of the \ewssf sample), scaled for cosmological flux, and re-projected onto a uniform $0.5$\,nm wavelength grid. Median fluxes were then extracted at each pixel, with uncertainties estimated via $350$ bootstrap realizations. Key emission lines (\halpha, \nii, \oiii, \hbeta) were deblended and measured in the stacked spectra using the \texttt{SlineFit} code.

\begin{figure}
\centering
\includegraphics[width=0.95\linewidth]{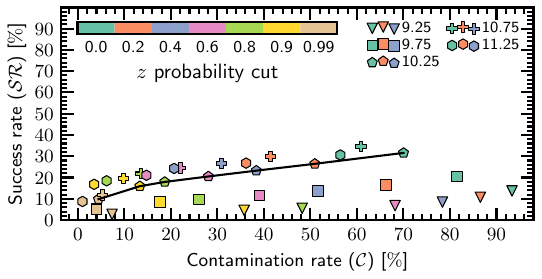}
\includegraphics[width=0.95\linewidth]{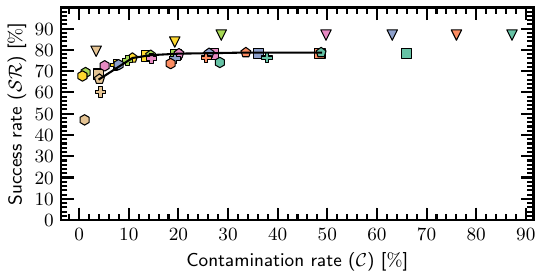}\\
\caption{Success rate (\srate) vs. contamination rate (\cont) for the mock \ewssf, colour-coded by the applied redshift probability threshold (\zprob). The top panel shows results for a flux threshold of $2 \times 10^{-17}$\flcgs, while the bottom panel corresponds to a higher threshold of $2 \times 10^{-16}$\flcgs. Different symbols represent galaxies in distinct stellar mass bins. As a reference, the black curves connect the points corresponding to the mass bin $\logten(M_*/M_\odot)\in [10, 10.5]$.}
\label{fig:srate_vs_cont}
\end{figure}

\begin{figure}
\centering
\includegraphics[width=0.95\linewidth]{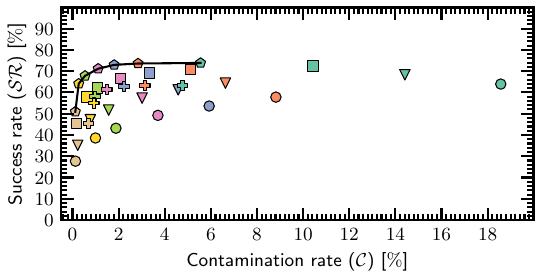}
\includegraphics[width=0.95\linewidth]{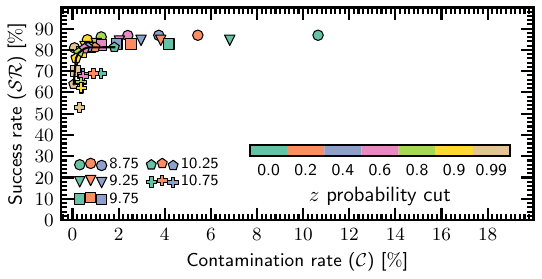}\\
\caption{Success rate (\srate) vs. contamination rate (\cont) for the mock \edssf, colour-coded by the applied redshift probability threshold (\zprob). The top panel shows results for a flux threshold of $2 \times 10^{-17}$\flcgs, while the bottom panel corresponds to a higher threshold of $5 \times 10^{-17}$\flcgs. The layout is the same as in Fig.\,~\protect\ref{fig:srate_vs_cont}.}
\label{fig:srate_vs_cont_EDS}
\end{figure}

Figure\,\ref{fig:srate_vs_cont} presents the success rate (\srate) as a function of contamination (\cont), colour-coded by the redshift probability threshold (\zprob), for the \ewssf. Top panels correspond to a flux limit of $2 \times 10^{-17}$ \flcgs; bottom panels show the fiducial case at $2 \times 10^{-16}$ \flcgs.

At the lowest flux limit, \srate remains below 37\% in all stellar mass bins, declining to 12\% in the lowest bin, regardless of the applied \zprob cut. As anticipated from Fig.\,\ref{fig:ha_distr_zprob}, higher \zprob thresholds exclude low-S/N galaxies, reducing \cont but also removing intrinsically fainter \halpha emitters and lowering \srate. This unfavourable trade-off makes this flux limit unsuitable for robust stacking: high-\zprob cuts lead to biased samples dominated by bright emitters, while low-\zprob cuts yield excessive contamination. Consequently, we do not recommend using a $2 \times 10^{-17}$ \flcgs limit for stacking purposes.

In contrast, the fiducial $2 \times 10^{-16}$ \flcgs sample shows improved performance. At \zprob = 0, \srate exceeds 70\% across all mass bins, albeit with contamination between 30-70\%. Increasing \zprob progressively lowers \cont while preserving high \srate up to \zprob = 0.9. For instance, in the [10, 10.5] bin, \cont drops from 50\% to 10\% while maintaining \srate at roughly 80\%.  Beyond \zprob = 0.9, the contamination continues to decrease, albeit more slowly, while the \srate also declines. We adopt \zprob = 0.99 to balance contamination control and completeness, limiting contamination to below 5\% while retaining a success rate of roughly 50–77\%.
. Intermediate flux thresholds ($5 \times 10^{-17}$ and $1 \times 10^{-16}$ \flcgs) perform better than the lowest limit but fall short of the efficiency achieved at the fiducial value and are not discussed further.

In summary, we adopt a selection based on \zprob $\geq 0.99$ and a flux limit of $2 \times 10^{-16}$ \flcgs in the main analysis. This combination ensures high completeness with minimal contamination, preserving the fidelity of stacked spectral measurements (see Sect.\,\ref{sec:science}).

We applied the same framework to the \edssf sample, which benefits from deeper flux sensitivity and extended spectral coverage via the blue grism, enabling the analysis of a broader redshift interval ($0.9 \leq z \leq 1.86$). To minimize evolutionary effects, we divided the sample into three redshift bins: [0.9, 1.2], [1.2, 1.52], and [1.52, 1.86], with the latter overlapping the \ewssf range. The selection optimization tests were conducted in this uppermost redshift bin.

We used the same grid of parameters as in the \ewssf analysis, but extended the stellar mass range down to $\logten(M_*/M_\odot)\in [8.5, 9]$, enabled by the lower flux threshold of $5 \times 10^{-17}$ \flcgs. Figure\,\ref{fig:srate_vs_cont_EDS} summarizes the results. The trends are qualitatively similar to the \ewssf case but show substantially better performance: at $2 \times 10^{-17}$ \flcgs, \cont remains below 18\% across all \zprob cuts, and \srate exceeds 80\%. At $5 \times 10^{-17}$ \flcgs, success rates reach 90\%. This robustness makes selecting a single criterion less critical.

Nevertheless, to ensure consistency with the benchmark population, we adopt a flux limit of $2 \times 10^{-17}$ \flcgs and a \zprob threshold of 0.4 for the \edssf sample. This choice provides the best balance between completeness and contamination across stellar mass bins.

We caution, however, that although simulations remain robust down to $2 \times 10^{-17}$ \flcgs, actual galaxies with \HE $< 26$ may fall below this sensitivity, potentially limiting redshift recovery in the real survey. As such, our contamination estimates may be slightly optimistic, and follow-up validation using real \Euclid data will be essential.

In conclusion, based on the selection criteria, stacked spectra for the \edssf were generated with an \halpha flux limit $\geq 2\times10^{-17}$ \flcgs and a redshift probability threshold of \zprob$\geq 0.4$, whilst for the \ewssf, stacked spectra were created with a stricter flux limit in \halpha of $\geq 2\times10^{-16}$ \flcgs and a redshift probability threshold of \zprob$\geq 0.99$.

\section{\label{app:scaling_relation} Scaling relations for EDS-SF$_\text{sim}$, in other redshift bins}
\begin{figure}
\centering
\includegraphics[width=\linewidth]{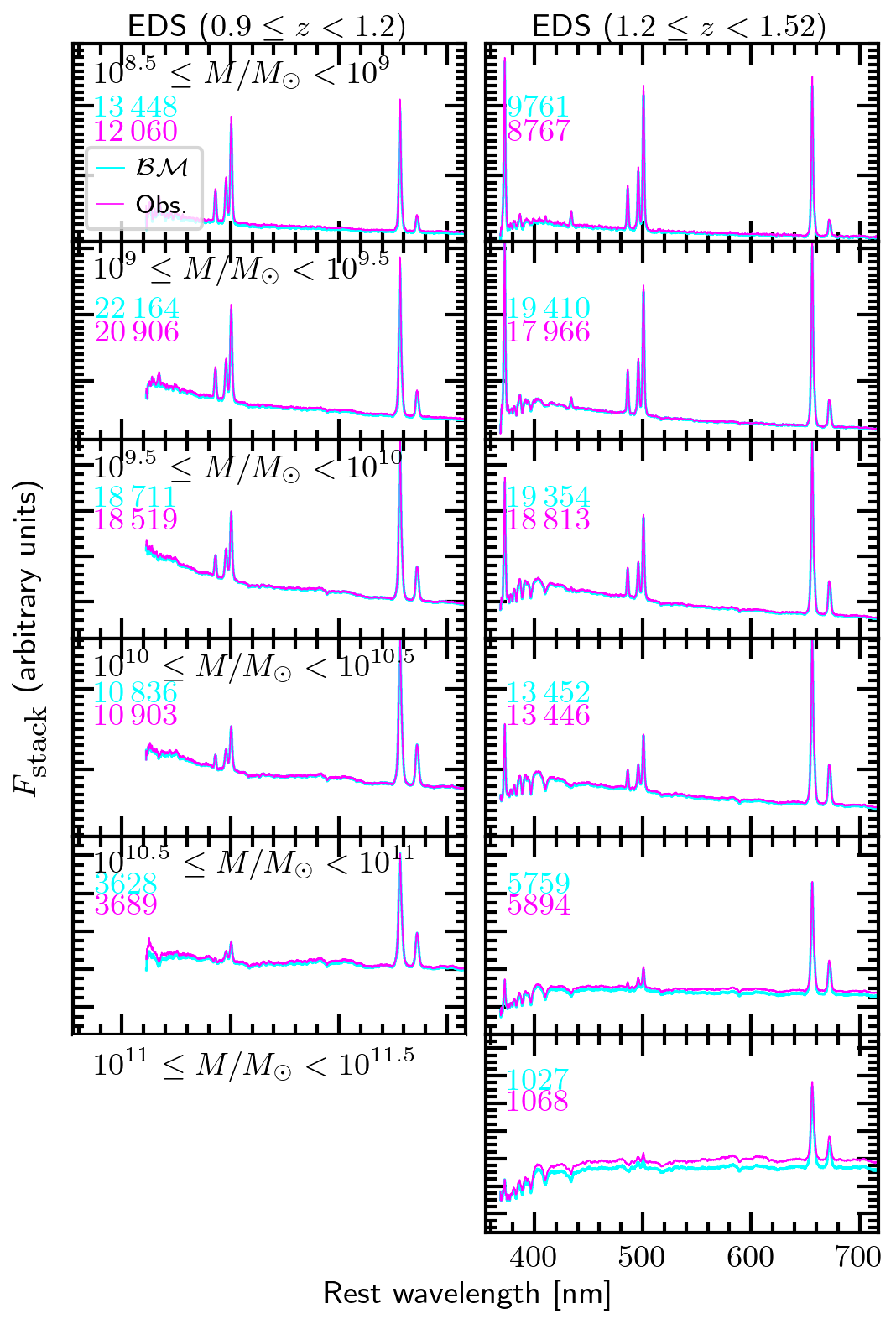}\\
\caption{Stacked mean spectra for the mock \edssf galaxies used for the scientific predictions in this paper, for the redshift bins $[0.9, 1.2]$ (left panels), and $[1.2,  1.52]$ (right panels).
The layout and notes are the same as in Fig.\,\ref{fig:stack_specs}.}
\label{fig:stack_specs_other_z}
\end{figure}
\begin{figure}
\centering
\includegraphics[width=\linewidth]{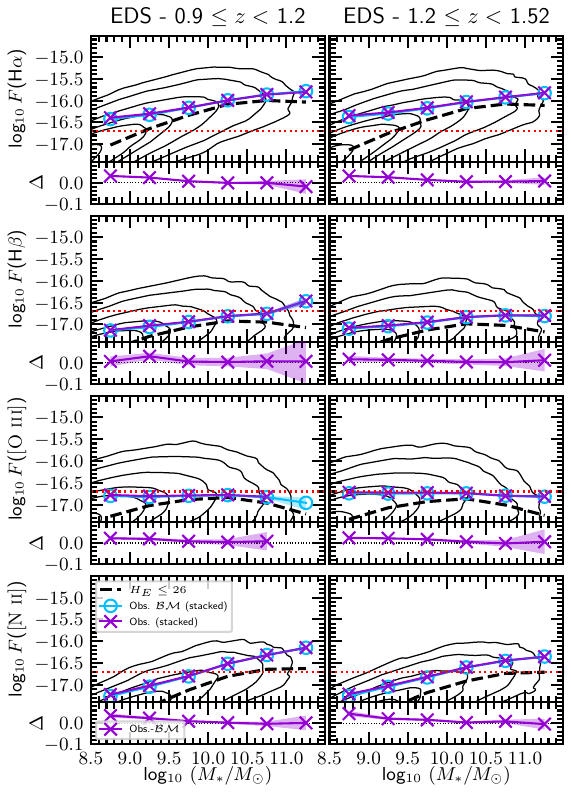}\\
\caption{Emission line fluxes (in units of \flcgs) measured on stacked mock spectra as a function of stellar mass for the \edssf selection corresponding to the redshift intervals $0.9 \leq z \leq 1.2$ (left panels), and $1.2 \leq z \leq 1.52$ (right panels). From top to bottom, panels show the fluxes of \halpha, \hbeta, \oiii, and \nii emission lines. Fluxes have been corrected for flux loss using the method by Cassata et al. (in preparation). The layout is the same as in Fig.\,\ref{fig:line_vs_M_stack_DEEP}.
}
\label{fig:line_vs_M_stack_DEEP_other_z}
\end{figure}
\begin{figure}
\centering
\includegraphics[width=0.95\linewidth]{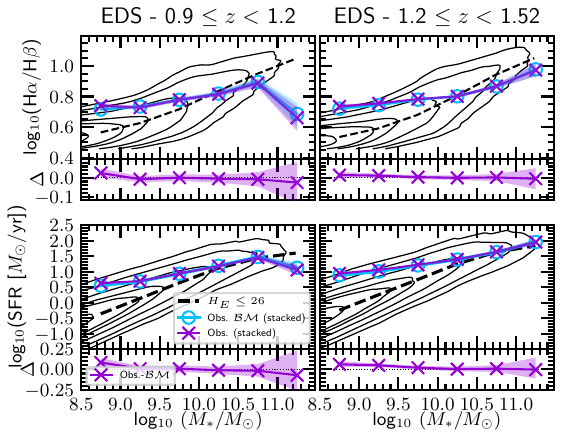}\\
\caption{The attenuation–$M_*$ relation (top panels) and the SFR–$M_*$ relation (bottom panels) derived from the \edssf stacked mock spectra for the \edssf selection corresponding to the redshift intervals $0.9 \leq z \leq 1.2$ (left panels), and $1.2 \leq z \leq 1.52$ (right panels). The attenuation is measured using the Balmer decrement (\halpha/\hbeta ratio), while the SFR is calculated from the dust-corrected \halpha luminosity, applying the \cite{Calzetti2000} extinction law and assuming case B recombination conditions (electron density of 100 cm$^{-2}$ and $10\,000$ K. 
The panel layout follows the structure of Fig.\,\ref{fig:line_vs_M_stack_DEEP}. Shaded regions represent the $1\sigma$ uncertainties in the measured quantities.
}
\label{fig:SFR_vs_M_stack_DEEP_other_z}
\end{figure}
\begin{figure}
\centering
\includegraphics[width=0.95\linewidth]{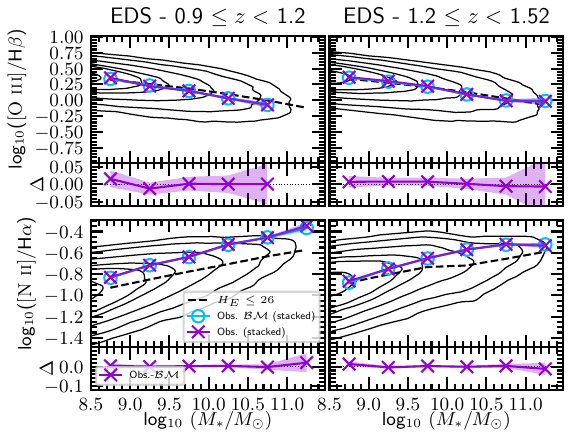}\\
\vspace{-0.9mm}
\hspace{2mm}
\includegraphics[width=0.93\linewidth]{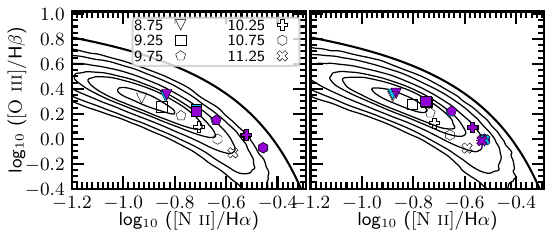}
\caption{The BPT diagram and corresponding stellar mass relations for the \oiii/\hbeta and \niib/\halpha ratios based on the \edssf stacked mock spectra. The top and middle panels show the \oiii/\hbeta–stellar mass \citep[i.e., the MEx diagnostic diagram,][]{Juneau2011}, and \niib/\halpha–stellar mass relations, respectively, with measurements from the stacked spectra in different stellar mass bins. The bottom panel displays the BPT diagram with symbols representing the positions of the stacked spectra for various mass bins. The black contours in all panels represent the \edssf parent sample cut at \HE $\leq$ 26, and the black dashed curves in the top and middle panels represent its median relations.
}
\label{fig:BPT_stack_DEEP_other_z}
\end{figure}
\begin{figure}
\centering
\includegraphics[width=\linewidth]{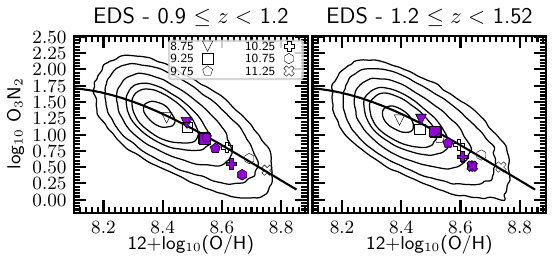}\\
\vspace{0.005mm}
\includegraphics[width=\linewidth]{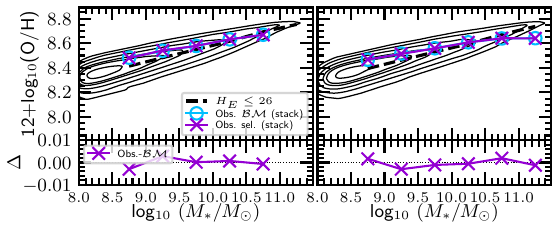}\\
\caption{The top panels show the O$_3$N$_2$ as a tracer of the gas-phase metallicity 12+$\logten$(O/H) parameter, in the redshift bins [0.9, 1.2], and [1.2, 1.52]. 
The O$_3$N$_2$ values for stacked mock spectra in this study are calibrated using the \citet{Curti2017} relation. The layout of the panels is consistent with the bottom panel of Fig.\,\ref{fig:BPT_stack_DEEP}.
The bottom panels show the mass-metallicity relation (MZR), showing the dependence of gas-phase metallicity (inferred from O$_3$N$_2$) on stellar mass. The layout matches Fig.\,\ref{fig:line_vs_M_stack_DEEP}.}
\label{fig:Z_stack_DEEP_other_z}
\end{figure}

In this appendix, we present the scaling relations derived from the \edssf stacked spectra in the lower redshift bins $0.9 \leq z \leq 1.2$ and $1.2 \leq z \leq 1.52$. These results are qualitatively consistent with those discussed in Sect.\,\ref{scaling_relations} for the highest redshift bin ($1.52 \leq z \leq 1.86$).
The emission line fluxes (Fig.\,\ref{fig:line_vs_M_stack_DEEP_other_z}), measured from the stacked mock spectra shown in Fig.\,\ref{fig:stack_specs_other_z}, along with derived physical quantities such as attenuation, the star-formation rate–stellar mass relation (Fig.\,\ref{fig:SFR_vs_M_stack_DEEP_other_z}), the BPT diagram (Fig.\,\ref{fig:BPT_stack_DEEP_other_z}), and gas-phase metallicity (Fig.\,\ref{fig:Z_stack_DEEP_other_z}), exhibit the same overall behaviour, with only modest variations attributable to the cosmological evolution encoded in the \mambo simulation. In particular, the fluxes of key emission lines (e.g., \halpha, \oiii, \nii) measured from stacked spectra track the corresponding benchmark samples with comparable accuracy, and the main sequence relations maintain their characteristic shapes and slopes.
We note slight differences in normalization and scatter that reflect the gradual evolution of galaxy properties over cosmic time, as expected from both simulations and observations. These differences remain well within the uncertainties and do not alter the qualitative conclusions drawn for the higher redshift bin.

\end{appendix}

\end{document}